\documentclass[11pt,a4paper]{article}
\usepackage{amsmath,scalerel, amsfonts,amssymb,mathtools,nicefrac,nccmath,cases}
\usepackage{mathrsfs}
\usepackage{microtype,booktabs,multirow}
\usepackage[usenames,dvipsnames,table]{xcolor}
\usepackage{dsfont}
\usepackage[shortlabels]{enumitem}
\usepackage[normalem]{ulem}
\usepackage{colortbl}
\usepackage{tikz}
\usepackage{verbatim}
\usetikzlibrary{shapes,arrows}
\usetikzlibrary{calc,shapes.callouts,shapes.arrows,bending,intersections}
\usetikzlibrary{shapes.geometric}
\usetikzlibrary{arrows.meta,arrows}
\usetikzlibrary{decorations.pathmorphing}
\usetikzlibrary{decorations.markings}
\usetikzlibrary{shapes.multipart}
\usetikzlibrary{tikzmark}
\usepackage{arydshln}
\usepackage{caption}
\usepackage{subcaption}
\usepackage{bm}

\usepackage[nosort]{cite}
\usepackage{colortbl}
\usepackage{amsthm}
\usepackage{soul}
\usepackage{bbm}
\usepackage[normalem]{ulem}
\usepackage{tikz}
\usetikzlibrary{3d,arrows.meta,decorations.pathmorphing,positioning}

\usepackage{svg}

\usepackage{tocloft}

\usepackage{xcolor}
\usepackage{lipsum}

\usepackage[percent]{overpic}

\usepackage[linktocpage=true]{hyperref}
\hypersetup{colorlinks=true,linkcolor=nicecolor,citecolor=nicecolor,urlcolor=nicecolor}
\definecolor{nicecolor}{rgb}{0.1, 0.3, 0.4}

\definecolor{blue}{rgb}{0.06, 0.3, 0.57}
\definecolor{Gray}{gray}{0.4}

\definecolor{nicecolor}{rgb}{0.1, 0.3, 0.4}
\definecolor{blue}{rgb}{0.06, 0.3, 0.57}
\definecolor{Gray}{gray}{0.4}
\colorlet{tableheadcolor}{gray!15} % Table header colour = 25% gray
\colorlet{tablerowcolor}{gray!7} % Table row separator colour = 10% gray

\def\hybrid{\topmargin -20pt    \oddsidemargin 0pt
	\headheight 0pt \headsep 0pt
	\textwidth 6.5in        % US paper
	\textheight 9in         % US paper
	\textwidth 6.25in       % A4 paper
	\textheight 9 in       % A4 paper
	\marginparwidth .875in
	\parskip 5pt plus 1pt 
	\jot = 1.5ex
}
\hybrid
\numberwithin{equation}{section}
\numberwithin{table}{section}
\usepackage{float}
\usepackage{tabularx}
\usepackage{multirow}
\usepackage{hhline}
\usepackage{stackengine}

\newcolumntype{D}{>{\centering\arraybackslash}X}
\newcolumntype{L}{>{$}l<{$}}
\newcolumntype{R}{>{$}r<{$}}
\newcolumntype{C}{>{$}c<{$}}
\usepackage{IEEEtrantools}
\newcommand{\beq}{\begin{equation}}  \newcommand{\eeq}{\end{equation}}
\newcommand{\bal}{\begin{aligned}}   \newcommand{\eal}{\end{aligned}}
\newcommand{\bea}{\begin{eqnarray}}  \newcommand{\eea}{\end{eqnarray}}
\def\beqa{\begin{eqnarray}}
\def\eeqa{\end{eqnarray}}

\newcommand{\bmat}{\left(\begin{array}}
\newcommand{\emat}{\end{array}\right)}

\usepackage{cancel}

\usepackage[section]{placeins} % TO FORCE FIGURES IN ASSOCIATED SECTIONS

\newcommand{\be}{\begin{equation}}
\newcommand{\ee}{\end{equation}}

\newcommand{\bbZ}{\mathbb{Z}}
\newcommand{\bbQ}{\mathbb{Q}}

\definecolor{goodgreen}{RGB}{0,158,115}

\definecolor{Gray}{gray}{0.95}
\definecolor{darkspringgreen}{rgb}{0.09, 0.45, 0.27}
\definecolor{darkseagreen}{rgb}{0.56, 0.74, 0.56}
\definecolor{darkmouthgreen}{rgb}{0.05, 0.5, 0.06}
\definecolor{darkcyan}{rgb}{0.0, 0.55, 0.55}

\def\d {{\rm d}}
\def\del          {\partial}

\def\ii           {{\rm i}}

\def\tr           {\mathop{\rm tr}}

\def\Im           {{\rm Im\hskip0.1em}}

\def\cala         {{\cal A}}

\def\calb         {{\cal B}}
\def\calc         {{\cal C}}
\def\cald         {{\cal D}}

\def\calf         {{\cal F}}
\def\calg         {{\cal G}}

\def\calh         {{\cal H}}
\def\cali         {{\cal I}}
\def\calj         {{\cal J}}

\def\calm         {{\cal M}}
\def\caln         {{\cal N}}
\def\calo         {{\cal O}}
\def\calp         {{\cal P}}
\def\calq         {{\cal Q}}
\def\calr         {{\cal R}}

\def\calt         {{\cal T}}

\def\calv         {{\cal V}}

\def\sfC         {\mathsf{C}}

\def\sfK         {\mathsf{K}}
\def\sfL         {\mathsf{L}}
\def\sfM         {\mathsf{M}}
\def\sfN         {\mathsf{N}}

\def\sfP         {\mathsf{P}}
\def\sfQ         {\mathsf{Q}}

\def\sfT         {\mathsf{T}}

\def\sfX         {\mathsf{X}}
\def\sfY         {\mathsf{Y}}

\def\sff         {\mathsf{f}}

\def\Pl      {{\text{\tiny P}}}

\def\lg           {\left\{}
\def\lq           {\left[}
\def\lt           {\left(}

\def\rg           {\right\}}
\def\rq           {\right]}
\def\rt           {\right)}

\usepackage{xcolor}
\definecolor{colorloc1}{RGB}{0,0,102}  
\definecolor{colorloc2}{RGB}{0,125,253} 
\usepackage[framemethod=default]{mdframed}
\newmdenv[skipabove=10pt,
skipbelow=7pt,
rightline=false,
leftline=true,
topline=false,
bottomline=false,
linecolor=colorloc1,
backgroundcolor=colorloc2!5,
innerleftmargin=4pt,
innerrightmargin=0pt,
innertopmargin=0pt,
leftmargin=2pt,
rightmargin=0pt,
linewidth=2pt,
innerbottommargin=0pt]{lbBox}

\begin{document}

\baselineskip=14pt
\parskip 5pt plus 1pt

\vspace*{-1.5cm}
\begin{flushright}    % Publication numbers
  {\small 
 % IFT-UAM/CSIC-21-36
  }
\end{flushright}

\vspace{2cm}
\begin{center}        % Main title

\textbf{\textsc{\huge 
Non-invertible symmetries
in the axiverse, and the imaginary wormholes
%\\\smallskip and the species scale
%Wormholes and the species scale \\\smallskip in the axiverse
}}\\[.3cm]

\end{center}

\vspace{0.5cm}
\begin{center}        % Authors
%{\large   Luca Martucci}
{\large  Daniele Licciardello \& Luca Martucci }
\end{center}

\vspace{0.15cm}
\begin{center}  
%${}^1$ 
\emph{Dipartimento di Fisica e Astronomia ``Galileo Galilei",  Universit\`a di Padova,} \\ 
%${}^2$
\emph{\& INFN Sezione di Padova, Via F. Marzolo 8, 35131 Padova, Italy}
%\\${}^3$\emph{Department of Physics, University of Basel, Klingelbergstrasse 82, CH-4056 Basel, Switzerland}
\\
%[2mm]${}^4$\emph{Hamburg}\\[.3cm]
\end{center}

\vspace{2cm}

%%%%%%%%%%%%%%%%%%%%%%%%%%%%%%%%%%%%%%%%%%%%%%%
%%%%%%%%%%%%%%%%%%%%%%%%%%%%%%%%%%%%%%%%%%%%%%%
%%%%%%%%%%%%%%%%%%%%%%%%%%%%%%%%%%%%%%%%%%%%%%%
%%%%%%%%%%%%%%%%%%%%%%%%%%%%%%%%%%%%%%%%%%%%%%%
%%%%%%%%%%%%%%%%%%%%%%%%%%%%%%%%%%%%%%%%%%%%%%%
%%%%%%%%%%%%%%%%%%%%%%%%%%%%%%%%%%%%%%%%%%%%%%%
%%%%%%%%%%%%%%%%%%%%%%%%%%%%%%%%%%%%%%%%%%%%%%%
%%%%%%%%%%%%%%%%%%%%%%%%%%%%%%%%%%%%%%%%%%%%%%%

\begin{abstract}

\noindent We study the symmetry structure of four-dimensional axiverse effective field theories with multiple axions coupled to abelian gauge sectors, including their extensions to broad classes of \({\cal N}=1\) models. We identify the invertible and non-invertible generalized symmetries, and discuss the associated symmetry-breaking mechanisms together with the resulting hierarchies of energy scales. In particular, we discuss the quantum-gravitational breaking of non-invertible axion shift symmetries predicted by the existence of wormholes and  the recently proposed Imaginary Distance Bound. In \({\cal N}=1\) axiverses, these wormhole-based arguments imply that towers of BPS EFT instantons play a distinguished role and generate infinitely many superpotential terms.

\end{abstract}

\thispagestyle{empty}
\clearpage

\setcounter{page}{1}

%%%%%%%%%%%%%%%%%%%%%%%%%%%%%%%%%%%%%%%%%%%%%%%
%%%%%%%%%%%%%%%%%%%%%%%%%%%%%%%%%%%%%%%%%%%%%%%
%%%%%%%%%%%                 %%%%%%%%%%%%%%%%%%%
%%%%%%%%%%%  DOCUMENT BODY  %%%%%%%%%%%%%%%%%%%
%%%%%%%%%%%                 %%%%%%%%%%%%%%%%%%%
%%%%%%%%%%%%%%%%%%%%%%%%%%%%%%%%%%%%%%%%%%%%%%%
%%%%%%%%%%%%%%%%%%%%%%%%%%%%%%%%%%%%%%%%%%%%%%%
%%%%%%%%%%%%%%%%%%%%%%%%%%%%%%%%%%%%%%%%%%%%%%%

\newpage

  \tableofcontents

\section{Introduction}

From a bottom-up viewpoint, light axion-like fields are motivated by several important questions in particle phenomenology. In particular, they furnish an appealing mechanism for solving the strong CP problem \cite{Peccei:1977hh,Weinberg:1977ma,Wilczek:1977pj}, have long been considered viable dark matter candidates \cite{Hu:2000ke}, and may also participate in inflationary dynamics \cite{Frieman:1995pm}.  
It has also been understood since the early 1980s that such light degrees of freedom arise naturally in string theory constructions; see \cite{Svrcek:2006yi} for a review. Indeed, the number $n_{\rm A}$ of axions present in the low-energy theory can easily be of order hundreds, or even thousands. Models with $n_{\rm A}\gg 1$ axions therefore exhibit a very rich phenomenology and are commonly referred to as the axiverse \cite{Arvanitaki:2009fg}; see  \cite{McAllister:2023vgy} for a recent review.

Before incorporating effects associated with a UV quantum gravity completion, it is useful to analyze axiverse models as low-energy effective field theories in their own right. A natural first step is to identify their structural properties and, in this respect, global symmetries provide particularly valuable information, as they strongly constrain the dynamics. Of course, global symmetries are expected to be broken, and hence can at most be regarded as approximate, once the theory is embedded into quantum gravity \cite{Misner:1957mt,Hawking:1982dj,Banks:2010zn,Harlow:2018tng}; see also \cite{Harlow:2022gzl,Reece:2023czb}. Within the Swampland program \cite{Vafa:2005ui,Ooguri:2006in}, reviewed for instance in \cite{Brennan:2017rbf,Palti:2019pca,vanBeest:2021lhn,Grana:2021zvf,Harlow:2022gzl,Agmon:2022thq,VanRiet:2023pnx,Reece:2023czb}, this expectation is usually formulated as the ``No Global Symmetry Conjecture''. Nevertheless, the emergence of global symmetries in suitable limits can provide a useful organizing principle for understanding the dynamics in those regimes, as well as the possible symmetry-breaking mechanisms.

A particularly powerful framework for this analysis is the generalized notion of symmetry introduced in \cite{Gaiotto:2014kfa}, in which global symmetries are encoded by extended topological operators. In this formulation, codimension-$(p+1)$ topological defects generate $p$-form symmetries, with ordinary global symmetries corresponding to $p=0$. This perspective has led to a much broader understanding of the symmetry structures in quantum field theory.  
Among the most important examples are non-invertible, or categorical, symmetries, generated by topological defects that do not possess inverses under fusion. These symmetries have been intensely studied in recent years; for reviews and further references, see \cite{Cordova:2022ruw,McGreevy:2022oyu,Freed:2022iao,Gomes:2023ahz,Schafer-Nameki:2023jdn,Brennan:2023mmt,Bhardwaj:2023kri,Shao:2023gho,Luo:2023ive,Carqueville:2023jhb,Costa:2024wks}. Their role in quantum gravity settings has also been investigated in several works, including \cite{Rudelius:2020orz,Heidenreich:2021xpr,McNamara:2021cuo,Cordova:2022rer,Heckman:2024obe,Rudelius:2024vmc,Basile:2025zjc,Gagliano:2025oqv,Apruzzi:2025byj}.

The goal of this paper is two-fold. First, we determine the maximal global symmetry structure of general four-dimensional axiverse models with no supersymmetry or minimal supersymmetry, initially leaving aside quantum gravity constraints. We focus in particular on continuous invertible symmetries and on quasi-continuous non-invertible ones, whose parameters can take arbitrarily dense rational values, since these impose the strongest constraints. In this context, quasi-continuous axion shift symmetries are especially important: even when ordinary invertible axion shift symmetries are absent, they can forbid the generation of a non-trivial axion potential and account for its natural smallness, as in \cite{Cordova:2022ieu,Choi:2022fgx}.

Our second goal is to begin a systematic analysis of quantum-gravitational symmetry breaking mechanisms. In particular, we study the consequences for axiverse models of the recent perspective on axion wormholes proposed in \cite{DiUbaldo:2026rly,Maldacena:2026jqd} and of the corresponding {\em Imaginary Distance Bound}.

We pursue the first objective by considering general axiverse theories containing an arbitrary number of axions together with an arbitrary number of abelian U(1) gauge fields. We also include Gauss-Bonnet and Pontryagin curvature-squared couplings, and comment on the possible effects of non-abelian gauge sectors and additional matter sectors. Our results build on and extend earlier work on the axion-Maxwell theory \cite{Choi:2022jqy,Cordova:2022ieu,Choi:2022fgx,Yokokura:2022alv}, making use of the framework developed for more general Gaillard–Zumino models in \cite{Apruzzi:2025byj}; see also \cite{Hong:2025qbw} for related work. We will provide a general construction of the relevant non-invertible topological operators. As in \cite{Choi:2022fgx}, see also \cite{Brennan:2020ehu,Sehayek:2026pvu}, this construction makes manifest their interplay with invertible higher-form symmetries and leads to hierarchies among the corresponding symmetry breaking scales.

As mentioned above, our discussion of symmetry-breaking effects induced by quantum gravity will significantly rely on the novel viewpoint on axion wormholes proposed in \cite{DiUbaldo:2026rly,Maldacena:2026jqd}. In the axiverse theories considered here, both invertible and non-invertible axion shift symmetries are tied to the presence of wormholes connecting two asymptotically flat regions, of the kind first constructed in \cite{Giddings:1987cg}. Such configurations may be reinterpreted as wormhole saddles with imaginary axion profiles, interpolating between imaginary Dirichlet boundary conditions imposed at the two asymptotic ends. The same interpretation extends to more general wormhole solutions \cite{Arkani-Hamed:2007cpn}; in what follows we will refer to this type of configurations as {\em imaginary wormholes}. The papers  \cite{DiUbaldo:2026rly,Maldacena:2026jqd}, providing complementary arguments, propose that the appearance of these saddles signals a breakdown of the original effective description and therefore points to the need for its modification, which in particular breaks the axion shift symmetries. More precisely, the total  distance between the asymptotic wormhole boundary conditions sets the Imaginary Distance Bound (IDB), below which the analytic continuation of the symmetry-breaking corrections must become relevant.

In this paper we assume the validity of the viewpoint proposed in \cite{DiUbaldo:2026rly,Maldacena:2026jqd} and explore its consequences for our axiverse models. As we will see, the strongest results can be obtained within the $\caln=1$ axiverse framework introduced in \cite{Lanza:2020qmt,Lanza:2021udy,Lanza:2022zyg}, which captures broad classes of string compactifications. In particular, a distinguished role in realizing the IDB will be played by a special class of BPS fundamental instantons, the {\em EFT instantons} as defined in \cite{Lanza:2021udy,Martucci:2024trp}. 

The paper is organized as follows. In Section~\ref{sec:aximodels0} we introduce the general four-dimensional axiverse effective field theories studied in this work, including their $\caln=1$ extension. In Section~\ref{sec:noninvsymm} we determine their invertible and non-invertible global symmetries, construct the corresponding topological operators, and describe the associated charged operators. In Sections~\ref{sec:physcons} and \ref{sec:axionat} we discuss the possible symmetry-breaking mechanisms and the corresponding hierarchies. In Section~\ref{sec:WHs} we discuss the implications of the  viewpoint proposed in \cite{DiUbaldo:2026rly,Maldacena:2026jqd} and the associated Imaginary Distance Bound. Finally, Section~\ref{sec:conclusions} contains our conclusions and possible future directions.

%%%%%%%%%%%%%%%%%%%%%%%%%%%%%%%%%%%%%%%%%%%%%%%%%%%%%%%%%%%%%%%%%%%%%%%%%%%%%%%%%%%%%%%%%%%%%%%%%%%%%%%%%%%%%%%%%%%%%%%%%%%%%%%%%%%%

\section{Axiverse models}
\label{sec:aximodels0}

In this section we will describe the axiverse effective field theories (EFTs) on which we will base our discussion. We will first consider non-supersymmetric models, and later describe the relevant features of their minimal supersymmetric extensions. In this section, we will not include any effective potential, or superpotential, for the axion sector. As we will see in Section \ref{sec:noninvsymm}, this requirement can be rephrased in terms of   the existence of generically non-invertible axion shift symmetries.\footnote{In this section we will just consider the most relevant fields  and couplings, but the results of Section \ref{sec:noninvsymm} hold also more generically, as long as the additional sector is not charged under the U(1) gauge fields, and any additional EFT coupling involves the axions only through their derivatives.  }

\subsection{Basic models}
\label{sec:aximodels}

Our models include $n_{\rm A}$ axions $a^i$, $i=1,\ldots,n_{\rm A}$, with integral peridicity
\be\label{unitperiodicity}
a^i\simeq a^i+1\,.
\ee
The  ``angular" variables $\vartheta^i$ often used to denote axions are related to our fields via $\vartheta^i=2\pi a^i$. Furthermore, we also include $n_{\rm V}$ U(1) gauge fields $A^I$, $I=1,\ldots,n_{\rm V}$, and the corresponding field strengths $F^I=\d A^I$.  We will assume the standard normalization 
\be
\frac1{2\pi}\oint F^I\in\mathbb{Z}\,. 
\ee

Our `minimal' axiverse EFT takes the form
\be\label{axionlag0} 
-\frac12 M^2\int\calg_{ij}\,\d a^i\wedge *\d a^j-\frac1{4\pi}\int f_{IJ}F^I\wedge *F^J-\frac1{4\pi} \sfK_{iIJ}\int a^i\,F^I\wedge F^J\,.
\ee
Here $M$ is a reference mass scale, so that the axion kinetic matrix $\calg_{ij}$ is dimensionless.\footnote{\label{foot:2pikin} In terms of the  $2\pi$-periodic axions $\vartheta^i\equiv 2\pi a^i\simeq \vartheta^i+2\pi$, the first term in \eqref{axionlag0} can be rewritten as $-\frac12 (\sff^2_\vartheta)_{ij}\int \d\vartheta^i\wedge *\d\vartheta^j$, with  $(\sff^2_\vartheta)_{ij}\equiv \frac{M^2}{(2\pi)^2}\calg_{ij}$. The square roots of the eigenvalues of the kinetic metric $(\sff^2_\vartheta)_{ij}$ determine the standard axion decay constants of the canonically normalized axion fields in the basis in which the kinetic metric is diagonal.} In a gravitational context, we will identify it with the Planck mass:
$M=M_\Pl$.  Similarly, the dimensionless kinetic matrix $f_{IJ}$ determines the U(1) gauge couplings. In the non-supersymmetric models discussed in this subsection, $f_{IJ}$ and $\calg_{ij}$ are constant, while in the supersymmetric models of Section \ref{sec:susymodels} they will be promoted to be field dependent. We assume that the spacetime $X$ admits a spin structure, so that the last term in \eqref{axionlag0} is compatible with \eqref{unitperiodicity} only if 
 \be\label{intK} 
\sfK_{iIJ}=\sfK_{iJI}\in\mathbb{Z}\,.
\ee

Non-abelian gauge sectors, such as QCD or GUT sectors, could be added to \eqref{axionlag0}. If they couple to the axions $a^i$ through theta-like terms, they generically have important effects. These effects can be straightforwardly taken into account at a later stage. Hence, to avoid overloading the presentation, we will mostly assume the absence of non-abelian sectors, only occasionally commenting on their effects, and relegating more details to the Appendix \ref{app:nonabelian}. 

We also implicitly allow for the inclusion of other sectors, as long as they only introduce derivative axion interactions. In particular, one can add matter charged under the abelian (and non-abelian) gauge groups. This will only be relevant for the discussions of Section \ref{sec:el1form} and part of  Section \ref{sec:physcons}, while it will not affect the rest of the paper.

While we will mostly work at the two-derivative EFT level, we will also consider  curvature-squared terms, and in particular terms involving the  Gauss-Bonnet and the Pontryagin density.   
The Gauss-Bonnet term takes the form
\be\label{GRtheta}
\int\gamma\,E_{\text{\tiny GB}}\,*1\,,
\ee
where $E_{\text{\tiny GB}}$ is the Gauss-Bonnet density 
\be\label{GBdensity} 
E_{\text{\tiny GB}}\equiv \frac1{32 \pi^2}\left(R_{abcd}R^{abcd}-4 R_{ab}R^{ab}+R^2\right)\,.
\ee
As for the kinetic matrices, here the coupling  $\gamma$ appearing in \eqref{GRtheta} is a  constant, while in the models of Section  \ref{sec:susymodels}  it will be allowed to  be  field dependent. Notice that, in presence of the Einstein Hilbert term, other curvature-squared corrections of the form $R^{ab}R_{ab}$ and $R^2$ can be reabsorbed by a metric redefinition (up to the possible generation of other types of four-derivative corrections). The Gauss-Bonnet density \eqref{GBdensity} is singled out by the special property that, for constant $\gamma$, its integral over a closed Euclidean space computes its Euler characteristic -- see e.g.\ \cite{Martucci:2024trp} for more comments about this point.\footnote{On spacetimes with boundaries one should also include appropriate boundary counterterms \cite{Myers:1987yn}, analogous to the Gibbons-Hawking counterterm \cite{Gibbons:1976ue}. However, this technicality  will not play any relevant role in the following and so we will keep it  implicit.}  

We also allow for  a possible axion-Pontryagin coupling 
\be\label{aRR}
-\frac1{192\pi}\tilde \sfK_i\int a^i\, \tr(R\wedge R) \,.
\ee 
Since  integral of the first Pontryagin class   $p_1(X)=-\frac1{8\pi^2}\tr(R\wedge R)$   over a spin four-manifold is a multiple of 48, consistency with \eqref{unitperiodicity} requires that 
\be 
\tilde\sfK_i\in\mathbb{Z}\,.
\ee

We will also exploit the description in which the axions $a^i$ are dualized to two-form gauge potentials  $\calb_{2,i}$. In Lorentzian signature, and at the classical level, the relation between these equivalent descriptions is provided by 
\be \label{eqn:H3}
\calh_{3,i}=-M^2\calg_{ij}*\d a^i\,,
\ee
where $\calh_{3,i}$ are the three-form field strengths of the two-form potentials $\calb_{2,i}$.
As recently emphasized in \cite{Witten:2026twr}, at the quantum level the duality works in a more subtle way. 

In the dual description, the last two  terms in \eqref{axionlag0} are  
replaced by
\be\label{B2kinetic} 
-\frac{1}{2M^2}\int\calg^{ij}\calh_{3,i}\wedge *\calh_{3,j}\,,
\ee
where $\calg^{ij}$ is the inverse of $\calg_{ij}$. In particular, the last term in \eqref{axionlag0} and the Pontryagin term \eqref{aRR} are encoded in  the modified Bianchi identities
\be\label{modBianchi} 
\d\calh_{3,i}=2\pi I_{4,i}\,,
\ee
where\footnote{\label{foot:nonabelian} The inclusion of non-abelian gauge sectors generically modifies this relation.  For instance, the addition of the term \eqref{nonaFF} in the axion formulation corresponds to  adding the term $-\frac1{16\pi^2}\widehat\sfK_{i}\,{\rm Tr}\left(F\wedge F\right)$ to the r.h.s\ of \eqref{I4forms}. 
}
\be\label{I4forms} 
I_{4,i}\equiv -\frac1{8\pi^2}\sfK_{iIJ}F^I\wedge F^J-\frac{1}{384\pi^2}\tilde\sfK_i\tr(R\wedge R)\,.
\ee
This implies that the two-form gauge potentials $\calb_{2,i}$ are locally defined by 
\be\label{BHrel} 
\d \calb_{2,i}=\calh_{3,i}-I^{(0)}_{3,i}\,,
\ee
where $I^{(0)}_{3,i}$ are the Chern-Simons three-forms such that
\be\label{anpol} 
\d I^{(0)}_{3,i}=I_{4,i}\,.
\ee

\subsection{Supersymmetric extension}
\label{sec:susymodels}

We will also consider minimally supersymmetric extensions of the axiverse models introduced in Section \ref{sec:aximodels}, which we now describe. These extensions will not play any role until Section \ref{sec:physcons}.   Thus, some readers may prefer to first read Section  \ref{sec:noninvsymm} and return to this subsection later.

In supersymmetric models the axions $a^i$ combine with  corresponding {\em saxions} $s^i$ into complex fields 
\be\label{chiral} 
t^i\equiv a^i+\ii s^i\,,
\ee
which are the lowest components of chiral multiplets. The spectrum also contains the supersymmetric fermionic partners, which, however, will be irrelevant for most of our purposes and will be explicitly taken into account only in Section \ref{sec:EFTsup}. 

The kinetic terms of the (s)axionic sector are determined by a K\"ahler potential $K$, which  will be assumed to be  invariant under axion shift symmetries, namely to depend on $t^i$ and $\bar t^i$ only through their saxionic combination $s^i=\Im t^i$. In this case the axion kinetic matrix appearing in \eqref{axionlag0} becomes saxion dependent, but does not depend on the axions, and is determined by the K\"ahler potential via  the relation
\be\label{calg1}
\calg_{ij}(s)\equiv\frac12\frac{\del^2 K}{\del 
 s^i\del s^j}\,.
\ee
The same matrix also determines the saxion kinetic terms\footnote{In the presence of other chiral fields, additional mixing terms could appear. Since these would not affect our main conclusions, they will be ignored in what follows.}
\be\label{skin} 
-\frac12 M^2\,\calg_{ij}\,\d s^i\wedge *\d s^j\,.
\ee
In a supergravity context, we will set $M=M_\Pl$. 

The gauge field kinetic terms in \eqref{axionlag0} are also saxion-dependent and are linked by supersymmetry to the non-derivative  axion  couplings in \eqref{axionlag0} 
\be\label{gaugecoupl}
f_{IJ}=\sfK_{IJi}\,s^i+\ldots\,,
\ee
where the ellipses denote possible additional contributions to the gauge couplings, which are either constant or depend on some hidden chiral fields. By supersymmetry, these contributions would require the inclusion of corresponding  $F^I\wedge F^J$  terms. However, they will not play any role in our discussion. 

Similarly, supersymmetry relates the Gauss-Bonnet term \eqref{GRtheta} and the Pontryagin term \eqref{aRR} -- see e.g.\ \cite{Martucci:2022krl} for more details in our same context.  In particular, the coupling $\gamma$ appearing in \eqref{GRtheta} necessarily takes the form
\be\label{gammaGB} 
\gamma= \frac{\pi}{12}\,\tilde\sfK_i s^i+\ldots\,,
\ee
where as in \eqref{gaugecoupl} the ellipses  are terms that are
either constant or at most dependent  on some hidden chiral fields. Again, by supersymmetry these possible contributions correspond to a modification of \eqref{aRR} that  will be irrelevant for our purposes.

 The self-consistency of the theory requires that the kinetic matrices are positive definite on the   saxionic domain, that is the set of possible values of $s^i$.  If the ellipses in \eqref{gaugecoupl} are assumed to be $\calo(1)$,  the perturbative regime requires that the saxions are `large' enough, in some appropriate sense. 
 
 Most of the results that we will discuss in Sections \ref{sec:noninvsymm}--\ref{sec:axionat} hold for any K\"ahler potential invariant under axion shift symmetries, and gauge couplings of the form \eqref{gaugecoupl}. However, we will later focus on quantum gravity models, which motivates us to restrict ourselves to  the more specific, though still quite broad, framework introduced in \cite{Lanza:2020qmt,Lanza:2021udy,Lanza:2022zyg} and further developed in \cite{Martucci:2022krl,Martucci:2024trp}. This framework covers large classes of string theory models and allows for the  identification of a well defined  large saxion regime. In this regime the ellipses  in \eqref{gaugecoupl} and \eqref{gammaGB} are subleading and the K\"ahler potential gets a leading contribution of the form
 \be\label{loPK} 
 K=-\log P({\bm s})\,,
 \ee
with $P({\bm s})$ a homogeneous function of positive {\em integral} degree $k\geq 1$:
\be\label{homP} 
P(\lambda {\bm s})=\lambda^k P({\bm s})\,.
\ee
The integrality of $k$ is just an assumption motivated by experience from string theory, with no clear bottom-up motivation. We emphasize  that the K\"ahler potential \eqref{loPK} can generically receive subleading corrections. Formally, these are corrections that vanish in the scaling limit $s^i\rightarrow \lambda s^i$ with $\lambda\rightarrow \infty$. As we will see,  for our purposes we can safely ignore these types of corrections.   

Note that this structure naturally leads one to consider conical saxionic domains, namely domains that are preserved under arbitrary rescalings $s^i\rightarrow \lambda s^i$, with $\lambda\in\mathbb{R}_{>0}$. We will denote this conical domain as $\Delta$ and 
refer to it as the {\em saxionic cone}. As in \cite{Lanza:2020qmt,Lanza:2021udy,Lanza:2022zyg}, motivated by string theory examples, one can consider  more structured saxionic cones, imposing that $\Delta$ is convex and rational  polyhedral. A prototypical example of $\Delta$ is the K\"ahler cone of a Calabi--Yau compactification in string theory. We refer to Section \ref{sec:susyexam} for more explicit examples and to Figure \ref{fig:saxionicones} for a visual illustration.  
In order to define more explicitly a saxionic cone satisfying these conditions, it is convenient to adopt a basis independent  notation, in which we introduce an $n_{\rm A}$-dimensional lattice $V_{\mathbb{Z}}\simeq \mathbb{Z}^{n_{\rm A}}$, which can be considered as the integral subset of a corresponding real vector space $V_{\mathbb{R}}\simeq \mathbb{R}^{n_{\rm A}}$. The saxions $s^i$ can be identified with the components of the vector ${\bm s}=\{s^i\}\in V_{\mathbb{R}}$, while the axions $a^i\simeq a^i+1$ are the components of the representative   ${\bm a}=\{a^i\}$ of an element of $V_{\mathbb{R}}/V_\mathbb{Z}$. The saxionic cone   $\Delta\subset V_{\mathbb{R}}$ is convex  if ${\bm s},{\bm s}'\in\Delta$ implies that ${\bm s}+{\bm s}'\in\Delta$, and is rational polyhedral if it is  generated by a set of integral vectors ${\bf e}\in V_\mathbb{Z}$. 
Alternatively, it satisfies these properties if 
\be\label{saxcone} 
\Delta=\{{\bm s}\in V_{\mathbb{R}}\ |\ \langle {\bf q},{\bm s}\rangle > 0, \forall {\bf q}\in\calc_{\rm I}\} \; ,
\ee
where    $\calc_{\rm I}\subset V^*_\mathbb{Z}$ is a subset of integral dual vectors ${\bf q}=\{q_i\}$ with $q_i\in\mathbb{Z}$, and $\langle \cdot ,\cdot\rangle$ is the canonical pairing between $V_\mathbb{R}$ and $V^*_{\mathbb{R}}$, namely
\be 
\langle {\bf q},{\bm s}\rangle\equiv q_i s^i\,.
\ee
Notice that the positivity of the gauge couplings \eqref{gaugecoupl}  for large saxions requires that the matrix $\sfK_{IJi}s^i$ is positive definite for any ${\bm s}\in \Delta$.

It is clear that $\calc_{\rm I}$ encodes the entire information of $\Delta$, but also  that different choices of $\calc_{\rm I}$ can  give the same  $\Delta$. For instance,  if both ${\bf q},{\bf q}'\in V^*_\mathbb{Z}$ have positive pairing with ${\bm s}\in V_{\mathbb{R}}$, then ${\bf q}+{\bf q}'$ also has positive pairing and could be added to $\calc_{\rm I}$ without changing $\Delta$.  We will then define $\calc_{\rm I}$ as the maximal subset of elements ${\bf q}\in V^*_\mathbb{Z}$ such that $\langle {\bf q},{\bm s}\rangle>0$ for any ${\bm s}\in\Delta$. Namely, we can  alternatively  identify $\calc_{\rm I}$ with the discretized cone dual to $\Delta$:
\be\label{CIcone} 
\calc_{\rm I}=\Delta^\vee\cap V^*_\mathbb{Z}\,.
\ee
See Figure \ref{fig:saxionicones} in Section \ref{sec:susyexam} for an illustration.

The choice of the symbol $\calc_{\rm I}$ was  introduced in \cite{Lanza:2020qmt,Lanza:2021udy} and was  motivated by the fact that $\calc_{\rm I}$ can be identified with the  set of possible mutually BPS instanton charges. These are the charges that can be  carried by possible  BPS instantons, which may modify the EFT by terms involving the chiral operators 
\be\label{caloqop} 
\calo_{\bf q}(x)\equiv e^{2\pi\ii\langle {\bf q},{\bm t}\rangle}\,.
\ee
The condition \eqref{saxcone} is equivalent to requiring that $|\calo_{\bf q}|<1$ for any  ${\bf q}\in\calc_{\rm I}$ and  ${\bm s}\in\Delta$. In particular, it implies that $\calo_{\bf q}$ is exponentially  suppressed in the scaling limit ${\bm s}\rightarrow \lambda{\bm s}$ with $\lambda\rightarrow\infty$.

Terms of this form can, for instance, be generated by an additional confining gauge sector, or by fundamental instantons that do not admit a field-theoretic interpretation, such as world-sheet instantons in the above-mentioned  Calabi--Yau compactifications, or brane instantons in more general string compactifications. At the rigid EFT level, there is nothing that forbids setting all such instanton contributions to zero. In Section \ref{sec:WHs}, we will discuss how, from a purely bottom-up perspective, this is no longer possible in a quantum gravity setting.

It will also be useful to summarize some aspects of the formulation in which the chiral multiplets are dualized  by linear multiplets -- see \cite{Lanza:2019xxg} for more details about the general formulation, and \cite{Lanza:2020qmt,Lanza:2021udy,Martucci:2022krl,Martucci:2024trp} for related  comments about the class of quantum gravity models that we will consider. The linear multiplets contain  the two-form potentials $\calb_{2,i}$ dual to the axions $a^i$, introduced at the end of Section \ref{sec:aximodels}.\footnote{At the quantum level, the duality between these two descriptions requires a counterterm involving the Gauss-Bonnet density \eqref{GBdensity} \cite{Donnelly:2016mlc,Witten:2026twr}. Such a term can be reabsorbed in  the subleading ellipses in \eqref{gammaGB}.} 
Furthermore, they also include real scalars  $\ell_i$, which  provide an alternative parametrization of the saxionic directions. They are related to the saxions $s^i$ by 
\be\label{dualsax} 
\ell_i\equiv -\frac12\frac{\del K}{\del s^i}\,.
\ee
We will refer to them as {\em dual saxions}. 
Note that $\ell_i$ identify the components of a vector  ${\bm\ell}\in V^*_\mathbb{R}$. 
We will denote by $\calp\subset V^*_\mathbb{R}$ the image of the saxionic cone $\Delta$ under the map \eqref{dualsax}:
\be\label{dualsaxdom} 
\calp\equiv \Im\Big\{ {\bm s}\in\Delta \ \mapsto \  {\bm\ell}= -\frac12\frac{\del K}{\del {\bm s}}\in V^*_\mathbb{R}\Big\}\,.
\ee
By using \eqref{loPK} and the homogeneity \eqref{homP}, it is easy to see that $\calp$ is also a cone, in the sense that if ${\bm\ell}\in \calp$, then $\lambda{\bm\ell}\in \calp$ for any $\lambda\in \mathbb{R}_{>0}$. For this reason we will refer to $\calp$ as the {\em dual saxionic cone} -- see Figure \ref{fig:saxionicones}  below for an example. 

Note also that, using \eqref{calg1} and  \eqref{homP}, one can easily derive the relations
\be\label{ellsrel} 
\ell_i=\calg_{ij}s^j\quad\Leftrightarrow\quad s^i=\calg^{ij}\ell_j\,,
\ee
where $\calg^{ij}$ is the inverse of $\calg_{ij}$. The inverse metric $\calg^{ij}$ appears in the kinetic terms of $\ell_i$ and $\calh_{3,i}$ --  see \cite{Lanza:2019xxg,Lanza:2021udy} for more details.  Note that, using again the homogeneity of $P({s})$, \eqref{ellsrel} implies that 
\be\label{ellsnorms} 
\ell_is^i=\|{\bm s}\|^2=\|{\bm \ell}\|^2=\frac{k}2\,.
\ee 
Here and in the following the norms of elements of $V_\mathbb{R}$ and $V^*_\mathbb{R}$ are computed by using the metrics  $\calg_{ij}$  and $\calg^{ij}$, respectively. That is, in the present case, $\|{\bm s}\|^2=\calg_{ij}s^is^j$  and $\|\ell\|^2=\calg^{ij}\ell_i\ell_j$. 

Finally, we recall that the dual description is more directly formulated in terms of the dual kinetic potential
\be\label{kinF} 
\calf({\bm\ell})=K+2\ell_i s^i\,.
\ee
For example, it provides the inverse of \eqref{dualsax},
\be\label{saxfromell} 
s^i=\frac12\frac{\del\calf}{\del\ell_i}\,,
\ee
and the inverse saxionic  metric,
\be\label{invcalG} 
\calg^{ij}=-\frac12\frac{\del^2\calf}{\del\ell_i\del\ell_j}\,.
\ee
The general relation \eqref{kinF}, that holds for general  K\"ahler potential that are invariant under axion shift symmetries,  becomes simpler by assuming the homogeneity condition \eqref{homP}. Indeed, \eqref{ellsnorms} implies that, omitting an  irrelevant additional constant, the dual saxion  kinetic potential \eqref{kinF} takes the form 
\be\label{calfP}
\calf({\bm\ell})=\log\tilde P({\bm\ell})\,,
\ee
  where $\tilde P({\bm\ell})\equiv 1/P({\bm s}({\bm\ell}))$. 
Notice that $\tilde P({\bm\ell})$ is homogeneous too:
\be\label{tildePhom} 
\tilde P(\lambda{{\bm\ell}})=\lambda^k\tilde P({{\bm\ell}})\,.
\ee

We emphazise that, given such a theory with homogeneity $k$, one may consider (s)axionic subsectors with homogeneity $k'<k$. One can identify these subsectors by considering limits in which some subset of saxions become much larger than the other, or some subset of dual saxions become much smaller than the others. If this limit gives a consistent  decoupling of the former set from the latter, we can use it to describe a theory with reduced homogeneity degree. In the following Subsection we will provide some examples of how this can work.

\subsection{Simple examples}
\label{sec:susyexam}

It is useful to discuss possible concrete realizations of the general class of models introduced in Section \ref{sec:susymodels}, as well as their possible string theory interpretation. We will be brief here; more details and other examples within the same framework can be found in \cite{Lanza:2021udy,Martucci:2022krl,Martucci:2024trp}.

The simplest example contains just one (s)axion $s^0$, with saxionic cone $\Delta =\{s^0>0\}$. In this case we necessarily have $K=-k\log s^0$, that is $P(s)=(s^0)^k$. The dual saxion is then given by $\ell_0=\frac{k}{2s^0}$, $\calp=\{\ell_0>0\}$ and the dual kinetic potential  is $\calf(\ell_0)=k\log\ell_0$, that is $\tilde P(\ell_0)=(\ell_0)^k$. For instance, the K\"ahler potential of heterotic compactifications on Calabi-Yau three-folds at weak string coupling contains such a contribution with $k=1$, where $\ell_0$ parametrizes the four-dimensional dilaton. 

In the large volume regime of these heterotic compactifications,  one can  add $h^{1,1}$  saxions $s^a$, which are the k\"ahler moduli that parameterize the (string frame) K\"ahler cone of the Calabi-Yau. These combine with the saxion $s^0$ introduced above to parametrize an $n_{\rm A}=h^{1,1}+1$ dimensional saxionic cone $\Delta$. This  can be represented as a fibration where the fiber is a one-dimensional cone parametrized by $s^0$, with $s^0>p_a s^a$, and the base is  the K\"ahler cone parametrized by the saxions $s^a$  \cite{Martucci:2022krl,Martucci:2024trp}. Here $p_a$ are constant integers and we can assume that $p_a s^a>0$. The K\"ahler potential is 
\be\label{hetK} 
K=-\log(s^0-\frac12p_a s^a)-\log\left(\kappa_{abc}s^as^b s^c\right)\,,
\ee
where $\kappa_{abc}$ are the Calabi-Yau triple intersection numbers. 
Hence, we have enlarged the homogeneity from $k=1$ to $k=4$. By considering the limit $s^0\gg |s^a|$, one can consistently focus on the one-dimensional subsector parametrized by $s^0$, thereby recovering the $k=1$ theory discussed above. On the other hand, the condition $s^0>p_a s^a$ generically obstructs taking a large (string-frame) volume limit in which all the saxions $s^a$ become much larger than $s^0$. Therefore, one cannot simply focus on the $s^a$ while keeping $s^0$ fixed.\footnote{See \cite{Kaufmann:2026fli,Kaufmann:2026mha} for a detailed analysis of similar obstructions in type IIB/F-theory models.} Nevertheless, other asymptotic limits can be considered, leading to different consistent saxionic subsectors. For instance, in heterotic models admitting an F-theory dual, one may focus on a saxionic subsector with effective homogeneity $k=3$, corresponding  to the  F-theory models discussed in the next paragraph -- see \cite[App.\,B]{Martucci:2024trp} for further details about this relation.

In F-theory compactification at large volume, the K\"ahler moduli are parametrized by a set of $n_{\rm A}=b_2(X)$ dual saxions $\ell_i$, where $X$ is the base   three-fold of the associated elliptically fibered Calabi-Yau four-fold. In the  kinetic potential \eqref{calfP} we must take 
\be 
\tilde P({\bm\ell})=\frac1{3!}\kappa^{ijk}\ell_i\ell_j\ell_k\,,
\ee
where $\kappa^{ijk}$ are the triple intersection numbers of $X$. Clearly, in this case we have $k=3$.
The saxionic cone $\Delta$ and dual saxionic $\calp$ is model dependent. 

\begin{figure}[ht]
\centering

\begin{subfigure}{0.42\textwidth}
\centering
\includegraphics[width=\textwidth]{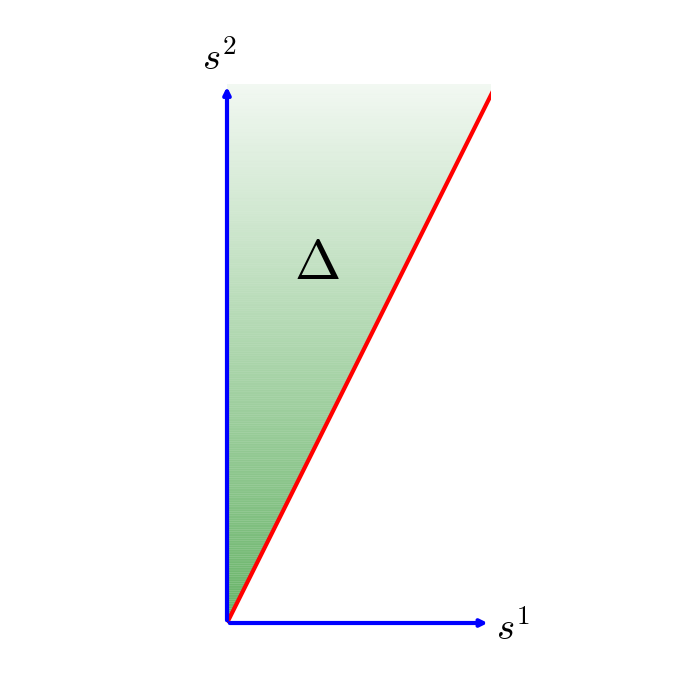}
\caption{Saxionic cone.}
\label{fig:delta}
\end{subfigure}
\hfill
\begin{subfigure}{0.55\textwidth}
\centering
\includegraphics[width=\textwidth]{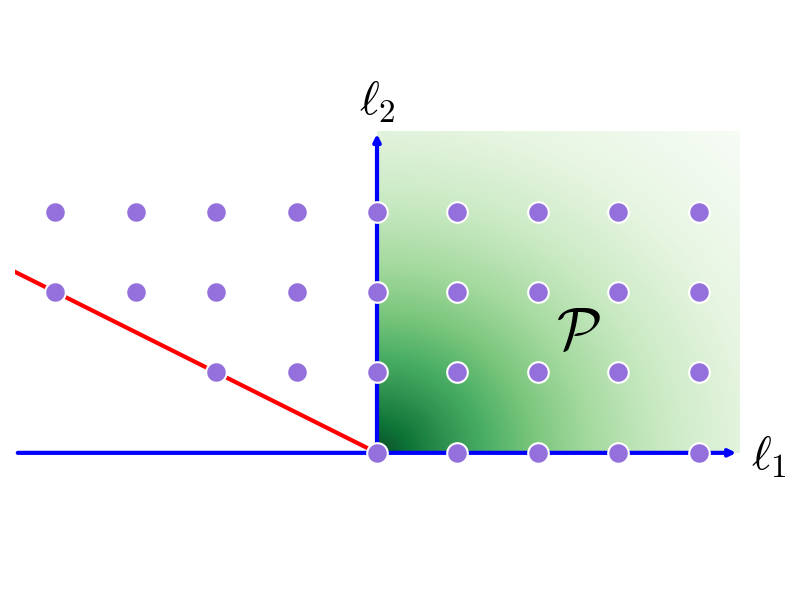}
\caption{Dual saxionic cone.}
\label{fig:calP}
\end{subfigure}

\caption{
Saxionic cone and dual saxionic cone of the  $\mathbb{P}^1\hookrightarrow X\rightarrow  \mathbb{P}^2$ F-theory model,
with twist parameter $p=2$. On the right, the bullets represent the set $\mathcal{C}_{\rm I}$ of BPS instanton charges.
}
\label{fig:saxionicones}
\end{figure}

For instance, consider the case in which $X$ is a $\mathbb{P}^1$ fibration over $\mathbb{P}^2$, with twist parameter $p\geq 0$. In this case we have two dual saxions $\ell_1,\ell_2$, parametrizing the dual saxionic cone $\calp=\{{\bm\ell}=(\ell_1,\ell_2)\in\mathbb{R}^2|\, \ell_1>0,\ell_2>0\}$ -- see Fig.~ \ref{fig:calP} -- and 
\be\label{P1fibrP}
\tilde P({\bm\ell})=3\ell_1^2\ell_2+3p\ell_1\ell_2^2+p^2\ell_2^3\,.
\ee
The corresponding saxions $s^1,s^2$ can be obtained from \eqref{saxfromell} and parametrize the saxionic cone $\Delta=\{{\bm s}=(s^1,s^2)\in\mathbb{R}^2|\, s^1> 0\,,\ s^2> p\,s^1\}$ -- see Figure \ref{fig:delta}. Notice that, even if \eqref{P1fibrP} has homogeneity $k=3$, we could consider the limit in which  $\ell_2\ll\ell_1$ and focus just on the subsector parametrized by $\ell_2$, with $\calf=\log\ell_2+\ldots$, so that we effectively have $k=1$. Indeed, the $\ell_2\rightarrow 0$ limit corresponds to an infinite field distance limit in which the F-theory model is dual to a weakly-coupled heterotic compactification of the type discussed above, with $\ell_2\equiv \ell_0$.
On the other hand, we cannot do the same by taking $\ell_1\rightarrow 0$, which is instead a finite distance limit.

%%%%%%%%%%%%%%%%%%%%%%%%%%%%%%%%%%%%%%%%%%%%%%%%%%%%%%%%%%%%%%%%%%%%%%%%%%%%%%%%%%%%%%%%%%%%%%%%%%%

\section{Non-invertible symmetries in the axiverse}
\label{sec:noninvsymm}

In this section, we discuss the invertible and non-invertible global symmetries, together with the corresponding topological defects and charged operators, of the axiverse models described in Section \ref{sec:aximodels0}. We follow a `descending' order, starting with the two-form symmetries and then turning to the one- and zero-form symmetries. This organization allows us to begin with the invertible symmetries and helps clarify the relations among the various defects. Supersymmetry plays no role in the present discussion. The results presented here can be regarded as an axiverse generalization of previous analyses of the axion-Maxwell model \cite{Choi:2022jqy,Cordova:2022ieu,Choi:2022fgx,Yokokura:2022alv,DelZotto:2024ngj}, see also \cite{Hidaka_2020,Hidaka:2020izy,Hidaka:2021kkf}, and \cite{Hong:2025qbw} for previous work considering axiverse settings. For the invertible symmetries discussed in Sections \ref{sec:winding} and \ref{sec:magnetic}, the generalization is essentially straightforward. By contrast, the axiverse  generalization of the non-invertible symmetries discussed in Sections \ref{sec:el1form} and \ref{sec:ax0form} is less obvious, and builds on and extends the techniques introduced in \cite{Martucci:2024trp} to describe the non-invertible symmetries of more general Gaillard-Zumino models \cite{Gaillard:1981rj}.

%%%%%%%%%%%%%%%%%%%%%%%%%%%%%%%%%%%%%%%%%%%%%%%%%%%%%%%%%%%%%%%%%%%%%%%%%%%%%%%%%%%%%%%%%%%%% 

\subsection{Winding two-form symmetries and vortex operators}
\label{sec:winding}

We start our discussion of the global axiverse symmetries by considering the {\em winding}  two-form symmetries. 
 They are associated with the one-form conserved currents 
\be\label{Jw} 
\calj_{(\rm w)}^i=2\pi\d a^i\,.
\ee
In our notation the Noether current  of a continuous $p$-form symmetry is given by a closed $(3-p)$-form $\calj$ -- such that $\d\calj=0$ -- rather than by its Hodge dual $(p+1)$-form.  The  winding  two-form  symmetries  are realized  by  codimension-three  topological operators 
\be\label{winding} 
\cald{}^{(\rm w)}_{\bm\beta}(\gamma)=\exp\left(\ii\beta_i \oint_\gamma\calj_{(\rm w)}^i\right)\,
\ee
which are supported on one-dimensional curves $\gamma$. The periodic parameters $\beta_i\simeq \beta_i+1$ label the corresponding U(1)$^{n_{\rm A}}$ element. 

The winding topological operator \eqref{winding}  acts on codimension-two extended operators, supported on two-dimensional surfaces $\Sigma$. These operators can be defined by excising a small tubular neighborhood of $\Sigma$ and imposing that the axions undergo an integral shift
\be\label{axionstrings} 
a^i\rightarrow a^i +e^i\quad,\quad e^i\in\mathbb{Z}\,,
\ee
around them. We will denote these operators as $\calv_{\bf e}(\Sigma)$, where the charge vector ${\bf e}=\{e^i\}$ takes values in the lattice $V_\mathbb{Z}\simeq \mathbb{Z}^{n_{\rm A}}$, and we will refer to them as {\em vortex} operator. The axion winding  \eqref{axionstrings} condition implies that the insertion of $\calv_{\bf e}(\Sigma)$ modifies the conservation equation $\d\calj^i_{\rm(w)}=0$ into 
\be\label{dJdelta} 
\d\calj^i_{\rm(w)}=2\pi e^i\delta_2(\Sigma)\,,
\ee
which encodes the contact term contribution appearing in Ward identities for the winding two-form symmetry.  
Hence, if $\gamma\simeq S^1$ is a circle that links $\Sigma$, we get
\be\label{caldwaction} 
\cald{}^{(\rm w)}_{\bm\beta}(\gamma)\calv_{\bf e}(\Sigma)=e^{2\pi\ii \langle {\bm \beta},{\bf e}\rangle}\calv_{\bf e}(\Sigma)\,,
\ee
where we have used the index-free notation for the pairing $\langle {\bm \beta},{\bf e}\rangle\equiv\beta_i e^i$ between ${\bf e}\in V_\mathbb{Z}\subset V_\mathbb{R}$ and ${\bm\beta}=\{\beta_i\}\in V^*_\mathbb{R}/V^*_\mathbb{Z}$, with $V^*_\mathbb{R}$ denoting the dual of $V_\mathbb{R}$, and $V_\mathbb{Z}$ and $V^*_\mathbb{Z}$ are the corresponding   integral lattices. 

Importantly, as emphasized e.g.\ in \cite{Choi:2022fgx}, the surface operator $\calv_{\bf e}(\Sigma)$  must support a non-trivial world-sheet sector. This is because of the anomaly inflow mechanism \cite{Callan:1984sa,Naculich:1987ci} -- see also \cite{Fukuda:2020imw,Fan:2021ntg,Heidenreich:2021yda,Martucci:2022krl}. The sum of the last term 
in \eqref{axionlag0}  and \eqref{aRR} can be rewritten as
\be 
-\int \calj^i_{(\rm w)}\wedge I^{(0)}_{3,i} \, ,
\ee
where $I^{(0)}_{3,i}$ are the Chern-Simons three-form defined in \eqref{anpol}. 
These are not invariant under infinitesimal gauge and local Lorentz transformation:  
\be 
\delta I^{(0)}_{3,i}=\d I^{(1)}_{2,i}\,.
\ee 
Hence, \eqref{dJdelta} produces  the following anomalous contribution  localized on $\Sigma$:
\be\label{vortexanomaly} 
\delta S_{\rm bulk}=-2\pi e^i\oint_\Sigma I^{(1)}_{2,i}\,.
\ee
This can alternatively be understood in the dual formulation introduced at the end of Section \ref{sec:aximodels}, in which the axion vortex supports the coupling $\exp(\ii e^i\int_\Sigma \calb_{2,i})$. Then \eqref{vortexanomaly} follows from \eqref{BHrel} and the gauge invariance of $\calh_{3,i}$.

The anomaly \eqref{vortexanomaly} must be canceled by a  world-sheet anomaly.\footnote{In a QFT setup in which gravity is not dynamical, we could allow for a non-vanishing 't Hooft gravitational anomaly, captured by the last term in \eqref{anpol}. However, having in mind the eventual coupling to dynamical gravity, for later convenience we keep this condition as one of the defining properties of  $\calv_{\bf e}$.} Hence, $\calv_{\bf e}$ must support a non-trivial world-sheet sector  including chiral fermions or  bosons, whose anomaly polynomial under bulk gauge and local Lorentz transformations is precisely given by $e^iI_{4,i}$. See  \cite[Section 3.2]{Martucci:2022krl} for more details in a very similar setting --  though considering dynamical axion strings rather than codimension-two operators -- including also the contribution of non-abelian gauge sectors as in Footnote \ref{foot:nonabelian}. Note that, for any anomaly polynomial $e^iI_{4,i}$,  the choice of the worldsheet degrees of freedom is not unique. However, we will not need to specify them and $\calv_{\bf e}$  will denote any such operator.    

\subsection{Magnetic one-form symmetries and 't Hooft lines}
\label{sec:magnetic}

The magnetic one-form symmetries of our axiverse model are generated by the two-form currents $F^I$. The corresponding U(1)$^{n_{\rm V}}$ one-form symmetries are realized by the invertible topological operator 
\be \label{eqn: magn ope}
\cald^{\rm(m)}_{\bm \beta}(\Sigma)=\exp\left(\ii\beta_I\oint_\Sigma F^I\right)\,, 
\ee
and labeled by the periodic parameters $\beta_I\simeq \beta_I+1$. The associated charged operators are provided by `t Hooft lines $H_{\bf m}(\gamma)$ carrying  magnetic charges $m^I$, whose insertion modifies the Bianchi identity $\d F^I=0$ into 
\be 
\d F^I=2\pi m^I\delta_3(\gamma)\,.
\ee 
These 't Hooft lines can  be defined by excising a small tubular neighborhood of $\gamma$ and imposing that 
\be\label{defthooft}
\oint_{S^2}F^I=2\pi m^I\,,
\ee
along it.
So, if we pick a two-sphere $\Sigma\simeq S^2$ linking $\gamma$ we have
\be\label{magnaction} 
\cald^{\rm(m)}_{\bm \beta}(\Sigma)H_{\bf m}(\gamma)=e^{2\pi\ii\langle {\bm \beta},{\bf m}\rangle }H_{\bf m}(\gamma)\,.
\ee
Here we are again using an index free notation for the canonical  pairing $\langle {\bm \beta},{\bf m}\rangle\equiv \beta_I m^I$ between ${\bf m}=\{m^I\}\in W_{\mathbb{Z}}$ and ${\bm\beta}=\{\beta_I\}\in W^*_\mathbb{R}/W^*_{\mathbb{Z}}$, where $W_{\mathbb{Z}}$  denotes the lattice of magnetic charges, $W_{\mathbb{R}}$ the corresponding real vector space, and $W^*_{\mathbb{Z}}$ and $W^*_{\mathbb{R}}$ the associated duals. 

Analogously to what  happens for the vortex operators in Section \ref{sec:winding}, and what discussed in \cite{Fukuda:2020imw,Fan:2021ntg,Choi:2022fgx} for the axion-Maxwell model, 't Hooft lines must support a non-trivial world-line sector. This basically comes from the Witten effect \cite{Witten:1979ey}, namely the fact that monopoles acquire electric charge under a $2\pi$ shift of the theta-angle, and can also be understood as an anomaly inflow  \cite{Cordova:2019jnf,Cordova:2019uob,Fukuda:2020imw}. 
In our framework, this inflow produces an anomalous behavior under integral axion shifts $a^i\rightarrow a^i+k^i$, $k^i\in\mathbb{Z}$. One  way to see it is to compute from \eqref{axionlag0}  the dual field strengths 
\be\label{dualG} 
G_I= -f_{IJ}*F^J-\sfK_{iIJ}a^iF^J\,.
\ee
An 't Hooft line $H_{\bf m}(\gamma)$ should contain the factor $\exp(\ii m^I\oint_\gamma\tilde A_I)$, where $\tilde A_I$ are the dual magnetic potentials, such that $\d\tilde A_I=G_I$. However, from \eqref{dualG} it is clear that the field strengths $G_I$,  and hence the magnetic potentials $\tilde A_I$, are not invariant under integral axion shifts. 
More precisely, 
\be 
\tilde A_I\quad\rightarrow\quad \tilde A_I -k^i\sfK_{iIJ}A^J
\ee
under $a^i\rightarrow  a^i+k^i$, and hence
$\exp(\ii m^I\oint_\gamma\tilde A_I)$  acquires an anomalous phase 
\be\label{1danom} 
\exp\left(-\ii m^Ik^i\sfK_{iIJ} \oint_\gamma A^J\right)\,,
\ee
which must be canceled by a non-trivial world-line sector. The choice of this sector is not unique, but a simple realization \cite{Jackiw:1975ep} is obtained by introducing a set of periodic world-line axions $\sigma_i\simeq \sigma_i+2\pi$, providing a many-body generalization of the particle-on-a-circle quantum mechanics \cite{Cordova:2019jnf,Gaiotto:2017yup}. These world-line axions
couple to the bulk fields through the world-line partition function 
\be\label{1dphase} 
\int [D\sigma]_\gamma\exp\left(\frac{\ii}{2}h^{ij}\int_\gamma D_A\sigma_i\wedge  *_\gamma D_A\sigma_j  -\ii \oint_\gamma a^i D_A\sigma_i\right)\ ,
\ee
where $h^{ij}$ is the world-line kinetic matrix and 
\be 
D_A\sigma_i\equiv \d\sigma_i-m^I\sfK_{iIJ}A^J\,.
\ee
Gauge invariance requires that $\sigma_i\rightarrow \sigma_i+ m^I\sfK_{iIJ}\lambda^J$ under the gauge transformation $A^I\rightarrow A^I+\d\lambda^I$. It is clear that, under the bulk axion shift $a^i\rightarrow a^i+k^i$, the second term in \eqref{1dphase} produces an anomalous phase that precisely cancels  \eqref{1danom}. We emphasise that the choice \eqref{1dphase} is not the only possible one. Furthermore, it is not even  uniquely defined by itself, since $h^{ij}$ is arbitrary.

\subsection{Non-invertible electric one-form symmetries}
\label{sec:el1form}

We now turn to the non-invertible symmetries present in our class of models. The results of the previous two subsections remain valid in the presence of fields charged under the U(1) gauge fields. Throughout this subsection, we assume that electrically charged particles are absent, since their presence would partially or completely break the electric one-form symmetries; see Section \ref{sec:elbreak}.

If the axions $a^i$ were {\em not} periodic, we could identify the two-form electric currents with the dual field strengths $G_I$ defined in \eqref{dualG}. However, as already emphasized in Section \ref{sec:magnetic}, the $G_I$ are not invariant under the axion periodicity and therefore cannot be used as gauge-invariant two-form currents.
As we are going to discuss, one can generalize the construction presented in \cite{Choi:2022fgx} for axion–Maxwell theory to show that most of the electric one-form symmetries are not lost: rather, a dense subset of them survives, albeit in a non-invertible form.  

More precisely, one can construct a topological defect $\cald^{\rm(e)}_{\bm\alpha}(\Sigma)$ for each ${\bm\alpha}=\{\alpha^I\}\in W_{\mathbb{Q}}/W_{\mathbb{Z}}$, where $W_{\mathbb{Z}}\simeq \mathbb{Z}^{n_{\rm V}}$ is the lattice of magnetic monopole charges, and $W_{\mathbb{Q}}$ is its $\mathbb{Q}$-span. In other words, $\cald^{\rm(e)}_{\bm\alpha}(\Sigma)$ is labeled by   $\alpha^I\in\mathbb{Q}$, with  $\alpha^I\simeq \alpha^I+1$. We will first define $\cald^{\rm(e)}_{\bm\alpha}(\Sigma)$ for any ${\bm\alpha}\in W_{\mathbb{Q}}/W_{\mathbb{Z}}$, and then argue that it is topological.

In order to define $\cald^{\rm(e)}_{\bm\alpha}(\Sigma)$, let us first introduce the $n_{\rm A}\times n_{\rm V}$ matrix $\sfQ({\bm\alpha})$ of components 
\be\label{sfQmatrix} 
\sfQ_{iI}({\bm\alpha})\equiv \sfK_{iIJ}\alpha^J\,.
\ee
Since $\sfK_{iIJ}\in\mathbb{Z}$ (see \eqref{intK}), it is clear that $\sfQ_{iI}({\bm\alpha})\in\mathbb{Q}$ for any ${\bm\alpha}\in W_{\mathbb{Q}}$ (and  $\sfQ_{iI}({\bm\alpha})\in\mathbb{Z}$ for any ${\bm\alpha}\in W_{\mathbb{Z}}$). As discussed in Appendix \ref{app:coprime fact}, one can then always pick a {\em right-coprime} factorization  
\be\label{QPNaxion} 
\sfQ({\bm\alpha})=\sfM\sfL^{-1}\,,
\ee
in terms of integral  matrices 
\be 
\sfM=\{\sfM_{iI}\}\in {\rm Mat}(n_{\rm A}, n_{\rm V},\mathbb{Z})\quad,\quad\sfL=\{\sfL^I{}_{J}\}\in {\rm Mat}(n_{\rm V},\mathbb{Z})\,.
\ee
An algorithm for explicitly identifying a right-coprime factorization is described in Appendix \ref{app:algor}. In components, \eqref{QPNaxion} reads $\sfQ_{iI}({\bm\alpha})=\sfM_{iJ}(\sfL^{-1})^J{}_I$, and we keep implicit the dependence of $\sfM$ and $\sfL$ on the axion shift vector ${\bm\alpha}$. As discussed in Appendix \ref{app:coprime fact}, being right-coprime, $\sfM$ and $\sfL$ are unique, up to an unimodular transformation 
\be\label{unimodML} 
\sfM\rightarrow \sfM\sfT\quad,\quad\sfL\rightarrow \sfL\sfT\,,\quad \text{with $\sfT\in{\rm GL}(n_{\rm V},\mathbb{Z})$}\,.
\ee
Notice that, in particular, if $\sfQ_{iI}({\bm\alpha})\in\mathbb{Z}$, then we can just pick $\sfM=\sfQ({\bm\alpha})$ and $\sfL=\mathds{1}$. 

 The defect $\cald^{\rm(e)}_{\bm\alpha}(\Sigma)$ can then be defined as follows,
\be\label{Del} 
\cald^{\rm(e)}_{\bm\alpha}(\Sigma)\equiv e^{\ii\alpha^I\oint_{\Sigma}\calj_I^{\rm (e)}}\int[\cald\phi\cald c]_\Sigma\exp\left[\ii\oint_\Sigma\left(\sfL^I{}_J\phi_I\d c^J+\sfM_{iI}a^i\d c^I+\phi_I F^I\right)\right]\,,
\ee
where 
\be\label{Jedef}
\calj_I^{\rm (e)}\equiv -f_{IJ}*F^J=G_I+\sfK_{iIJ}a^iF^J\,,
\ee
and 
we have  introduced $n_{\rm V}$ compact scalar fields $\phi_I\simeq \phi_I+1$ and $n_{\rm V}$ U(1) one-form vector fields $c^I$, all living on $\Sigma$. Notice that the two-form current $ \calj_I^{\rm (e)}$ is gauge invariant but not conserved:  
\be \label{elcons}
\d\calj_I^{\rm (e)}=\sfK_{iIJ}\d a^i\wedge F^J\,.
\ee 
The analogous defect constructed in \cite{Choi:2022fgx} for the axion-Maxwell model can be regarded as the $n_{\rm V}=n_{\rm A}=1$ subcase of \eqref{Del}.

It is clear that \eqref{Del} is well defined, since all the coefficients appearing in the path-integrated exponential  are properly quantized. 
As in \cite{Choi:2022fgx}, a simple heuristic way to check that \eqref{Del} is topological and provides a realization of the electric one-form symmetry is to try to integrate out the world-sheet fields $\phi_I$ and $c^I$. The corresponding equations of motion $\sfL^I{}_J\d c^J+F^I|_\Sigma=0$ and $\sfL^J{}_I\d\phi_J+\sfM_{iI}\d a^i|_\Sigma=0$ can be `naively' solved by setting  $\d c^I=-(\sfL^{-1})^I{}_JF^J|_\Sigma$ and $\phi_I=-\sfQ_{iI}({\bm\alpha}) a^i|_\Sigma$. This solution is naive because it does not respect the appropriate quantization conditions of the world-sheet fields. But by plugging it into \eqref{Del} and taking \eqref{Jedef} into account, it produces precisely the naive topological  operator $\exp\left(\ii\alpha^I\oint_\Sigma G_I\right)$, supporting the idea that \eqref{Del} does exactly what we need for our purposes. This conclusion can be confirmed by a more conceptual derivation of \eqref{Del} from a half higher  gauging \cite{Damia:2022bcd,Kaidi:2023maf,Choi:2022fgx}, carried out in Appendix \ref{app:halfsg}, along the lines of what was discussed in \cite{Choi:2022fgx} for the axion-Maxwell model.

For our purposes, it is sufficient to mention that one must perform a half higher gauging with respect to a discrete subgroup of U(1)$^{n_{\rm V}}\times$U(1)$^{n_{\rm A}}$. Insight into the structure of this subgroup is provided by studying the fusion  $\cald^{\rm(e)}_{\bm\alpha}\times \overline\cald^{\rm(e)}_{\bm\alpha}$, which also highlights the non-invertible nature of this symmetry operator. This fusion is computed in Appendix \ref{app:conddef} and here we just quote the result 
\begin{equation}\label{elfusion}
\begin{aligned}
\cald^{\rm(e)}_{\bm\alpha} \times  \overline\cald^{\rm(e)}_{\bm\alpha} &= \sum_{[{\bm\eta}] \in H^1(\Sigma, \Gamma_\sfL)} \exp \left\{ \ii\sfQ_{iI} \oint_\Sigma    \eta^I \cup\calj_{(\rm w)}^i \right\}\sum_{[\tilde{\bm\eta}] \in H^0(\Sigma,\Gamma_\sfL^*)}\, \exp \left\{ \ii \, (\sfL^{-1})^{I}{}_J\oint_\Sigma  \tilde\eta_I F^J \right\} \,,
\end{aligned}
\end{equation}
where for  simplicity we have omitted the overall normalization factor. 
In \eqref{elfusion} we have introduced the finite subgroups $\Gamma_\sfL\equiv W_\bbZ/(\sfL W_\bbZ)$ and $\Gamma^*_\sfL\equiv W^*_\bbZ/(\sfL^{\rm t} W^*_\bbZ)$. So, for instance, an element $[\tilde{\bm\eta}]\in\Gamma^*_\sfL$ corresponds to an integral  vector $\tilde{\bm\eta} \in W_\bbZ$, of components $\tilde\eta_I$,  modulo the identification $\tilde\eta_I\simeq \tilde\eta_I+k_J\sfL^J{}_I$ for any $k_I\in\mathbb{Z}$. 

It is clear that \eqref{elfusion} can be regarded as a condensation defect \cite{Roumpedakis:2022aik} involving a finite number of winding two-form and magnetic one-form symmetries.   
The fact that the fusion \eqref{elfusion} does not yield the identity reflects the non-invertible nature of \eqref{Del}.  We also remark that the operator \eqref{Jedef} does not depend on the specific coprime factorization we choose. Indeed, as already  emphasized, different  coprime factorizations are related by unimodular transformations \eqref{unimodML}, which can be reabsorbed by the  field redefinition  $c^I \rightarrow (\sfT^{-1})^I_{\; J} c^J$. Furthermore, one can explicitly check that \eqref{Del} is invariant under ${\bm\alpha}\rightarrow {{\bm\alpha}+{\bf m}}={{\bm\alpha}}$ for any ${\bf m}\in W_\mathbb{Z}$, consistently with the fact that it should depend only on $\alpha^i$ mod $\mathbb{Z}$; see Appendix \ref{app:shiftinv} for details. Once again, it is crucial that \eqref{QPNaxion} is right-coprime.

We also observe that, if more general  $\alpha^I\in\mathbb{Q}$ are such that $\sfQ_{iI}({\bm\alpha})\in\mathbb{Z}$, one can again set  $\sfM_{iJ}=\sfK_{iIJ}\alpha^J$ and $\sfL^I{}_J=\delta^I{}_J$, and exactly  integrate out $\phi_I$ as in the previous paragraph. This means that in these cases \eqref{Del} reduces to the  invertible operator 
\be\label{invelectr} \exp\left(\ii\oint_\Sigma\left[\alpha^I\calj_I^{\rm(e)}-\sfQ_{iI}({\bm\alpha})a^iF^I\right]\right)=\exp\left(\ii\alpha^I\oint_\Sigma G_I\right)\,,
\ee 
which is well defined precisely because $\sfQ_{iI}({\bm\alpha})\in\mathbb{Z}$,  and non-trivial if ${\bm\alpha}\notin W_{\mathbb{Z}}$.\footnote{In fact, the operator \eqref{invelectr} is well defined for any set of real parameters $\alpha^I\in\mathbb{R}$ such that  $\sfQ_{iI}({\bm\alpha})=\sfK_{iIJ}\alpha^J\in\mathbb{Z}$. This condition selects the  group of {\rm invertible} electric one-form symmetries, which of course strongly depends on $\sfK_{iIJ}$. As a trivial example, if $\sfK_{iIJ}\equiv 0$, then we have a U$(1)^{n_{\rm V}}$ group of  invertible one-form symmetries, and no non-invertible ones. As another simple case, suppose that  $n_{\rm A}=1$ and $\sfK_{IJ}\equiv \sfK_{1IJ}$ is non-degenerate. Then the invertible electric one-form symmetries form the finite subgroup $(\sfK^{-1}W^*_\mathbb{Z})/W_\mathbb{Z}$. More general cases can be analyzed similarly.}

Finally, let us   discuss the action of the topological defect  \eqref{Del} on various operators. First of all, it acts in an invertible way on Wilson lines 
\be\label{Wline} 
W_{\bf n}(\gamma)\equiv \exp\left(\ii\, n_I\oint_\gamma A^I\right)\,,
\ee
by measuring their electric charge. Namely, if we pick a two-sphere $\Sigma\simeq S^2$ linking $\gamma$, we have 
\be\label{elDaction} 
\cald_{\bm\alpha}(\Sigma)W_{\bf n}(\gamma)=e^{- 2\pi\ii \langle {\bf n},{\bm\alpha}\rangle}W_{\bf n}(\gamma)\,.
\ee
This comes just from the exponential in \eqref{Del} containing $\oint_\Sigma\calj^{\rm(e)}_I$, which acts on Wilson lines like $\oint_\Sigma G_I$. The action on local operators is clearly trivial, since a local operator cannot be linked with a codimension-two defect $\cald^{\rm (e)}_{{\bm\alpha}}(\Sigma)$. By contrast, the action on 't Hooft lines and vortex operators is more intricate, as already happens in the simpler axion-Maxwell case \cite{Choi:2022fgx}. We will not discuss the details of this non-trivial action in the present work.

\subsection{Non-invertible axion shift symmetries}
\label{sec:ax0form}

We finally turn to  the non-invertible zero-form symmetries associated with arbitrary rational shifts of the axions. This kind of symmetries were previously studied in the context of the axion-Maxwell model   \cite{Choi:2022jqy,Cordova:2022ieu,Choi:2022fgx,Yokokura:2022alv}, see also \cite{Putrov:2023jqi,Copetti:2023mcq,DelZotto:2024ngj}, and for the more general Gaillard-Zumino models in \cite{Apruzzi:2025byj}. As in  Section \ref{sec:el1form}, we will focus only on the defect operator and its properties, postponing the discussion of physical implications to Sections \ref{sec:physcons}-\ref{sec:WHs}. 

In the absence of the axion-gauge couplings in \eqref{axionlag0} and axion-Pontryagin couplings \eqref{aRR}, the action is invariant under a real shift of the axion, ${\bm a} \rightarrow {\bm a} \, + {\bm\alpha}$ for ${\bm\alpha} \in V_\mathbb{R}$. The associated conserved and gauge invariant currents are
\begin{equation} \label{eqn:jshift}
    \calj^{\rm(a)}_i=-M^2 \calg_{ij} * \d a^j=\calh_{3,i}\,,
\end{equation}
where in the second equation we have used the dual formulation, cf.~\eqref{eqn:H3}. However, non-vanishing $\sfK_{iIJ}$ and $\tilde\sfK_i$ in \eqref{axionlag0} and \eqref{aRR} break the closure relation \eqref{eqn:jshift} and hence, at first glance, appear to break the continuous axion shift symmetries. As we will now explain, a dense subgroup of {\em rational} axion shift symmetries nevertheless survives, albeit in a non-invertible form.  

For the time being, let us focus on the coupling between the axions and the abelian gauge sector by setting $\tilde\sfK_i=0$. We will return to the axion-Pontryagin coupling later. Then the exterior derivative of  \eqref{eqn:jshift} gives 
\begin{equation} \label{closedcurrents}
    \d \calj^{\rm(a)}_i = -\frac{1}{4 \pi} \sfK_{iIJ} \, F^I \wedge F^J \,.
 \end{equation}
Our aim is to  identify a codimension-one topological operator $\cald^{\rm(a)}_{\bm\alpha}(\Sigma)$,  that implements the axion shift ${\bm a} \rightarrow {\bm a} \, + {\bm\alpha}$ for any $ {\bm\alpha} = \{ \alpha^i \} \in V_{\bbQ} / V_{\bbZ}$, that is for any $\alpha^i\in\mathbb{Q}$ with $\alpha^i\simeq \alpha^i+1$.  
In order to define this operator, let us first introduce the $n_{\rm V}\times n_{\rm V}$ matrix $\sfC({\bm\alpha})=\{\sfC_{IJ}({\bm\alpha})\}$ of components 

\begin{equation}\label{CIJalpha}
    \sfC_{IJ}({\bm\alpha}) \equiv -\alpha^i \sfK_{iIJ} \, .
\end{equation}
Since $\sfK_{iIJ}\in\mathbb{Z}$, the matrix $\sfC({\bm\alpha})$ has rational entries, i.e. $\sfC_{IJ}({\bm\alpha})\in\mathbb{Q}$, for every ${\bm\alpha}\in V_{\mathbb{Q}}$. We can therefore directly apply the construction of the $\sfC$-defect developed in \cite{Apruzzi:2025byj}, which in turn generalizes the strategy originally proposed in \cite{Choi:2022jqy,Cordova:2022ieu}.

First of all, we have to pick a {\em right-coprime} factorization (see Appendix \ref{app:coprime fact})
\begin{equation}\label{0formCfact}
    \sfC({\bm\alpha}) = \sfP \sfN^{-1}\,,
\end{equation}
in terms of integral matrices $\sfP$ and $\sfN$, as was done for the electric one-form symmetries discussed  in Section \ref{sec:el1form}. In this case, $\sfP = \{ \sfP_{IJ} \}$ is also a square matrix. Obviously, $\sfP$ and $\sfN$ depend on ${\bm\alpha}$, but we keep this dependence  implicit. 
As explained in \cite{Apruzzi:2025byj}, given the right-coprime factorization
\eqref{0formCfact}, on any oriented three-dimensional submanifold $\Sigma$ one can
define a corresponding {\em minimal} \cite{Hsin:2018vcg} topological QFT (TQFT)
with $\Gamma^{(1)}_\sfN$ one-form symmetry, where
$\Gamma_\sfN \equiv W_{\mathbb{Z}}/(\sfN W_{\mathbb{Z}})$ (cf.\ comments below
\eqref{elfusion}),  and with corresponding anomaly fixed by $\sfP$.
One can couple this theory to
$(\sfN^{-1}F)^I \equiv (\sfN^{-1})^I{}_J F^J$, regarded as a background flat
connection for the $\Gamma^{(1)}_\sfN$ one-form symmetry.
We will denote the resulting partition function by $\cala^{(\sfN,\sfP)}_\Sigma[\sfN^{-1}F]$.

If ${\bm\alpha}\in V_{\mathbb{Q}}$ were such that $\sfP=\mathds{1}$, then $\cala^{(\sfN,\mathds{1})}_\Sigma[\sfN^{-1}F]$ would simply be the partition function of the three-dimensional Chern--Simons theory
\be
\frac{1}{4\pi}\sfN^{IJ}\oint_\Sigma a_I\wedge f_J
+\frac{1}{2\pi}\oint_\Sigma a_I\wedge F^I\,,
\ee
where $\sfN^{IJ}\equiv \sfN^I{}_K\delta^{KJ}$ and $a_I$ are world-volume U(1) gauge fields. However, it is not a priori obvious that vectors ${\bm\alpha}\in V_{\mathbb{Q}}$ with this property even exist, and one should generically use more complicated TQFTs, which nevertheless always admit a description in terms of abelian Chern--Simons theories \cite{Belov:2005ze}. Alternatively, $\cala^{(\sfN,\sfP)}_\Sigma[\sfN^{-1}F]$ can be identified more directly with the partition function of the four-dimensional TQFT defined on an auxiliary four-dimensional manifold $Y$ with boundary $\partial Y=\Sigma$. This theory is described by the  action 
\begin{equation} \label{eqn: mintqft 0f}
    \frac{1}{4 \pi} (\sfN^{\rm t} \, \sfP)_{IJ} \int_Y B^I \wedge B^J +\frac{1}{2 \pi} \sfN^I_{\; J} \int_Y B^J \wedge \tilde{F}_I  + \frac{1}{2 \pi} \oint_\Sigma \tilde{A}_I \wedge F^I \, ,
\end{equation}
where $\tilde{A}_I$ and $B^I$ are dynamical one- and two-form potentials living on Y.

The topological defect realizing a generic rational axion shift symmetry ${\bm a} \rightarrow {\bm a} \, + {\bm\alpha}$ for any $ {\bm\alpha} = \{ \alpha^i \} \in V_{\bbQ}$ can  then be obtained by stacking this TQFT on the  codimension-one operator $\exp\left({\ii \alpha^i \oint_{\Sigma}\calj_i^{\rm (a)}}\right)$: 
\begin{equation} \label{eqn: 0form defec}
    \cald^{\rm(a)}_{\bm\alpha}(\Sigma)\equiv  e^{\ii \alpha^i \oint_{\Sigma}\calj_i^{\rm (a)}} \cala^{(\sfN,\sfP)}_\Sigma[\sfN^{-1}F] \,.         
\end{equation}
In the following, we will analyze some simple properties of this defect. For the explicit construction, which shows that the defect is topological, we refer to \cite{Apruzzi:2025byj}.

One first simple but rather important property is the independence of the defect on the specific right-coprime factorization. This can for instance be checked by using \eqref{eqn: mintqft 0f}. Any two coprime factorizations are related  by a unimodular transformation $\sfT$, as discussed in Appendix \ref{app:coprime fact}, which can always be  reabsorbed by the field redefinition $B^I \rightarrow (\sfT^{-1})^I_{\; J} B^J$. Furthermore, one can check that \eqref{eqn: 0form defec} is invariant under ${\bm\alpha}\rightarrow{\bm\alpha}+{\bf m}$ for any ${\bf m}\in V_{\mathbb{Z}}$, consistently with the fact that \eqref{eqn: 0form defec} depends on ${\bm\alpha}$ only through its equivalence class in $V_{\mathbb{Q}}/V_{\mathbb{Z}}$ -- see Appendix \ref{app:shiftinv}.

One can also easily show that if we take ${\bm \alpha}$ such that $\sfC_{IJ}({\bm\alpha}) \in \bbZ$ then the defect assumes a simpler form. In particular, we can choose the factorization $\sfP= \sfC$ and $\sfN = \mathds{1}$. In this way, we can exactly integrate out the fields in \eqref{eqn: mintqft 0f} and find that \eqref{eqn: 0form defec} reduces to

\begin{equation}
    exp\lt \ii \oint_\Sigma \lq \alpha^i \calj^{\rm(a)}_i - \frac1{4\pi}\sfC_{IJ}({\bm\alpha}) \, A^I \wedge F^J \rq  \rt \, .
\end{equation}
Notice that the condition $\sfC_{IJ}({\bm\alpha})\in\mathbb{Z}$ can be satisfied not only for $\alpha^i\in\mathbb{Z}$ but also, depending on the form of $\sfK_{iIJ}$, for more general values $\alpha^i\in\mathbb{Q}$. In the former case the operator is trivial, whereas in the latter the corresponding defect is non-trivial, showing that a non-trivial invertible subgroup of U$(1)^{n_{\rm V}}$ survives.

The topological defect realizes the expected axion shift. Since the axions are periodic, they can enter any local operator only through factors of the form  $\exp\lq {2 \pi\ii \, \langle {\bf q} ,{\bm a}(x)\rangle}\rq$, with $\langle {\bf q} ,{\bm a}\rangle \equiv q_i a^i$ and $q_i\in\mathbb{Z}$. Picking a three-sphere  $\Sigma\simeq S^3$ that surrounds  the point $x$, we get   
\begin{equation}
    \cald^{\rm(a)}_{\bm\alpha}(\Sigma) e^{2 \pi \ii \langle {\bf q} ,{\bm a}(x)\rangle} = e^{2 \pi \ii \langle {{\bf q},{\bm \alpha}} \rangle} e^{2 \pi \ii \langle {\bf q} ,{\bm a}(x)\rangle}\, .
\end{equation}
From this equation it is clear that \eqref{eqn: 0form defec} acts in an invertible way on the charged local operators $\exp\lq {2 \pi\ii \, \langle {\bf q} ,{\bm a}(x)\rangle}\rq$. On the other hands,   when one analyzes the action on extended operators,  its non-invertible nature shows up. In particular 't Hooft lines are mapped to non-genuine line operators, because under the axionic shift the dual field strengths $G_I$ (see \eqref{dualG}) are mapped to $G_I+\sfC_{IJ}({\bm\alpha})F^J$, which are generally not properly quantized to be identified with U(1) field strengths. See \cite{Apruzzi:2025byj} for more details, as well as \cite{Choi:2022jqy,Cordova:2022ieu,Choi:2022fgx,Yokokura:2022alv,DelZotto:2024ngj} for the analogous phenomenon in the simpler axion-Maxwell model.

The non-invertible nature of \eqref{eqn: 0form defec} also shows up in the fusion $\cald^{\rm(a)}_{\bm\alpha} \times \overline \cald^{\rm(a)}_{\bm\alpha}$. 
In the particular cases in which  $\sfC({\bm\alpha})$ admits the simple factorization with $\sfP = \mathds{1}$, this fusion gives the condensation defect \cite{Apruzzi:2025byj} 
\begin{equation} \label{eqn:axicond}
    \begin{split}
        \cald^{\rm(a)}_{\bm\alpha} \times  \overline\cald^{\rm(a)}_{\bm\alpha} =  \sum_{{\bm\eta} \in H^1(\Sigma,\Gamma^*_\sfN)} & \exp \lq \ii (\sfN^{-1})^I{}_J \oint_\Sigma \eta_I \cup F^J +\pi \ii \delta^{IJ} \oint_\Sigma \eta_I \cup \beta_J(\bm{\eta})\rq\, ,
    \end{split}     
\end{equation} 
where $\bm{\beta}$ is the Bockstein image of $\bm\eta$, and for simplicity we have again omitted the overall normalization. For our purposes, it is important to notice that this condensation defect is non-trivial and involves the  magnetic one-form symmetries with parameters in $\sfN^{-1{\rm t}}\Gamma^*_{\sfN}\equiv (\sfN^{-1{\rm t}}W^*_\mathbb{Z})/W^*_{\mathbb{Z}}$. This shows that the existence of $\cald^{\rm(a)}({\bm\alpha})$ requires the presence of this subgroup of magnetic one-form symmetries. The same conclusion extends to more general $\sfC({\bm\alpha})$, as can also be seen from the half-space gauging construction of $\cald^{\rm(a)}_{\bm\alpha}$ \cite{Apruzzi:2025byj}, which explicitly makes use of the subgroup $\Gamma^*_{\sfN}$ of magnetic one-form symmetries.

Let us now reconsider the possibility of a non-vanishing axion-Pontryagin coupling \eqref{aRR}. (According to \cite{Martucci:2022krl}, such a coupling is actually expected to be necessarily  present in $\caln=1$ quantum gravity theories with a non-trivial gauge sector.) This term modifies the non-conservation equation \eqref{closedcurrents} by adding the term
\begin{equation}
     - \frac1{192\pi} \tilde \sfK_i \tr(R\wedge R)\,,
\end{equation}
to its right-hand side. In the presence of a non-flat metric, this appears to render the defect
\eqref{eqn: 0form defec} non-topological. However, as discussed in \cite{Putrov:2023jqi}, this type of violation can be canceled by an appropriate modification of \eqref{eqn: 0form defec}, which restores its topological nature. Briefly, one can add to the defect the counterterm
\begin{equation}
    e^{\frac{\ii}{48} \alpha^i\tilde{\sfK}_i \int_\Sigma {\rm GCS}(\omega)} \,,
\end{equation}
where ${\rm GCS}(\omega)$ is the gravitational Chern-Simons three-form, satisfying $\d{\rm GCS}(\omega)=\frac{1}{4\pi}\tr(R\wedge R)$. This term, together with the minimal TQFT supported on the defect, introduces a framing anomaly \cite{Witten:1988hf} with an overall rational coefficient. This anomaly can then be canceled by stacking the defect with a decoupled TQFT that has an appropriate rational chiral central charge \cite{Putrov:2023jqi}. 

Finally, as discussed more explicitly in Appendix \ref{app:nonabelian}, we note that the inclusion of non-abelian sectors generically removes part of the possible non-invertible axion shift symmetries. In what follows, we will continue to assume that these sectors are either absent, or that we have restricted attention to axions that do not couple to them.

%%%%%%%%%%%%%%%%%%%%%%%%%%%%%%%%%%%%%%%%%%%%%%%%%%%%%%%%%%%%%%%%%%%%%%%%%

\section{Symmetry breaking of higher-form symmetries}
\label{sec:physcons}

The models described in Section \ref{sec:aximodels0} should generally be regarded as low-energy effective field theories. Hence, the global symmetries discussed in Section \ref{sec:noninvsymm} may either uplift to global symmetries of the corresponding UV-complete theory or be approximate symmetries emerging in the infrared. Both possibilities can arise when gravity is decoupled. In a quantum gravity context, however, it is widely believed that the former possibility can never occur \cite{Misner:1957mt,Banks:2010zn,Harlow:2018tng}.

The purpose of this section is to discuss the possible mechanisms for breaking higher-form symmetries. Since higher-form symmetries act on extended operators, their breaking cannot occur without introducing new degrees of freedom into the EFT. The breaking of each higher-form symmetry can therefore be associated with the appearance of a corresponding charged sector, characterized by a symmetry-breaking energy scale below which the charged objects cannot be excited.
 
For each case, we first consider the simpler non-supersymmetric setting of Section \ref{sec:aximodels} -- see \cite{Brennan:2020ehu,Cordova:2022ieu,Choi:2022fgx,Cordova:2022fhg,Sehayek:2026pvu} for related discussions of the simpler axion-Maxwell model -- and then discuss what additional insights can be gained in the supersymmetric extensions of Section \ref{sec:susymodels}. Throughout, we will also strive to distinguish clearly between conclusions that rely on conjectural quantum gravity constraints and those that hold more generally.

\subsection{Winding symmetry breaking and axion strings}
\label{sec:windingbreak}

We proceed in the same order as in Section \ref{sec:noninvsymm}, and first consider the invertible winding two-form symmetries. Their breaking requires the existence of {\em dynamical} axion strings, namely finite-tension counterparts of the axion vortex operators $\calv_{\bf e}$ discussed in Section \ref{sec:winding}. We denote by $\cali^{\rm(w)}\subset V_\mathbb{Z}$ the set of axion charges populated by physical strings.

From \eqref{caldwaction}, a dynamical string with charge vector ${\bf e}\in\cali^{\rm(w)}$ breaks the winding symmetries with parameters $\beta_i\simeq \beta_i+1$ whenever $\langle{\bm\beta},{\bf e}\rangle\notin\mathbb{Z}$. Hence, for fixed ${\bm\beta}$, we can identify the following scale at which the corresponding winding two-form symmetry is certainly broken:
\be\label{wenergyscale} 
M^{\rm(w)}_{\bm\beta}\equiv
\min\left\{
\sqrt{2\pi\calt_{\bf e}}\ \big|\ 
{\bf e}\in\cali^{\rm(w)}\,,\ 
\langle{\bm\beta},{\bf e}\rangle\notin\mathbb{Z}
\right\}\,,
\ee
where $\calt_{\bf e}$ denotes the tension of the lightest string of charge ${\bf e}$. A rough motivation for \eqref{wenergyscale} is the following. At the EFT level, one can consider the nucleation of string loops. For a string not to be resolved below the EFT cutoff $\Lambda$, one should have $2\pi\calt_{\bf e}>\Lambda^2$, and the loop radius should satisfy $R>\Lambda^{-1}>1/\sqrt{2\pi\calt_{\bf e}}$. An estimate of the loop mass is then given by  $2\pi\calt_{\bf e}R>\sqrt{2\pi\calt_{\bf e}}$. This suggests that the winding symmetry in the direction ${\bm\beta}$ can only emerge at energies below $M^{\rm(w)}_{\bm\beta}$. Denoting by $E^{\rm(w)}_{\bm\beta}$ the actual  symmetry-breaking scale, we get the upper bound
\be\label{Ewbreak} 
E^{\rm(w)}_{\bm\beta}\lesssim M^{\rm(w)}_{\bm\beta}\,.
\ee
In particular, the winding symmetry may be broken at energies much smaller than $M^{\rm(w)}_{\bm\beta}$ if the string is not fundamental, namely if it admits a solitonic description within an intermediate four-dimensional EFT; see for instance \cite{Brennan:2020ehu}. On the other hand, in the absence of supersymmetry, little can be said about $\calt_{\bf e}$, and hence about $M^{\rm(w)}_{\bm\beta}$, without knowing something about the UV completion.

In the presence of supersymmetry, more can be said. The tension of BPS strings is fixed by supersymmetry to be \cite{Lanza:2019xxg}
\be\label{stringten} 
\calt_{\bf e}=M^2 e^i\ell_i\,,
\ee
where $\ell_i$ are the dual saxions introduced in \eqref{dualsax}, and $M=M_\Pl$ in the presence of gravity. 
Anti-BPS strings instead have tension $\calt_{\bf e}=-M^2 e^i\ell_i$. For a given string charge vector ${\bf e}\in\cali^{\rm(w)}$, we expect BPS strings to be the lightest and most stable ones. Therefore, using \eqref{stringten} in \eqref{wenergyscale} gives a more explicit expression for the upper bound \eqref{Ewbreak}.\footnote{In the presence of massless saxions, the interpretation of \eqref{stringten} is affected by the saxionic backreaction, which should be appropriately understood as an RG flow along the string \cite{Lanza:2020qmt,Lanza:2021udy}.}

In practice, applying \eqref{wenergyscale} requires knowing the spectrum of axion strings, not only their tensions. In quantum gravity models, further information follows from standard quantum gravity expectations. First, the absence of exact global symmetries \cite{Misner:1957mt,Banks:2010zn,Harlow:2018tng} implies that the set of string charges $\cali^{\rm(w)}$ must generate the entire lattice $V_\mathbb{Z}$. Furthermore, the Weak Gravity Conjecture (WGC) \cite{ArkaniHamed:2006dz} for multiple gauge fields \cite{Cheung:2014vva} -- see \cite{Harlow:2022gzl} for a review and more references -- implies that the convex hull of the vectors
\be 
{\bf w}_{\bf e}\equiv\frac{M^2_\Pl{\bf e}}{\calt_{\bf e}}\in V_\mathbb{R}
\ee
with ${\bf e}\in \cali^{\rm(w)}$ should contain a ball of radius $\gamma\sim\calo(1)$, as measured by the metric $\calg_{ij}$ appearing in \eqref{axionlag0}. This follows from \eqref{B2kinetic}, with $M=M_\Pl$, since $\calg_{ij}$ defines the couplings of the dual two-form gauge fields. In particular, the lightest strings are expected to satisfy the WGC bound
\be\label{stringWGB} 
\frac{M^2_\Pl\calq_{\bf e}}{\calt_{\bf e}}
=\|{\bf w}_{\bf e}\|
\geq \gamma\,,
\qquad
\calq^2_{\bf e}\equiv \calg_{ij}e^ie^j=\|{\bf e}\|^2\,.
\ee 
Hence, for such strings, $\calt_{\bf e}\leq M^2_\Pl \calq_{\bf e}/\gamma$, which gives the bound
\begin{equation}
    M^{\rm(w)}_{\bm\beta}
    \leq
    M_\Pl
    \min\left\{
    \sqrt{\frac{2\pi\|{\bf e}\|}{\gamma}}\ \big|\ 
    {\bf e}\in\cali^{\rm(w)}\,,\ 
    \langle{\bm\beta},{\bf e}\rangle\notin\mathbb{Z}
    \right\}\,.
\end{equation}

The above WGC considerations should hold independently of supersymmetry, but they can be sharpened in the supersymmetric models described in Section \ref{sec:susymodels}. Using \eqref{ellsrel} and \eqref{ellsnorms} in \eqref{stringten}, the Cauchy--Schwarz inequality gives
\be\label{stringWGC} 
\calt_{\bf e}
=
M^2_\Pl \calg_{ij}e^is^j
\leq
M^2_\Pl \|{\bf e}\|\|{\bm s}\|
=
\sqrt{\frac{k}{2}}\,M^2_\Pl\calq_{\bf e}\,.
\ee
Thus, in models with K\"ahler potential \eqref{loPK}, the BPS string tension \eqref{stringten} satisfies a WGC bound \eqref{stringWGB} with $\gamma_{\text{\tiny BPS}}=\sqrt{2/k}$. A more detailed discussion on this bound, including the subtleties in the definition of string tension, can be found  \cite{Lanza:2020qmt,Lanza:2021udy}\footnote{As discussed in \cite{Lanza:2020qmt}, \eqref{stringWGC} holds also for more general no-scale models, with $K^{i{\bar\jmath}}K_i K_{\bar\jmath}=k$, and not just those with K\"ahler potential \eqref{loPK}.}; see also the recent discussion in \cite{Etheredge:2026rio}.

\begin{figure}[ht]
\centering
\begin{subfigure}{0.42\textwidth}
\centering
\includegraphics[width=\textwidth]{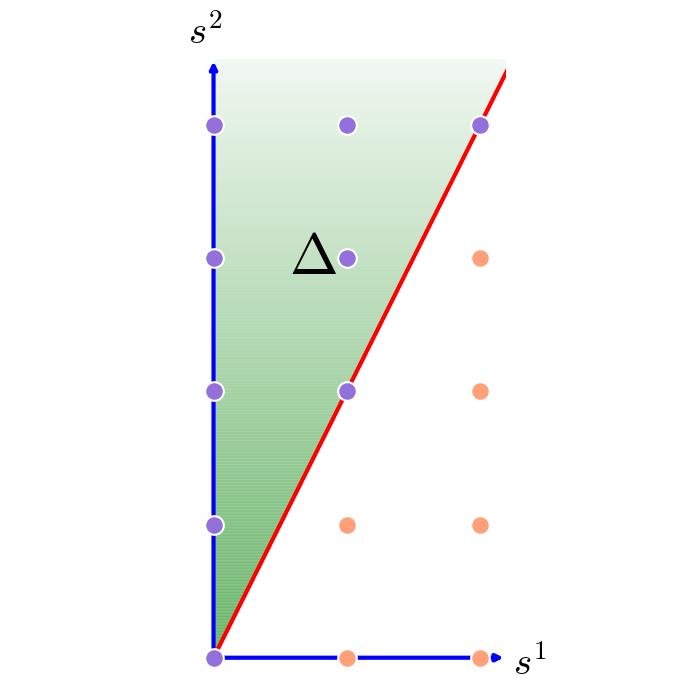}
\end{subfigure}
\hfill
\caption{The violet and orange bullets represent the set $\calc_{\rm S}$ of BPS string charges of the $\mathbb{P}^1\hookrightarrow X\rightarrow \mathbb{P}^2$ F-theory model, with twist parameter $p=2$, discussed in Section \ref{sec:susyexam}. The violet bullets represent the subset $\calc^{\text{\tiny EFT}}_{\rm S}$ of EFT string charges, while the orange bullets represent BPS but non-EFT string charges.}
\label{fig:BPSstrings}
\end{figure}

As argued in \cite{Lanza:2020qmt,Lanza:2021udy}, there is a distinguished subclass of BPS axion strings, the {\em EFT strings}, which are genuinely gravitational in nature and have special properties. In particular, they can saturate the WGC bound with $\gamma_{\text{\tiny BPS}}=\sqrt{2/k}$. More precisely, the set $\calc^{\text{\tiny EFT}}_{\rm S}$ of EFT string charges can be identified with the discretization of the closure of the saxionic cone $\Delta$ introduced in Section \ref{sec:susymodels}:
\be\label{CSEFT} 
\calc^{\text{\tiny EFT}}_{\rm S}
=
\overline\Delta\cap V_\mathbb{Z}\,.
\ee
See Figure \ref{fig:BPSstrings} for an illustration. Since ${\bm s}\in\Delta$, the bound \eqref{stringWGC} can be saturated by taking ${\bm s}$ aligned, or arbitrarily close, to the direction identified by ${\bf e}$.

There may exist other BPS strings, filling out a larger set of charges $\calc_{\rm S}\supset \calc^{\text{\tiny EFT}}_{\rm S}$. In concrete string theory compactifications, BPS strings typically arise from higher-dimensional branes or bundles localized on internal supersymmetric cycles. EFT strings are distinguished by their association with ``movable'' internal configurations, which can  recombine and explore the full configuration space \cite{Lanza:2021udy}. In particular, string theory realizations suggest that only EFT strings generically satisfy completeness, namely, that for each  ${\bf e}\in\calc^{\text{\tiny EFT}}_{\rm S}$ there exists a physical string carrying charge ${\bf e}$. This  would also guarantee that the winding symmetries are fully broken by EFT strings alone, without the need to invoke additional axion strings, whether BPS or not. From the quantum gravity viewpoint, non-EFT strings appear accidental rather than necessary.

Since EFT strings are fundamental, their tension directly sets the corresponding winding symmetry-breaking scale. We may therefore identify the {\em quantum gravity} winding symmetry-breaking scale with
\be\label{wenebound} 
E^{\rm(w)}_{\text{\tiny QG},{\bm\beta}}({\bm\ell})
\equiv
\min\left\{
M_\Pl \sqrt{2\pi \langle{\bm\ell},{\bf e}\rangle}\ \big|\ 
{\bf e}\in\calc^{\text{\tiny EFT}}_{\rm S}\,,\ 
\langle{\bm\beta},{\bf e}\rangle\notin\mathbb{Z}
\right\}\,,
\ee
where we have emphasised the dependence of $E^{\rm(w)}_{\text{\tiny QG},{\bm\beta}}$ on the  moduli space position. 
An additional piece of information comes from the relation between EFT strings and the species scale $M_{\rm sp}$ \cite{Dvali:2007hz,Dvali:2007wp,Dvali:2009ks,Dvali:2010vm}, namely the scale at which any EFT description of the gravitational model is expected to break down. As proposed in \cite{Martucci:2024trp} -- see also \cite{Bedroya:2024ubj} -- this relation is encoded in
\be
\label{speciesbound} 
M^2_{\rm sp}\leq 2\pi \calt^{\text{EFT string}}_{\bf e}\leq M^2_\Pl\,.
\ee  
Applying \eqref{speciesbound} to \eqref{wenebound}, we obtain
\be
\label{EQGbounds} 
M_{\rm sp}
\leq
E^{\rm(w)}_{\text{\tiny QG},{\bm\beta}}
\leq
M_\Pl\,.
\ee

To summarize, the chain of arguments above suggests that, except for accidental winding-symmetry breaking due to non-EFT strings, genuinely quantum-gravitational breaking occurs at energies above the species scale, as expected. In the presence of non-EFT strings, one may instead have breaking at a lower scale $E^{\rm(w)}_{{\bm\beta}}<E^{\rm(w)}_{\text{\tiny QG},{\bm\beta}}$, possibly even below the species scale, but this should not be interpreted as originating from quantum gravity. It would be interesting to develop this idea further, for instance by adapting to our $\caln=1$ setting the framework recently proposed in \cite{Aoufia:2026mqb}.

Finally, we recall that, as in \cite{Martucci:2024trp}, completeness and quantum consistency of EFT strings can be used to argue that, in quantum gravity models,
\be\label{GBbound} 
r({\bf e})\geq 1
\quad \Rightarrow\quad
\langle \tilde\sfK,{\bf e}\rangle
\equiv
\tilde\sfK_i e^i
\geq 1\,,
\qquad
\forall{\bf e}\in\calc_{\rm S}^{\text{\tiny EFT}}\,,
\ee
where $r({\bf e})$ is the rank of the abelian and non-abelian gauge sector detected by the EFT string. In the models of Section \ref{sec:susymodels},
\be
r({\bf e})= {\rm rank}\,\langle\sfK_{IJ},{\bf e}\rangle\,,
\ee 
where the matrix $\langle\sfK_{IJ},{\bf e}\rangle\equiv \sfK_{IJi}e^i$ is determined by the couplings \eqref{gaugecoupl}.

%%%%%%%%%%%%%%%%%%%%%%%%%%%%%%%%%%%%%%%%%%%%%%%%%%%%%%%%%%%%%%

\subsection{Magnetic one-form symmetry breaking}
\label{sec:magn}

Similarly to axion strings, dynamical magnetic monopoles, which are the finite mass counterparts of the 't Hooft lines $H_{\bf m}$ introduced in Section \ref{sec:magnetic}, provide the natural mechanism for breaking the invertible magnetic one-form symmetries. We denote by $\cali^{\rm(m)}\subset W_\mathbb{Z}$ the set of magnetic charges populated by dynamical monopoles. From \eqref{magnaction}, we see that the magnetic one-form symmetry associated with a given choice of  $\beta_I$ is certainly broken at the energy scale
\be\label{Mmag} 
M^{\rm(m)}_{\bm\beta}\equiv
\min\left\{
M_{\bf m}\ \big|\ 
{\bf m}\in\cali^{\rm(m)}\,,\ 
\langle{\bm\beta},{\bf m}\rangle \notin\mathbb{Z}
\right\}\,,
\ee
where $M_{\bf m}$ is the mass of the lightest monopole of charge ${\bf m}$. As for \eqref{wenergyscale}, $M^{\rm(m)}_{\bm\beta}$  is expected to provide only an upper bound on the actual magnetic one-form symmetry-breaking scale $E^{\rm(m)}_{\bm\beta}$:
\be 
E^{\rm(m)}_{\bm\beta}\lesssim M^{\rm(m)}_{\bm\beta}\,.
\ee
The scale $E^{\rm(m)}_{\bm\beta}$ can be significantly smaller than $M^{\rm(m)}_{\bm\beta}$ for non-fundamental monopoles that admit a solitonic uplift within an intermediate four-dimensional EFT. For instance, in the SU(2) model giving the 't Hooft-Polyakov \cite{tHooft:1974kcl,Polyakov:1974ek} uplift of the Dirac monopole, the magnetic one-form symmetry is broken at the $W$-boson mass $E^{\rm(m)}\simeq gv$, where $g\ll 1$ is the gauge coupling and $v$ sets the scalar VEV, while the monopole mass is $M^{\rm(m)}\simeq v/g\gg E^{\rm(m)}$.

Unlike axion strings, monopoles in $\caln=1$ models completely break the bulk supersymmetry. Therefore, the computation of their mass depends a priori on details of the UV completion. In quantum gravity models, one may try to invoke the magnetically dual version of the WGC. This would put an upper bound of order $\sqrt{f_{IJ}m^Im^J}\,M_\Pl$ on the lightest monopoles, and hence on \eqref{Mmag}. In the weak-coupling regime assumed here, however, this scale is much larger than the Planck mass and therefore does not provide useful information. Some universal features of \eqref{Mmag} could perhaps be extracted from explicit string-theory constructions. In any case, as in the axion-string case, we expect some subclass of monopoles to stand out as genuinely gravitational, and to set the scale of the quantum-gravitational breaking of the magnetic one-form symmetries. We leave a more detailed investigation of these aspects to future work.

%%%%%%%%%%%%%%%%%%%%%%%%%%%%%%%%%%%%%%%%%%%%%%%%%%%%%

\subsection{Electric one-form symmetry breaking and hierarchies}
\label{sec:elbreak}

Let us finally turn to the electric one-form symmetry described in Section \ref{sec:el1form}. Its breaking is associated with the presence of electrically charged matter, which can be regarded as the finite-mass counterpart of the Wilson lines \eqref{Wline}. As in the other cases, one should identify the set of electric charges $\cali^{\rm(e)}\subset W^*_\mathbb{Z}$ populated by physical particles.

In this case, it is natural to identify the symmetry-breaking scale associated with a given non-invertible topological operator \eqref{Del} with the mass of the lightest particle charged under it. Recalling \eqref{elDaction}, for any ${\bm\alpha}\in W_\mathbb{Q}/W_\mathbb{Z}$ we can then define
\be\label{Mel} 
E^{\rm(e)}_{\bm\alpha}\equiv
\min\left\{
M_{\bf n}\ \big|\ 
{\bf n}\in\cali^{\rm(e)}\,,\ 
\langle {\bf n},{\bm\alpha}\rangle \notin\mathbb{Z}
\right\}\,,
\ee
where $M_{\bf n}$ is the mass of the lightest particle of charge ${\bf n}$.

As in \cite{Choi:2022fgx}, the non-invertibility of the electric one-form symmetries introduces an interesting correlation between their breaking scales and the symmetry-breaking scales discussed above; see also \cite{Brennan:2020ehu,Sehayek:2026pvu}. Following \cite{Choi:2022fgx}, this correlation can be understood in terms of the condensation defect \eqref{elfusion}, whose definition relies on the magnetic one-form and winding two-form symmetries. More precisely, preserving the electric one-form symmetry in a given direction ${\bm \alpha}\in W_\mathbb{Q}/W_\mathbb{Z}$ requires that the winding and magnetic symmetries appearing in its condensation defect remain unbroken. Their parameters take values in the following finite groups: 
\be\label{QGamma} 
{\bm\beta}^{\rm(w)}
\in
\sfQ\Gamma_\sfL({\bm\alpha})
\subset
V^*_\mathbb{Q}/V^*_\mathbb{Z}\,,
\qquad
{\bm\beta}^{\rm(m)}
\in
\sfL^{-1{\rm t}}\Gamma^*_\sfL({\bm\alpha})
\subset
W^*_\mathbb{Q}/W^*_\mathbb{Z}\,,
\ee
where $\Gamma_\sfL({\bm\alpha})$ and $\Gamma^*_\sfL({\bm\alpha})$ were introduced in Section \ref{sec:el1form}, and here we emphasize their   dependence on ${\bm \alpha}\in W_\mathbb{Q}/W_\mathbb{Z}$. Moreover, the group   $\sfQ\Gamma_\sfL({\bm\alpha})$ is the image of $\Gamma_\sfL({\bm\alpha})$ into $V^*_\mathbb{Q}$ mod $V^*_\mathbb{Z}$, under the linear map $\sfQ:W_\mathbb{Z}\rightarrow V^*_\mathbb{Q}$ defined by the matrix $\sfQ$ introduced in \eqref{sfQmatrix}.  
 Importantly, the finite groups appearing in \eqref{QGamma} do not depend on the particular coprime factorization \eqref{QPNaxion} one chooses, but only on $\sfQ(\bm\alpha)$. Hence, we obtain the hierarchy
\begin{equation} \label{EEbound} 
\begin{split}
E^{\rm(e)}_{\bm\alpha}
& \lesssim
\min\left\{
E^{\rm(w)}_{{\bm\beta}^{\rm(w)}}\,,
E^{\rm(m)}_{{\bm\beta}^{\rm(m)}}
\ \big|\ 
{\bm\beta}^{\rm(w)}\in \sfQ\Gamma_\sfL({\bm\alpha})\,,
{\bm\beta}^{\rm(m)}\in \sfL^{-1{\rm t}}\Gamma^*_\sfL({\bm\alpha})
\right\}
\\
&\lesssim
\min\left\{
M^{\rm(w)}_{{\bm\beta}^{\rm(w)}}\,,
M^{\rm(m)}_{{\bm\beta}^{\rm(m)}}
\ \big|\ 
{\bm\beta}^{\rm(w)}\in \sfQ\Gamma_\sfL({\bm\alpha})\,,
{\bm\beta}^{\rm(m)}\in \sfL^{-1{\rm t}}\Gamma^*_\sfL({\bm\alpha})
\right\}\,.
\end{split}
\end{equation}

This hierarchy looks intricate, but it has a clear physical interpretation. Let us first focus on the appearance of $M^{\rm(m)}_{{\bm\beta}^{\rm(m)}}$. This can be understood by recalling that a monopole of charge vector ${\bf m}\in W_{\mathbb{Z}}$, like the corresponding 't Hooft loop, must generically support a world-line sector as in \eqref{1dphase}. This means that it couples to the bulk gauge fields through the combination $-m^I\sfK_{iIJ}A^J$. This coupling breaks the electric one-form symmetry with parameters $\alpha^I\in\mathbb{Q}$ mod $\mathbb{Z}$ such that $\sfK_{iIJ}\alpha^Jm^I\equiv \sfQ_{iI}({\bm\alpha})m^I$ is not an integer, namely such that  $\sfQ({\bm\alpha}){\bf m}\notin V^*_{\mathbb{Z}}$. Using the right-coprime factorization \eqref{QPNaxion}, it is easy to see that this is equivalent to requiring $\sfL^{-1}({\bm\alpha}){\bf m}\notin W_{\mathbb{Z}}$. From \eqref{magnaction}, we see that the monopoles satisfying this condition are precisely those that break the magnetic one-form symmetries with $\beta^{\rm (m)}_I=(L^{-1})^J{}_I k_J$ mod $\mathbb{Z}$ for some $k_J\in\mathbb{Z}$, that is, with ${\bm\beta}^{\rm (m)}\in (\sfL^{-1{\rm t}}W^*_{\mathbb{Z}})/W^*_{\mathbb{Z}}= \sfL^{-1{\rm t}}\Gamma^*_\sfL$. Hence \eqref{EEbound} simply means that $E^{\rm (e)}_{{\bm\alpha}}$ is bounded by the mass of the lightest monopole that breaks the non-invertible electric one-form symmetry identified by ${\bm\alpha}\in V_{\mathbb{Q}}$.

A similar argument explains the appearance of $M^{\rm(w)}_{{\bm\beta}^{\rm(w)}}$. Indeed, as already remarked, an axion string of charge vector ${\bf e}\in V_\mathbb{Z}$ must contain a chiral world-sheet sector that couples to the bulk gauge fields $A^I$ and produces an anomaly that cancels \eqref{vortexanomaly}. One simple possibility is provided by a set of world-sheet chiral bosons $\phi_I$ that shift under the bulk gauge symmetries, with corresponding covariant derivative $D\phi_I=\d\phi_I-e^i\sfK_{iIJ}A^J$; see e.g.\ \cite{Naculich:1987ci}. This minimal coupling breaks the electric one-form symmetry with parameters $\alpha^I\in\mathbb{Q}$ mod $\mathbb{Z}$ such that $e^i\sfK_{iIJ}\alpha^J\equiv e^i\sfQ_{iI}({\bm\alpha})\notin \mathbb{Z}$, namely such that $\sfQ^{\rm t}({\bm\alpha}){\bf e}\notin W^*_{\mathbb{Z}}$. From \eqref{caldwaction}, it is clear that  the strings that satisfy this condition are precisely those that break the winding symmetries with ${\beta}^{\rm(w)}_i=\sfQ_{iI}({\bm\alpha}) k^I$ mod $\mathbb{Z}$ for some $k^I\in\mathbb{Z}$, that is, with ${\bm{\beta}}^{\rm(w)}\in \sfQ\Gamma_\sfL({\bm\alpha})$. Thus \eqref{EEbound} implies that $E^{\rm (e)}_{{\bm\alpha}}$ is also bounded by the energy scale set by the lightest axion string that breaks the non-invertible electric one-form symmetry identified by ${\bm\alpha}$.

These observations give a clean physical interpretation of the bounds \eqref{EEbound}, make manifest that they do not depend on the choice of coprime factorization \eqref{QPNaxion}, and provide a non-trivial  self-consistency check of the overall framework. Of course, the non-invertible electric one-form symmetry can be broken at energy scales which are smaller than those set by the above monoples and strings, for instance by the appearance of charged matter.

The field-theoretic result \eqref{EEbound} is universal, but its concrete realization can be strongly model dependent. Introducing quantum gravity input provides further information. First, as in \cite{Choi:2022fgx,Rudelius:2020orz,Heidenreich:2021xpr}, the expected absence of global symmetries \cite{Misner:1957mt,Banks:2010zn,Harlow:2018tng} requires the set of physical charges $\cali^{\rm(e)}$ to generate the entire lattice $W^*_\mathbb{Z}$, thereby realizing the completeness hypothesis \cite{Polchinski:2003bq,Banks:2010zn}. Furthermore, the WGC \cite{ArkaniHamed:2006dz}, in the multifield formulation of \cite{Cheung:2014vva}, imposes upper bounds on $E^{\rm(e)}_{\bm\alpha}$; see \cite{Harlow:2022gzl} for a review and further references. It is also interesting to observe that, in our ${\cal N}=1$ framework, among the possible axion strings, the non-gravitational ones appear to give the strongest contribution to the upper bound \eqref{EEbound}. Indeed, as recalled in Section~\ref{sec:windingbreak}, the species scale sets the lower bound \eqref{speciesbound} on the tension of EFT strings. Hence, their contribution to the right-hand side of \eqref{EEbound} lies beyond the ultimate quantum-gravity EFT cutoff. It would be interesting to understand whether a similar conclusion holds for the monopole contribution to \eqref{EEbound}. However, since monopoles are not BPS in minimally supersymmetric models, addressing this point probably requires an analysis of concrete string theory constructions.

%%%%%%%%%%%%%%%%%%%%%%%%%%%%%%%%%%%%%%%%%%%%%%%%%%%%%%%%%%%

\section{Axion symmetry breaking and naturalness}
\label{sec:axionat}

Compared to the higher-form cases, the breaking of the zero-form symmetries discussed in Section \ref{sec:ax0form}  is  qualitatively very different,\footnote{We thank Luca Vecchi for useful discussions related to these aspects.} since it  can be encoded in a modification of the local four-dimensional Lagrangian. In particular, in absence of non-minimal supersymmetry, 
once the  axion shift symmetries have been broken,  nothing generically prevents the generation of a non-trivial axion potential $V(a)$, e.g.\  by quantum corrections. This means that the  axion shift symmetries of Section \ref{sec:ax0form}  cannot precisely be regarded as emergent at low energies. Rather, they hold (approximately) at intermediate energy scales between the energy  scales $E^{\rm (a)}_{\text{\tiny IR}}$ set by  $V(a)$ -- say given by the corresponding axion mass -- and a symmetry breaking scale $ E^{\rm (a)}_{\text{\tiny UV}}$ that characterizes symmetry breaking effects that are suppressed at energies much lower than $E^{\rm (a)}_{\text{\tiny UV}}$, say associated with higher-derivative operators. 

So, in order to talk about approximate shift symmetry, we need  a hierarchy $E^{\rm (a)}_{\text{\tiny IR}}\ll E^{\rm (a)}_{\text{\tiny UV}}$, which is a priori not guaranteed. In such a case, for energies   $E^{\rm (a)}_{\text{\tiny IR}}\ll E \ll E^{\rm (a)}_{\text{\tiny UV}}$, the symmetry breaking strength is parametrized by powers of $E/E^{\rm (a)}_{\text{\tiny UV}}\ll 1$ and $E^{\rm (a)}_{\text{\tiny IR}}/E\ll 1$. The key point is that the axion shift symmetry is restored for $E^{\rm (a)}_{\text{\tiny IR}}\rightarrow 0$ and $E^{\rm (a)}_{\text{\tiny UV}}\rightarrow \infty$.  It is then {\em technically  natural} \cite{tHooft:1979rat} to have $E^{\rm (a)}_{\text{\tiny IR}}\ll E^{\rm (a)}_{\text{\tiny UV}}$. 

On the other hand, this technical naturalness requires  a mechanism that ensures that $E^{\rm (a)}_{\text{\tiny IR}}\ll E^{\rm (a)}_{\text{\tiny UV}}$ is small in the first place. In particular, in order to address this point, one should take into account  the possible non-trivial constraints which, as in \cite{Choi:2022fgx}, can be associated with the non-invertible nature of the axion shift symmetries. As emphasized in Section \ref{sec:ax0form}, these zero-form  symmetries cannot exist in the absence of magnetic one-form symmetries. This connection is  easily seen by considering the condensation defect \eqref{eqn:axicond}. This shows that for a given ${\bm\alpha}\in V_\mathbb{Q}$, defined mod $V_{\mathbb{Z}}$, the topological defect $\cald^{\rm(a)}_{\bm\alpha}$ defined in \eqref{eqn: 0form defec} requires that the magnetic one-form symmetries with parameters ${\bm\beta}\in \sfN^{-1{\rm t}}\Gamma^*_\sfN({\bm\alpha})\in W^*_\mathbb{Q}/W^*_\mathbb{Z}$ are not broken. We recall that $\Gamma^*_\sfN({\bm\alpha})\equiv W^*_\mathbb{Z}/(\sfN^{\rm t}W^*_\mathbb{Z})$, and we have emphasized the dependence of this discrete group  on ${\bm\alpha}$. By consistency, for any  shift symmetry  ${\bm \alpha}\in V_\mathbb{Q}/V_{\mathbb{Z}}$, we get the following upper bounds 
\begin{equation} \label{Eabound}
\begin{aligned}
  E^{\rm (a)}_{\text{\tiny UV},{\bm\alpha}}&\lesssim  \min\{E^{\rm (m)}_{\bm\beta}\ |\  {\bm\beta}\in \sfN^{-1{\rm t}}\Gamma^*_\sfN({\bm\alpha})\}\\
  &\lesssim M^{\rm (a)}_{\text{\tiny UV},{\bm\alpha}}(\text{mon})\equiv\min\{M^{\rm (m)}_{\bm\beta}\ |\  {\bm\beta}\in \sfN^{-1{\rm t}}\Gamma^*_\sfN({\bm\alpha})\}\,.  
  \end{aligned}
  \end{equation}

As for the bound \eqref{EEbound},  \eqref{Eabound} admits a clear physical interpretation. Recall that, as for the 't Hooft lines discussed in Section \ref{sec:magnetic}, monopoles must support a world-line theory like \eqref{1dphase}. Hence, a monopole of charge vector ${\bf m}\in W_\mathbb{Z}$ couples to the bulk axions through the combination $-a^i\sfK_{iIJ}m^J$. This coupling breaks the bulk axion shift symmetries with parameters $\alpha^i\in\mathbb{Q}$ such that $-\alpha^i\sfK_{iIJ}m^J\equiv \sfC_{IJ}({\bm\alpha})m^J$ is not an integer, that is, such that $\sfC({\bm\alpha}){\bf m}\notin W^*_{\mathbb{Z}}$. By using the right-coprime factorization \eqref{0formCfact}, it is easy to see that this is equivalent to requiring that $\sfN^{-1}({\bm\alpha}){\bf m}\notin W_{\mathbb{Z}}$. The monopoles whose charge vector satisfies this condition are precisely the monopoles that completely  break the magnetic one-form symmetries  associated with ${\bm\beta}\in (\sfN^{-1{\rm t}}W^*_{\mathbb{Z}})/W^*_{\mathbb{Z}}\equiv \sfN^{-1{\rm t}}\Gamma^*_\sfN$. Hence \eqref{Eabound} just means that $E^{\rm (a)}_{\text{\tiny UV},{\bm\alpha}}$ is bounded by the mass of the lightest monopole  that breaks the non-invertible shift symmetry identified by ${\bm\alpha}\in V_{\mathbb{Q}}$.   

This physical interpretation of \eqref{Eabound} also suggests that,  in absence of other symmetry breaking sources, we could set $E^{\rm (a)}_{\text{\tiny UV},{\bm\alpha}}=M^{\rm (a)}_{\text{\tiny UV},{\bm\alpha}}(\text{mon})$. Furthermore,  there should exist a mechanism that communicates the magnetic one-form symmetry breaking to $V(a)$, and hence to    $E^{\rm (a)}_{\text{\tiny IR},{\bm\alpha}}$. Such a mechanism was identified in \cite{Fan:2021ntg},  which discusses how  quantum loops  of dynamical monopole  can induce a non-trivial axion potential. In our framework, this must originate precisely from the interaction between   the bulk axions and the monopole world-line sector that was the key to explain \eqref{Eabound}. The relevance of this mechanism in the context of non-invertible axion symmetries has been already  emphasized in  \cite{Cordova:2022ieu,Cordova:2022fhg,Choi:2022fgx}.  Even if we have not worked out a precise form of this potential in our more general axiverse models, it should share the same qualitative features highlighted in \cite{Cordova:2022ieu,Cordova:2022fhg,Choi:2022fgx}.  
In particular, at weak gauge coupling,  the monopole generated axion potential should be exponentially suppressed, hence realizing the technically natural hierarchy $E^{\rm (a)}_{\text{\tiny IR}}(\text{mon})\ll E^{\rm (a)}_{\text{\tiny UV}}(\text{mon})$. %Notice that, in some cases, this type of monopole generated potential  could also be understood from the ultraviolet point of view as the familiar axion potential generated by non-abelian instantons \cite{Fan:2021ntg,Cordova:2022ieu}, and can hence be understood in purely rigid field theory terms. 

In any case, as already emphasized, nothing excludes other symmetry breaking contributions which may compete with the monopole ones. So, we may regard the possible IR symmetry breaking scale induced by monopole loops as a setting the lower bound
\be 
E^{\rm (a)}_{\text{\tiny IR},{\bm\alpha}}\gtrsim E^{\rm (a)}_{\text{\tiny IR},{\bm\alpha}}(\text{mon})\,.
\ee
Since at weak coupling $E^{\rm (a)}_{\text{\tiny IR}}(\text{mon})$ should be exponentially suppressed, it is still natural to assume  $E^{\rm (a)}_{\text{\tiny IR}}\ll E^{\rm (a)}_{\text{\tiny UV}}$.

Note that, a priori, in axiverse models, these types of   monopole effects could not be sufficient to break all  axion shift symmetries. For instance, this clearly happens if the number of axions is much larger than the total rank of the gauge sector. Moreover, in  minimally supersymmetric models, monopoles cannot be BPS and hence do not seem sufficient to generate a superpotential, and hence a potential. So,  the leading contribution to the above symmetry breaking scales  could actually have a different origin. In particular, among the various possibilities, it would be important to have a criterion  for estimating  symmetry breaking contributions that are genuinely quantum gravitational in nature and to verify under which conditions they can be interpreted  as a technically natural violation of the  axion shift symmetries. In the next section we will discuss such a criterion.

%%%%%%%%%%%%%%%%%%%%%%%%%%%%%%%%%%%%%%%%%%%%%%%%%%%%%%%%%%%%%%%%%%%%%%%%%%%%%%%%%%%%%%%

\section{Axion symmetry breaking and imaginary wormholes}
\label{sec:WHs}

Since the 1980s, wormholes have provided a window into the non-perturbative quantum-gravitational properties of low-energy effective field theories. In particular, as in the seminal paper of Giddings and Strominger (GS) \cite{Giddings:1987cg}, they arise naturally in axion models, and their contribution to the effective field theory has been interpreted as a purely bottom-up indication that axion shift symmetries should be broken in quantum gravity; see e.g.\ \cite{Abbott:1989jw,Coleman:1989zu,Kallosh:1995hi} and \cite{Hebecker:2018ofv} for a review and a more complete list of references.\footnote{In particular, see  \cite{Montero:2015ofa,Brown:2015iha,Heidenreich:2015nta,Hebecker:2016dsw,Hertog:2017owm,Shiu:2018wzf,Daus:2020vtf,Andriolo:2020lul,Andriolo:2022rxc,Loges:2023ypl,Eichhorn:2024rkc} for a sample of works discussing the role of axion wormholes in investigating  quantum gravity constraints on  EFTs.}
Wormholes also raise various puzzles and technical subtleties. For instance, their proposed interpretation \cite{Coleman:1988cy} involves an average over EFT couplings corresponding to different `$\alpha$-vacua' -- see also the more recent \cite{Marolf:2020xie} -- which seems to conflict with what one encounters in string theory models; see e.g.\ \cite{McNamara:2020uza}. In particular, in the AdS/CFT context they lead to the factorization puzzle \cite{Maldacena:2004rf,Arkani-Hamed:2007cpn}. Moreover, the stability of the simple Giddings-Strominger wormhole and some of its generalizations has been verified only recently \cite{Loges:2022nuw,Jonas:2023qle,Hertog:2024nys,Marolf:2025evo}; see also \cite{Held:2026huj}.

Recently, \cite{DiUbaldo:2026rly,Maldacena:2026jqd} proposed a novel strategy to extract important physical information from axion wormholes, as well as from more general ones. In the absence of a cosmological constant, axion wormholes connect two asymptotically flat Euclidean spacetimes. One can then work with Dirichlet boundary conditions, fixing the asymptotic values of the axions, and of any other scalar fields, on each side of the wormhole.
The strategy of \cite{DiUbaldo:2026rly,Maldacena:2026jqd} is based on considering what happens if one analytically continues these Dirichlet boundary conditions to imaginary values. The key observation is that, even though for real asymptotic moduli the wormhole contributions are exponentially suppressed, they become more important under this analytic continuation, eventually giving a divergent contribution at sufficiently large imaginary distance.
In \cite{DiUbaldo:2026rly,Maldacena:2026jqd}, it is argued that this divergence signals a fundamental pathology that needs to be cured by some additional microscopic contribution to the EFT. This led \cite{DiUbaldo:2026rly,Maldacena:2026jqd} to propose an {\em imaginary distance bound} (IDB) on the maximal imaginary proper distance that can be traveled by the asymptotic moduli before these modifications become important. More specifically, in the case of imaginary axion distances, the most natural modifications arise from fundamental instanton contributions.

Although the logic proposed in \cite{DiUbaldo:2026rly,Maldacena:2026jqd} is general, the properties of the wormhole and, therefore, of the associated EFT modification implied by the IDB, depend strongly on the EFT from which one starts. In particular, the wormhole solutions considered in \cite{DiUbaldo:2026rly,Maldacena:2026jqd}  assume that the relevant moduli have trivial potential. From a standard EFT point of view, in non-supersymmetric or minimally supersymmetric models this condition can be guaranteed only in the presence of appropriate global symmetries.

This is precisely what happens in our axiverse models. Indeed, in a non-gravitational framework, the axion shift symmetries discussed in Section \ref{sec:ax0form} obstruct the presence of a non-trivial potential or superpotential for the axions. Thus, in a certain sense, these symmetries are responsible for the existence of the specific types of wormholes that appear once gravity becomes dynamical. According to \cite{DiUbaldo:2026rly,Maldacena:2026jqd}, these wormholes  in turn signal the quantum-gravitational inconsistency of the symmetries one starts with. The proposal of \cite{DiUbaldo:2026rly,Maldacena:2026jqd} provides a recipe for extracting from  wormholes more quantitative, though still coarse-grained, information about the symmetry-breaking modifications of the EFT required by quantum gravity.

%In other words, the EFT axion shift symmetry must be interpreted as a symmetry of the ensemble-averaging interpretation of the gravitational path integral. In connected saddles with a single asymptotically flat boundary, the axion shift symmetry is a symmetry of the coarse-grained interpretation of the gravitational path integral.

In the following we will discuss the application of the proposal of \cite{DiUbaldo:2026rly,Maldacena:2026jqd} to our axiverse models. In  these models, the wormhole geometry is the same as in \cite{Giddings:1987cg}, see Figure \ref{fig:WH}. We can first describe it as the gluing of two half-wormholes, which have identical metrics:
\be\label{WH} 
\d s^2=\frac{\d r^2_\pm}{1-\frac{L^4}{r^4_\pm}}+r^2_\pm\d\Omega^2\,.
\ee 
Here $\d\Omega^2$ is the volume element of a round $S^3$ of radius $1$, and the radial coordinates $r_\pm\in[L,\infty)$ cover the upper and lower half-wormholes, respectively. It is clear that $L$ represents the radius of the wormhole neck.

One can also introduce a global radial coordinate $\tau$, related to the radii $r_\pm$ on the two half-wormholes by
\be\label{coordchange} 
\cos\left(2\pi M^2_{\text{\tiny P}}L^2\tau\right)= \frac{L^2}{r^2_\pm}\,,
\ee
where $\tau\in (-\tau_\infty,\tau_\infty)$ with
\be\label{tauinf} 
\tau_\infty\equiv\frac1{4M^2_{\text{\tiny P}}L^2}\,.
\ee
The coordinates $r_+$ and $r_-$ cover the intervals $\tau\in[0,\tau_\infty)$ and $\tau\in(-\tau_\infty,0]$, respectively, with $r_\pm=L$ corresponding to $\tau=0$.
The normalization of $\tau$ has been chosen for later purposes, and matches the choice made in \cite{Martucci:2024trp}. 
In this parametrization, the metric is smooth and shows more clearly that the wormhole solution has topology $\mathbb{R}\times S^3$.\footnote{See \eqref{ymetric} below, written in terms of  the rescaled coordinate \eqref{ytaur}.}

In the non-symmetric models of Section \ref{sec:aximodels}, the relevant multi-axion wormholes are those discussed in \cite{Montero:2015ofa,Bachlechner:2015qja}, and their consequences are immediate generalizations of those already discussed in \cite{DiUbaldo:2026rly,Maldacena:2026jqd}; see also \cite{Etheredge:2026rio}. In this respect, the minimally supersymmetric models of Section \ref{sec:susymodels} allow us to discuss some interesting new features.
As shown in \cite{Martucci:2024trp}, these models admit a universal class of wormhole solutions, which can be directly related to a specific subclass of fundamental BPS instantons. In this section we will revisit and extend the results of \cite{Martucci:2024trp} from the perspective proposed by \cite{DiUbaldo:2026rly,Maldacena:2026jqd}.\footnote{Note that the metric \eqref{WH} assumes a vanishing cosmological constant. In the absence of supersymmetry, this is of course an additional strong assumption that has to face the standard cosmological constant problem. On the other hand, it is automatically guaranteed in supersymmetric axiverse models that preserve the (non-invertible) axion shift symmetries and do not spontaneously break supersymmetry.
%Indeed, the (non-invertible) axion shift symmetry implies that any superpotential $W$ cannot depend on the chiral fields $t^i=a^i+\ii s^i$. Hence, the F-flatness conditions $D_iW|_{\rm vacuum}=(W\del_i K)|_{\rm vacuum}=0$ require that $W|_{\rm vacuum}=0$. This in turn implies that $V|_{\rm vacuum}=M^3_\Pl e^{K}\left(|DW|^2-3|W|^2\right)|_{\rm vacuum}=0$ \cite{Cremmer:1982en}. Of course, one still needs to assume that $W|_{\rm vacuum}=0$. This is, however, a much milder and natural assumption, thanks to the non-renormalization properties of the superpotential.
}

%%%%%%%%%%%%%%%%%%%%%%%%%%%%%%%%%%%%%%%%%%%%%%%%%%%%%%%%%%%%%%%%%%%%%%%%%%%%%%%%%%%%%%%%%%%%%%%%%%%%%

\subsection{Purely axionic wormholes} 
\label{sec:axioWH}

As a warm-up, we start with the purely axionic wormholes appearing in the non-supersymmetric models described in Section \ref{sec:aximodels}. This will allow us to review some of the main points of \cite{DiUbaldo:2026rly,Maldacena:2026jqd}, adapt them to this simpler class of axiverses, and introduce some ingredients that we will be useful also later. The relevant wormholes can be obtained as simple generalizations of the Giddings--Strominger wormhole \cite{Giddings:1987cg}, and we will therefore only recall their main properties; see \cite{Montero:2015ofa,Bachlechner:2015qja} for more details. We will use the conventions and notation of \cite{Martucci:2024trp}.

It is convenient to first use the dual formulation in terms of the two-form potentials $\calb_{2,i}$. The relevant action is given by the Euclidean version of \eqref{B2kinetic}, with $M=M_\Pl$, together with the Euclidean Einstein-Hilbert term:\footnote{The Gibbons-Hawking term \cite{Gibbons:1976ue} turns out to be irrelevant for our purposes, so we will not write it explicitly.}
\be\label{EEH} 
S_{\text{E}}=-\frac12 M^2_\Pl\int R*1+\frac1{2M^2_\Pl}\int \calg^{ij}\calh_{3,i}\wedge *\calh_{3,j}\,.
\ee
In order to preserve the SO(4) symmetry of \eqref{WH}, the field strengths must necessarily take the form
\be\label{H3ansatz}
\calh_{3\,i}=\frac1{\pi}q_i\,{\rm vol}_{S^3}\,,
\ee
with $q_i$ constants. Notice that this choice satisfies $\d\calh_{3,i}\equiv 0$, and is consistent with \eqref{modBianchi}, since we have set $F^I=0$ and the $SO(4)$ symmetry implies that $\tr(\calr\wedge\calr)\equiv0$. The Einstein equation then sets
\be\label{eqn:LQ rel} 
L^4=\frac{\|{\bf q}\|^2}{6\pi^2M^4_{\text{\tiny P}}}\,,
\ee
with
\be 
\|{\bf q}\|^2\equiv \calg^{ij}q_iq_j\,.
\ee
One can also use the traced Einstein equation $R\,*1=-\frac{1}{M^4_{\text{\tiny P}}}\calg^{ij}\calh_{3\,i}\wedge *\calh_{3\,j}$ to rewrite the on-shell value of \eqref{EEH} as
\be\label{WHonshell}
\left.S_{\rm E}\right|_{\text{\tiny WH}}=\frac1{M_{\text{\tiny P}}^2}\int\calg^{ij}\calh_{3,i}\wedge *\calh_{3,j}=2\pi\int_{-\tau_\infty}^{\tau_\infty}\d\tau\, \|{\bf q}\|^2 \,.
\ee
Hence, recalling \eqref{eqn:LQ rel} and \eqref{tauinf}, one gets
\be\label{WHonshell2} 
S^{\text{\tiny WH}}_{\bf q}\equiv \left.S_{\rm E}\right|_{\text{\tiny WH}}=6\pi^3M^2_{\text{\tiny P}}L^2\,.
\ee
The topological term \eqref{GRtheta}, which does not modify the equations of motion, provides an additional constant term $-\gamma \int E_{\text{\tiny GB}}=\gamma$ to the on-shell action, as in \cite{Giddings:1987cg}.

Note that the highest curvature is localized around the neck of the wormhole and is of order $L^{-2}$. Thus, the validity of the semiclassical approximation is guaranteed by taking the charges $q_i$ sufficiently large. Another ingredient that makes the wormhole solution robust is that it is charged under the axion zero-form symmetry discussed in Section \ref{sec:ax0form}. Indeed, evaluating the topological defect \eqref{eqn: 0form defec} along any $S^3$ slice of the wormhole gives
\be\label{DVEV} 
\langle \cald^{\rm(a)}_{\bm\alpha}(S^3)\rangle_{\text{\tiny WH}} =e^{2\pi\ii\langle {\bf q},{\bm\alpha}\rangle}\,,
\ee
for any ${\bm\alpha}\in V_\mathbb{Q}/V_{\mathbb{Z}}$, where we have oriented $S^3$ so that $\int_{S^3}\calh_{3,i}=2\pi q_i$. Hence, the constants $q_i$ can be interpreted as wormhole charges under the axion shift symmetry.
Requiring \eqref{DVEV} to be invariant under $\alpha^i\simeq \alpha^i+1$ makes it clear that one must impose ${\bf q}\in V^*_\mathbb{Z}$, that is, $q_i\in\mathbb{Z}$. These conclusions have a more standard counterpart when the gauge sector in \eqref{axionlag0} and the Pontryagin term \eqref{aRR} are not included, so that the axion shift symmetries become invertible. Our discussion shows how non-invertible global symmetries provide the proper framework to extend these considerations to our more general axiverse models.\footnote{\label{foot:nonabWHs} The inclusion of non-abelian sectors can change the story. Take, for instance, the single gauge sector considered in Appendix \ref{app:nonabelian}, and focus on the unbroken zero-form symmetries with parameters as in \eqref{restralpha}. On the r.h.s.\ of \eqref{DVEV}, we now have $\exp(-2\pi\ii \tilde\alpha^a\langle {\bf q},\tilde{\bm v}_\alpha\rangle)$. For generic $\tilde\alpha^a\in\mathbb{Q}$ this differs from $1$, unless ${\bf q}$ is proportional to $\widehat{\bf K}$. This means that the wormhole charge is a priori conserved only modulo jumps along the $\widehat{\bf K}$ directions. We postpone a more detailed discussion of these charge non-conservation effects to future work. Alternatively, one may restrict to an axion sector that does not couple to the non-Abelian sector in the first place.}

\begin{figure}[ht]
\centering
\begin{subfigure}{0.58\textwidth}
\centering
\includegraphics[width=\textwidth]{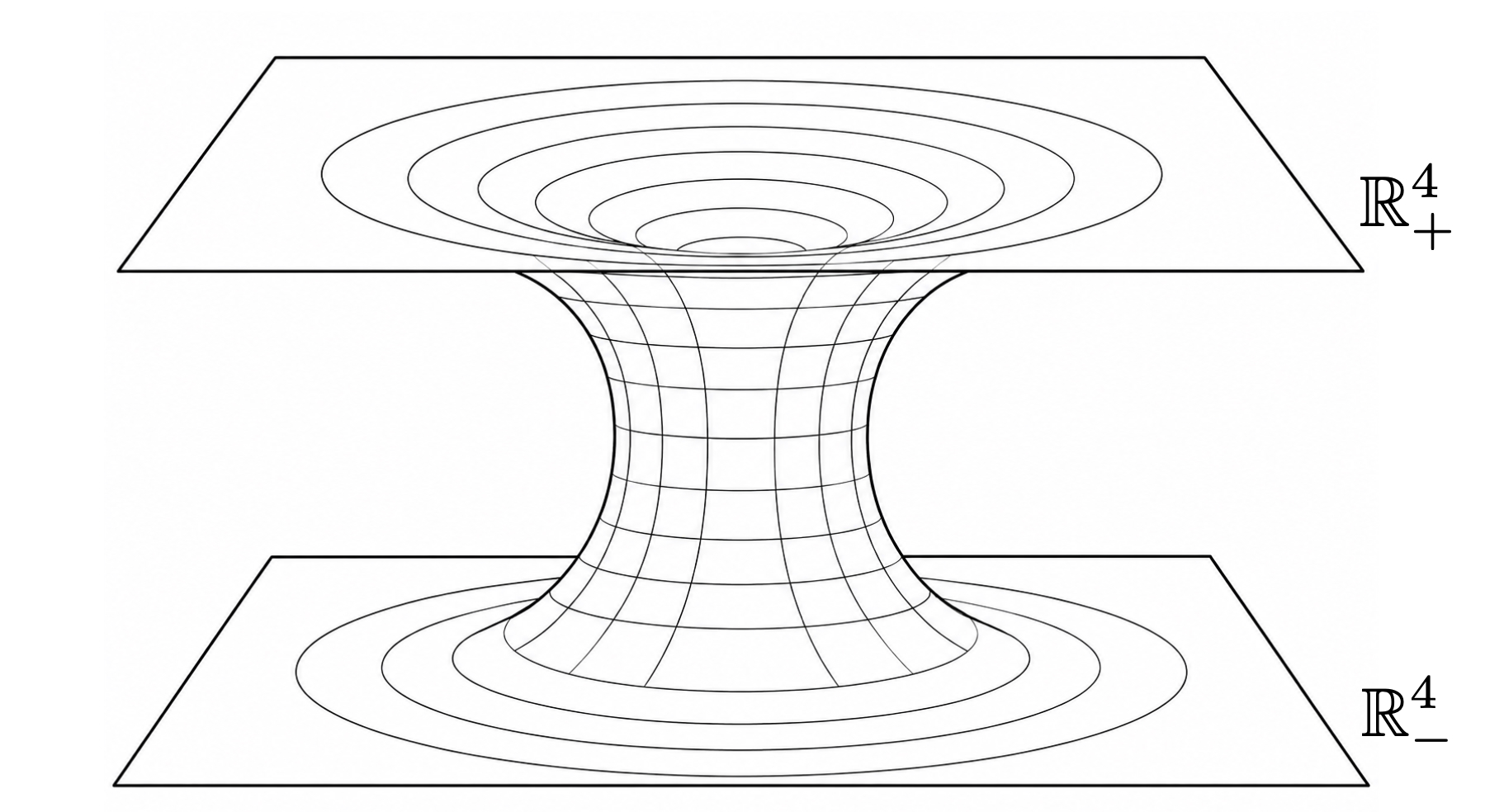}
\end{subfigure}
\hfill
\caption{}
\label{fig:WH}
\end{figure}

As illustrated in Figure \ref{fig:WH}, the wormhole solution can be regarded as connecting two flat spacetimes $\mathbb{R}^4_+$ and $\mathbb{R}^4_-$ supporting asymptotic zero-form symmetry charges. In \eqref{DVEV}, we picked an $S^3$ orientation compatible with the orientation of the bulk global coordinates. From the point of view of two asymptotic observers on $\mathbb{R}^4_\pm$ that use the same orientation, the upstairs spacetime carries asymptotic charge $q_i$, while the downstairs spacetime carries asymptotic charge $-q_i$. Furthermore, the wormhole solution has at least eight bosonic zero modes $x^\mu_\pm\in \mathbb{R}^4_\pm$, representing the positions of the wormhole mouth on each of its two sides \cite{Witten:2026twr}.

One can alternatively impose Dirichlet boundary conditions, in which the axions $a^i$ have fixed asymptotic values $a^i_{\pm\infty}\equiv a^i(\tau_{\pm\infty})$. The transition from one type of boundary condition to the other has recently been discussed in detail in \cite{Witten:2026twr}. Adapted to our case, it amounts to adding to the Euclidean action the boundary terms
\be 
2\pi\ii\langle{\bf q}, \Delta{\bm a}_\infty\rangle= 2\pi\ii q_i\, \Delta a^i_\infty\quad,\quad \Delta a^i_\infty\equiv  a^i_{+\infty} - a^i_{-\infty}\,,
\ee
and summing over all charges $q_i$. Factoring out the integrals over the zero modes $x^\mu_\pm\in \mathbb{R}^4_\pm$, the resulting partition function takes the form
\be 
\int_{\mathbb{R}^4_+}\d^4x_+\int_{{\mathbb{R}}^4_-}\d^4 x_- \, Z_{\text{\tiny WH}}[\Delta{\bm a}_\infty] \; ,
\ee
with
\be\label{WHpartfunct} 
Z_{\text{\tiny WH}}[\Delta{\bm a}_\infty]=e^{-\gamma}\sum_{{\bf q}\in V^*_\mathbb{Z}}c_{\bf q}\, \exp\left(-S^{\text{\tiny WH}}_{\bf q}+2\pi\ii\langle{\bf q}, \Delta{\bm a}_\infty\rangle  \right)\,,
\ee
where the $e^{-\gamma}$ factor comes from the Gauss-Bonnet term, and $c_{\bf q}$ are coefficients incorporating other corrections to the classical on-shell contribution \eqref{WHonshell2}, e.g.\ from the one-loop determinant or higher-derivative terms. We assume that $|c_{\bf q}|$ are not too large, so that most of the terms in \eqref{WHpartfunct} are exponentially suppressed, making the infinite sum rapidly converge.

Now, following \cite{DiUbaldo:2026rly}, the key point is that \eqref{WHpartfunct} exhibits divergences if we continue $a^i_{\pm\infty}$ to imaginary values. To see this more explicitly, let us use \eqref{WHonshell} and \eqref{tauinf} to rewrite \eqref{WHonshell2} as
\be\label{WHq1} 
S^{\text{\tiny WH}}_{\bf q}=4\pi \langle {\bf q},{\bm\varphi}_{\text{\tiny IDB}}({\bf q})\rangle\,,
\ee 
with
\be\label{varphiIDB} 
\varphi^i_{\text{\tiny IDB}}({\bf q})\equiv\frac\pi2\sqrt{\frac32}\,\frac{\calg^{ij}q_j}{\|{\bf q}\|}\,.
\ee
Notice that ${\bm\varphi}_{\text{\rm IDB}}({\bf q})$ depends only on the direction of the vector ${\bf q}$, and not on its magnitude, and that ${\bm\varphi}_{\text{\rm IDB}}(-{\bf q})=-{\bm\varphi}_{\text{\rm IDB}}({\bf q})$. Let us then fix any   charge vector ${\bf q}\in V^*_{\mathbb{Z}}$, and focus on the contributions to \eqref{WHpartfunct} corresponding to aligned charge vectors ${\bf q}'=k{\bf q}$:
\be\label{restrsum} 
\begin{aligned}
Z^{{\bf q}}_{\rm WH}[\Delta{\bm a}_\infty]\equiv  e^{-\gamma}\sum_{k\in\mathbb{Z}}\cala_{k{\bf q}}\, \exp\left(-2\pi k\big[2\langle {\bf q},{\bm\varphi}_{\text{\rm IDB}}({\bf q}) -\ii\Delta{\bm a}_\infty\rangle \big]  \right)\subset Z_{\rm WH}[\Delta{\bm a}_\infty]\,.
\end{aligned}
\ee
It is now clear that, even though most of the terms in \eqref{restrsum} are exponentially suppressed for real $\Delta{\bm a}_\infty$, they become of order one if one adds to the asymptotic values $a^i_{\pm\infty}$ purely imaginary contributions $\Im a^i_{\pm\infty}\equiv - \varphi^i_{\pm\infty}$ and takes the limit
\be\label{IDBlimit} 
\Delta\varphi^i_{\infty}\equiv \varphi^i_{+\infty}-\varphi^i_{-\infty}\quad\rightarrow\quad  2\varphi^i_{\text{\tiny IDB}}({\bf q})\,.
\ee 
Hence, $Z^{{\bf q}}_{\text{\tiny WH}}[\Delta{\bm a}_\infty -\ii \Delta{\bm \varphi}_\infty]$ badly diverges in the limit \eqref{IDBlimit}.
In \cite{DiUbaldo:2026rly,Maldacena:2026jqd}, such a divergence is interpreted as the coarse-grained manifestation of a more fundamental pathology, which signals the need to take into account additional ingredients from the UV completion of the EFT.

Of course, the same argument can be repeated for any ${\bf q}$. For any such choice, the imaginary half-displacement measured by the axionic metric $\calg_{ij}$ has the same norm
\be 
\|{\bm\varphi}_{\text{\tiny IDB}}({\bf q})\|=D_{\text{\tiny IDB}}\quad,\quad \forall {\bf q}\in V^*_{\mathbb{Z}}\,,
\ee
with
\be\label{IDBD} 
D_{\text{\tiny IDB}}\equiv \frac{\pi}{2}\sqrt{\frac32}\,.
\ee
Furthermore, since the metric $\calg_{ij}$ is invertible, the images ${\bm\varphi}_{\text{\tiny IDB}}({\bf q})$ of all possible charge vectors ${\bf q}\in V^*_{\mathbb{Z}}$ densely fill the sphere of radius $D_{\text{\tiny IDB}}$ in $V_{\mathbb{R}}$. The proposal of \cite{DiUbaldo:2026rly,Maldacena:2026jqd} then implies that, in a consistent quantum gravity theory, there should exist some effect that makes the low-energy approximation invalid at or before a critical imaginary displacement $\Delta{\bm\varphi}_{\rm cr}$ with
\be\label{IDB} 
\|\Delta{\bm\varphi}_{\rm cr}\|\leq 2D_{\text{\tiny IDB}}\,.
\ee
This is nothing but the IDB of \cite{DiUbaldo:2026rly,Maldacena:2026jqd}.\footnote{The different numerical factor with respect to \cite{Maldacena:2026jqd} comes from the normalization of the kinetic metric, $\calg^{\rm there}_{ij}=2\calg^{\rm here}_{ij}$.}

It is instructive, and useful for later generalizations, to also review the derivation of the IDB by working directly in the axion formulation, as in \cite{Maldacena:2026jqd}. Setting the U(1) field strengths to zero, the relevant two-derivative terms in the Euclidean action are
\be\label{EEH1} 
S^{\rm (a)}_{\text{E}}=-\frac12 M^2_\Pl\int R*1+\frac{1}{2}M^2_\Pl\int \calg_{ij}\d a^i\wedge *\d a^j\,.
\ee
Classical saddles can be immediately obtained from the fixed-charge solution described above through the Euclidean counterpart of \eqref{eqn:H3}:
\be \label{EaHrel}
\d a^i= \frac\ii{M^2_{\Pl}}\calg^{ij}*\calh_{3\, j}= - \ii \calg^{ij}q_j\d\tau\,,
\ee
where we have used \eqref{H3ansatz}, \eqref{coordchange}, and \eqref{WH}, together with the orientation fixed by the global coordinates. Thus, we can set
\be\label{axionprof} 
a^i(\tau)\equiv a_0^i -\ii\varphi^i(\tau)\,,
\ee
with $\varphi^i(\tau)$ satisfying
\be\label{varphieq} 
\frac{\d\varphi^i}{\d\tau}= \calg^{ij}q_j\,.
\ee
We then recover the well-known result \cite{Arkani-Hamed:2007cpn} that, in the axion formulation, one obtains a saddle point involving an imaginary axion displacement. For this reason, we  refer to these saddles as {\em imaginary} wormholes.

Since $\calg^{ij}$ is constant, \eqref{varphieq} can be immediately integrated to
\be 
\varphi^i(\tau)=\calg^{ij}q_j\tau\,,
\ee
where we have fixed the integration constant so that $a^i(\tau)$ is real at $\tau=0$, namely at the smallest radius $r=L$.\footnote{\label{foot:symBC} Less symmetric boundary conditions are also possible, but they do not affect the main implications  and will not be considered.} Recalling \eqref{tauinf}, \eqref{eqn:LQ rel}, and \eqref{varphiIDB}, we see that this gives a straight line in the imaginary axion directions, connecting the extreme imaginary values $\varphi^i_{\pm\infty}=\pm \varphi^i_{\text{\tiny IDB}}({\bf q})$.
As already noted, $\varphi^i_{\text{\tiny IDB}}({\bf q})$ depends only on the direction of ${\bf q}$, and not on its norm $\|{\bf q}\|$. The same conclusion holds for the entire imaginary axion profile, while $\|{\bf q}\|$ only fixes the minimal wormhole radius $L$ through \eqref{eqn:LQ rel}. In this dual classical solution, the charges $q_i$ need not be quantized, and therefore one obtains a continuous family of wormholes, parametrized by $L$, that share the same straight imaginary axion trajectory. Furthermore, the straight imaginary axion trajectory can have any direction and is constrained only to have total length $\|\Delta{\bm\varphi}_\infty\|=2D_{\text{\tiny IDB}}$ in Planck units.

As emphasized in \cite{Maldacena:2026jqd}, these wormhole saddles give a divergent contribution to the path integral. Indeed, for any imaginary axion direction identified by a given direction of ${\bf q}$ in the above discussion, the on-shell value of \eqref{EEH1} vanishes and therefore does not generate an exponential suppression,\footnote{As in \eqref{WHpartfunct} and \eqref{restrsum}, the Gauss-Bonnet term \eqref{GRtheta} contributes only an irrelevant overall factor $e^{-\gamma}$.} and
furthermore one must integrate over the parameter $L$, which can take arbitrarily large values. This divergence is the counterpart of the divergence, observed above, of $Z^{{\bf q}}_{\text{\tiny WH}}[\Delta{\bm a}_\infty -\ii \Delta{\bm \varphi}_\infty]$ in the limit \eqref{IDBlimit}, and provides a dual motivation for the IDB \eqref{IDB}.

The most natural realization of the IDB, suggested also by the structure of \eqref{WHpartfunct}, is through the appearance of EFT corrections that break the (non-invertible) axion shift symmetries. These can be interpreted as insertions of fundamental instantons, which produce corrections proportional to
\be\label{instcorr} 
e^{-S^{\rm inst}_{\bf q}+2\pi\ii \langle {\bf q},{\bm a}\rangle}\,.
\ee
The IDB suggests that, as a {\em minimal} necessary condition, there should exist a minimal set of instanton corrections that become of order one before the imaginary axion crosses the IDB \eqref{IDB}. Let us denote the set of corresponding charges by $\cali\subset V^*_\mathbb{Z}$. Notice that if ${\bf q}$ belongs to $\cali$, then the CPT conjugate $-{\bf q}$ belongs to $\cali$ as well. Then, {\em for any} vector ${\bm\varphi}_{\text{\tiny IDB}}\in V_\mathbb{R}$ of length $\|{\bm\varphi}_{\text{\tiny IDB}}\|=D_{\text{\tiny IDB}}$, there should exist  at least one charge vector ${\bf q}\in \cali$ such that
\be\label{AWGC0} 
S^{\rm inst}_{\bf q}\leq 2\pi \langle {\bf q},{\bm\varphi}_{\text{\tiny IDB}}\rangle\,.
\ee
For any fixed ${\bf q}\in\cali$, the r.h.s.\ of \eqref{AWGC0} is maximized by choosing ${\bm\varphi}_{\text{\tiny IDB}}$ as in \eqref{varphiIDB}, namely the one associated with a wormhole of charge ${\bf q}$. One concludes \cite{DiUbaldo:2026rly,Maldacena:2026jqd} that each instanton in the set $\cali$ satisfies the axion WGC bound \cite{ArkaniHamed:2006dz} with a specific coefficient:
\be\label{AWGC1} 
S^{\rm inst}_{\bf q}\leq 2\pi D_{\text{\tiny IDB}}\|{\bf q}\|\,,
\ee
which can also be written as  $S^{\rm inst}_{\bf q}\leq \frac12 S^{\text{\tiny WH}}_{\bf q}$.\footnote{See also \cite{Montero:2015ofa} for previous arguments leading to the bound $S^{\rm inst}\leq \frac12 S_{\text{\tiny WH}}$.} In terms of the kinetic metric  $(\sff^{2}_\vartheta)_{ij}$, introduced in Footnote \ref{foot:2pikin}, the axion WGC bound \eqref{AWGC1} takes the form  
\be 
S^{\rm inst}_{\bf q}\leq D_{\text{\tiny IDB}}M_{\Pl}\sqrt{(\sff^{-2}_\vartheta)^{ij}q_iq_j}\,,
\ee 
where $(\sff^{-2}_\vartheta)^{ij}$ is the inverse of $(\sff^{2}_\vartheta)_{ij}$.

On the other hand, the condition \eqref{AWGC0} is stronger than \eqref{AWGC1} and, as already mentioned in \cite{Maldacena:2026jqd}, gives the axion counterpart of the convex hull condition \cite{Cheung:2014vva} for particles. For each ${\bf q}\in \cali$, define the corresponding charge-to-action vector ${\bm z}_{\bf q}\in V_{\mathbb{R}}$ by
\be 
z^i_{\bf q}\equiv \frac{2\pi \calg^{ij}q_j}{S^{\rm inst}_{\bf q}}\,.
\ee
Then the condition \eqref{AWGC0} can be rewritten as
\be\label{zvarphicond} 
{\bm z}_{\bf q}\cdot {\bm \varphi}_{\text{\tiny IDB}}\geq 1\,,
\ee
 where $\cdot$ denotes the inner product on $V_\mathbb{R}$ defined by $\calg_{ij}$. This condition is equivalent to requiring that the convex hull ${\rm conv}\{{\bm z}_{\bf q}\}_{{\bf q}\in\cali}\subset V_{\mathbb{R}}$ contains the ball of radius $1/D_{\text{\tiny IDB}}$.

As already emphasized, this convex hull condition is only a minimal requirement implied by the IDB. In fact, the imaginary wormhole  can encode additional information,
providing the typical value of corrections of the form \eqref{instcorr} under a suitable averaging or coarse-graining procedure \cite{Maldacena:2026jqd}. The divergence of \eqref{restrsum} then suggests that the set $\cali$ satisfying the convex hull condition should in fact contain infinitely many charges, populating all of $V^*_{\mathbb{Z}}$, or infinite subsets thereof in every direction, e.g.\ realizing a tower or sublattice version of the axion WGC \cite{Heidenreich:2015nta,Heidenreich:2016aqi}.

Given the absence of supersymmetry, it is difficult to interpret the above claims in terms of concrete UV-complete realizations. In the next subsection, we discuss the implications of imaginary wormholes and IDB for the minimally supersymmetric models of Section \ref{sec:susymodels}, which do not suffer from this problem.

\subsection{IDB in the supersymmetric axiverse}
\label{sec:susyIDB}

Fixed-charge wormholes in the supersymmetric models described in Section \ref{sec:susymodels} were studied in detail in \cite{Martucci:2024trp}. Here, we revisit some of those findings from the perspective suggested in \cite{DiUbaldo:2026rly,Maldacena:2026jqd}. In particular, the non-invertible axion shift symmetries discussed in Section \ref{sec:ax0form} provide a sharper characterization of the perturbative EFTs of Section \ref{sec:susymodels}, leading to a cleaner understanding of the role of wormholes and their effects. In what follows, we recall only the aspects of \cite{Martucci:2024trp} that are most relevant for the present work; further details can be found therein.

One of the main differences with respect to the purely axionic case considered in Section \eqref{sec:axioWH} is the presence of massless saxions, which significantly affect the solution.  It is convenient to use dual saxions \eqref{dualsax}. The corresponding contribution to the Euclidean action 
\be\label{Eellaction} 
\frac12M_{\text{\tiny P}}^2\int_\calm \calg^{ij}\d \ell_i\wedge *\d \ell_j\,.
\ee
must then be added to \eqref{EEH}, where now  $\calg^{ij}$ depends on the the dual saxions -- see \eqref{invcalG}. In the fixed-charge picture, the three-forms $\calh_{3,i}$ are still given by \eqref{H3ansatz}, while the dual saxion profiles  $\ell_i(\tau)$ must obey appropriate equations. In particular, they satisfy the energy conservation condition
\be\label{encons} 
\frac12\calg^{ij}({\bm\ell})\left(\dot\ell_i\dot\ell_j-q_iq_j\right)=E\,,
\ee
where the constant `energy' $E<0$ is related to the wormhole radius $L$ by
\be\label{WHL} 
L^4\equiv \frac{|E|}{3\pi^2M^4_{\text{\tiny P}}}\,.
\ee
 We will focus on  solutions in which ${\bm \ell}(\tau)$ describes the same trajectory along the two half-wormholes, connecting ${\bm\ell}_*$ at $\tau=0$ (i.e.\ at $r=L$) to ${\bm\ell}_\infty$ at $\tau=\pm\tau_\infty$ (i.e.\ at $r_\pm=\infty$). This means that ${\bm\ell}|_{\tau=0}={\bm\ell}_*$ and $\frac{\d{\bm\ell}}{\d\tau}|_{\tau=0}={\bm 0}$. This fixes 
\be\label{EandL} 
E=-\frac12\|{\bf q}\|_*^2\quad\Rightarrow\quad L^4=\frac{\|{\bf q}\|^2_*}{6\pi^2M^4_{\text{\tiny P}}}\ ,
\ee
with $\|{\bf q}\|^2_*\equiv \calg^{ij}({\bm\ell}_*)q_iq_j$. This shows  that the wormhole radius is determined by the charges as well as by the value of the dual saxions at $r=L$.

In \cite{Martucci:2024trp} it is argued that the conditions imposed by the existence of wormhole solutions naturally select, as a preferred subset of wormhole charge vectors ${\bf q}\in V^*_\mathbb{Z}$, those for which either ${\bf q}$ or $-{\bf q}$ belongs to the dual saxionic cone $\calp$. We denote this set of charges by\footnote{The definition \eqref{CWH} of $\mathcal{C}_{\text{\tiny WH}}$ is slightly simpler than the one given in \cite{Martucci:2024trp}, where $\mathcal{C}{\text{\tiny WH}}$ also includes charge vectors belonging to the finite-distance boundary of $\mathcal{P}$. For our purposes, this extension is inessential and only introduces unnecessary complications.} 
\be\label{CWH} 
\calc_{\text{\tiny WH}} = \calp \cap V^*_{\mathbb{Z}}  \,,
 \ee
which  can be  regarded as the discretization of $\calp$. 
 Clearly, for each admissible wormhole with charge vector  ${\bf q}\in \calc_{\text{\tiny WH}}$, there exists a corresponding  conjugate solution carrying charge vector $-{\bf q}$. Hence, we can focus on charge vectors  ${\bf q}\in \calc_{\text{\tiny WH}}$.

\begin{figure}[ht]
\centering
\begin{subfigure}{0.60\textwidth}
\centering
\includegraphics[width=\textwidth]{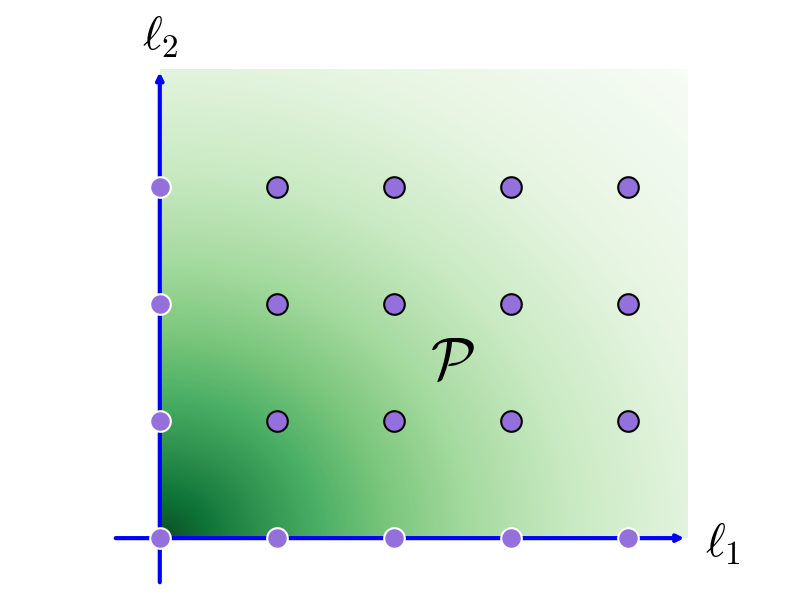}
\end{subfigure}
\hfill
\caption{The   bullets represent the set $\calc^{\text{\tiny EFT}}_{\rm I}$ of EFT instanton charges  of the $\mathbb{P}^1\hookrightarrow X\rightarrow  \mathbb{P}^2$ F-theory model, with twist parameter $p=2$, discussed in Section \ref{sec:susyexam}. The black-encircled  bullets represent the subset $\calc_{\text{\tiny WH}}$ of 
wormhole charges. Cf.\ also Fig.~\ref{fig:saxionicones}.
}
\label{fig:EFTWHcharges}
\end{figure}

While for each ${\bf q}\in \calc_{\text{\tiny WH}}$ there can exist infinitely many dual saxionic profiles, satisfying different boundary conditions, for our purposes we can just focus on the universal class of  `homogeneous' solutions \cite{Martucci:2024trp}, whose dual saxions  rescale homogeneously along the direction identified by ${\bf q}$:  
\be\label{ellansatz} 
{\bm \ell}(\tau)=\tilde\ell(\tau)\,{\bf q}\,.
\ee
Our boundary conditions now read ${\tilde\ell}|_{\tau=0}=\tilde\ell_*$ and  $\frac{\d{\tilde\ell}}{\d\tau}|_{\tau=0}= 0$. Using ${\bm\ell}_*=\tilde\ell_*{\bf q}$ and the homogeneity of $\calg^{ij}({\bm\ell})$, and recalling \eqref{ellsnorms} and \eqref{tauinf}, one gets the useful relations
\be\label{Ltau} 
L^4=\frac{k}{12\pi^2\tilde\ell_*^2M^4_{\text{\tiny P}}}
\quad~~~\Rightarrow\quad~~~ \tau_\infty=\frac\pi2 \sqrt{\frac{3}{k}}\,\tilde\ell_* \ ,
\ee
where we recall that $k$ is the homogeneity degree in \eqref{homP} and \eqref{tildePhom}.
The profile \eqref{ellansatz} can be obtained by integrating  \eqref{encons}, again exploiting the homogeneity of $\calg^{ij}({\bm\ell})$:
\be\label{ellsol}
\tilde\ell(\tau)=\tilde\ell_*\cos\left(\frac{\tau}{\tilde\ell_*}\right)=\tilde\ell_*\cos\left(\frac\pi2 \sqrt{\frac{3}{k}}\,\frac{\tau}{\tau_\infty}\right)\,.
\ee

Even for more general fixed-charge wormhole solutions, the on-shell Euclidean action combining  \eqref{EEH} and \eqref{Eellaction} is still given by \eqref{WHonshell}. In the presence of the dual saxions, $\|{\bf q}\|^2$ is not constant, but for the straight profiles  \eqref{ellansatz}, and exploiting again the homogeneity \eqref{tildePhom}, it reduces to $\frac12 k \tilde\ell^{-2}(\tau) $. By using \eqref{ellsol}, we then get 
\be\label{WHonshell3} 
S^{\text{\tiny WH}}_{\bf q}\equiv \left.S_{\rm E}\right|_{\text{\tiny WH}}=\frac{2\pi k}{\tilde\ell^2_*}\int^{\tau_\infty}_{0}\frac{\d\tau}{\cos^2\left(\frac\pi2 \sqrt{\frac{3}{k}}\,\frac{\tau}{\tau_\infty}\right)}\,.
\ee

Note that the  dual saxionic trajectory  \eqref{ellansatz} described by \eqref{ellsol} is left invariant  by any simultaneous positive rescaling of the charge vector ${\bf q}\rightarrow \lambda{\bf q}$ (preserving its integrality), if accompanied by a corresponding rescaling $\tau\rightarrow \tau/\lambda$ and $\tilde\ell_*\rightarrow \tilde\ell_*/\lambda$. On the other hand, under this rescaling the wormhole radius {\em does} change: $L\rightarrow \sqrt{\lambda} L$, consistently with the rescaling of the parameter $\tau$, and of the corresponding endpoint \eqref{tauinf}. It is easy to see that this scaling symmetry holds also for more general, non-homogeneous, solutions. Hence this scaling argument produces, for  each dual saxionic profile 
\be\label{Lindell} 
{\bm\ell}(\tau)={\bm\ell}_*\cos\left(\frac\pi2 \sqrt{\frac{3}{k}}\,\frac{\tau}{\tau_\infty}\right)\,,
\ee
an infinite family of wormholes parametrized by discretized values of $L$.

It is clear that, since we are assuming  ${\bf q}\in \calc_{\text{\tiny WH}}$, the  trajectory \eqref{ellansatz} is contained in the dual saxionic cone $\calp$ as long as $\tilde\ell(\tau)>0$. This happens only if $k>3$. In this case the integral in \eqref{WHonshell3} converges and the on-shell action becomes
\be\label{WHonshell4} 
S^{\text{\tiny WH}}_{\bf q}=\frac{2\pi k}{\tilde\ell_*}\tan\left(\frac\pi2 \sqrt{\frac{3}{k}}\right)\quad~~~~~~\text{(if $k>3$)}\,.
\ee
On the other hand, the profile described by \eqref{ellsol} reaches the tip ${\bm\ell}=0$ of $\calp$, and then enters in the specular cone $-\calp$,  at the critical radial coordinates $\tau=\pm\tau_{\rm cr}$ with
\be\label{taucr} 
\tau_{\rm cr}=\,\frac\pi2\tilde\ell_*=\,\tau_\infty\sqrt{\frac{k}{3}}\,.
\ee
Clearly, if $k< 3$ then   $\tau_{\rm cr}<\tau_\infty$, and the dual saxion trajectory \eqref{ellansatz} passes through the tip of the dual saxionic cone $\calp$, which is always at infinite field space distance. Furthermore, the on-shell action \eqref{WHonshell3}  clearly diverges. For this reason in \cite{Martucci:2024trp}  the solutions were considered degenerate and physically irrelevant, as traditionally  done in this context -- see e.g.\ \cite{Arkani-Hamed:2007cpn} for a general discussion. The marginally degenerate case $k=3$ can be  regularized and in fact played a special role \cite{Martucci:2024trp}.   It will also be important in the present paper, but for the moment let us just take notice that, as it is, it seems unphysical as well.        

However, following the logic of \cite{Maldacena:2026jqd}, these wormhole solutions may still contain relevant physical information even if  $k\leq 3$. This possibility becomes more apparent by passing to the dual imaginary wormholes, as we did in Section \ref{sec:axioWH}, in which the axions acquire a non-trivial imaginary profile of the form \eqref{axionprof}. Equation \eqref{varphieq} continues to hold in the present case, provided one takes into account that $\calg^{ij}$ now depends on the dual saxions. Applying it to \eqref{ellsol}, and using \eqref{ellsrel} together with homogeneity, \eqref{varphieq} reduces to
\be\label{varphieq2} 
\frac{\d\varphi^i}{\d\tau}=\frac{s^i_*}{\tilde\ell_*\cos^2\left(\frac{\tau}{\tilde\ell_*}\right)}\,,
\ee
where $s^i_*=s^i(\tau=0)$.
Integrating this equation and imposing that the axions be real at the minimal radius $r=L$ (see Footnote \ref{foot:symBC}), we obtain
\be\label{varphi} 
{\bm\varphi}(\tau)= \,{\bm s}_*\tan\left(\frac\pi2 \sqrt{\frac{3}{k}}\,\frac{\tau}{\tau_\infty}\right) \,.
\ee
 Notice that \eqref{varphi} is also invariant under the scaling symmetry  discussed above. In this dual classical saddle, the radius $L$ is not required to take discrete values. Hence, we obtain a continuous family of wormhole saddles supporting the same dual saxionic and imaginary axionic profiles \eqref{Lindell} and \eqref{varphi}, respectively.

We have not yet addressed the degeneration issue, which now also involves \eqref{varphi}. The key point, as in \cite{Maldacena:2026jqd}, is that after including the imaginary axion directions, nothing particularly dramatic  appears to happen at $\tau=\pm\tau_{\rm cr}$ in the  extended moduli space  $\calm$  parametrized by $(\ell_i,\varphi^j)$.
To see this, let us first write the analytically continued metric on $\calm$:
\be\label{calmmetric} 
\d s^2_\calm\equiv\calg^{ij}\d\ell_i\d\ell_j-\calg_{ij}\d\varphi^i\d\varphi^j\,.
\ee
So, as a first check, observe that the  trajectory in the extended moduli space $\calm$ along half wormhole covers a {\em finite} time-like distance even for  $k\leq 3$, which coincides with $D_{\text{\tiny IDB}}$ as defined in \eqref{IDBD}:
\be\label{genID} 
\int^{\tau_\infty}_{0}\sqrt{-\d s^2_\calm}=D_{\text{\tiny IDB}}\,.
\ee
This result follows more directly from  \eqref{varphieq}, \eqref{encons}, \eqref{WHL}, and \eqref{tauinf}, and actually holds in full generality \cite{Arkani-Hamed:2007cpn}. Indeed, it is 
at the origin of the formulation of the IDB provided in \cite{Maldacena:2026jqd}.

Furthermore, the divergence of \eqref{varphi} at $\tau=\pm\tau_{\rm cr}$ can be avoided  by replacing the $\calm$ coordinates $\varphi^i$ with the null coordinates 
 \be\label{upm} 
u^i_\pm\equiv s^i\pm \varphi^i\,,
\ee
where we recall that the saxions $s^i$ are related to the dual saxions by \eqref{ellsrel}. In terms of the saxions, the trajectory \eqref{Lindell} corresponds to 
\be\label{stau} 
{\bm s}=\frac{{\bm s}_*}{\cos\left(\frac\pi2 \sqrt{\frac{3}{k}}\,\frac{\tau}{\tau_\infty}\right)}\,.
\ee
Combined with \eqref{varphi}, this implies that the null $\calm$ coordinates \eqref{upm} have profiles:
\be\label{upmflow}
{\bm u}_+(\tau)= {\bm s}_*\,\tan\left(\frac\pi4 \sqrt{\frac{3}{k}}\,\frac{\tau_{\rm cr}-\tau}{\tau_\infty}\right)\quad,\quad
{\bm u}_-(\tau)={\bm s}_*\,\tan\left(\frac\pi4 \sqrt{\frac{3}{k}}\,\frac{\tau_{\rm cr}+\tau}{\tau_\infty}\right)\,.
\ee
Hence,  the new field space coordinates $u^i_\pm$ also move along a straight line in the direction identified by $s^i_*$, but remain finite, and actually vanish, at $\tau=\pm\tau_{\rm cr}$, repectively. Moreover, in  the Eddington-Finkelstein-like coordinates  $(\ell_i,u^j_\pm)$,  the metric \eqref{calmmetric} becomes 
\be\label{calmmetric2} 
\d s^2_\calm =\d u^i_\pm\d\ell_i -\calg_{ij}({\bm\ell})\,\d u^i_\pm \d u^j_\pm\,.
\ee
Since $\calg_{ij}(\lambda{\bm\ell})=\lambda^2\calg_{ij}({\bm\ell})$, the components $\calg_{ij}({\bm\ell}(\tau))$ vanish at $\tau=\pm\tau_{\rm cr}$, but the metric \eqref{calmmetric2} remains non-degenerate. This suggests that these wormhole saddles should also be considered for $k\leq 3$.

In the axion formulation, the relevant two-derivative terms contributing to the on-shell action are given by the sum of \eqref{EEH1} and \eqref{Eellaction}. As in the non-supersymmetric models discussed in Section \ref{sec:axioWH}, it is straightforward to show, using the traced Einstein equations, that their on-shell contribution vanishes, in agreement with the more general argument of \cite{Maldacena:2026jqd}. Consequently, they do not provide any exponential suppression of the corresponding path integral contribution. As argued in \cite{Maldacena:2026jqd}, the presence of the arbitrary parameter $L$ then suggests an overall divergent contribution, signaling the existence of non-perturbative EFT corrections. As in the purely axionic case, this divergence should provide the semiclassical counterpart of the sum of an infinite number of fixed-charge wormholes. In order to make the connection more explicit we notice that, even for more general (non-homogeneous) wormholes, by using \eqref{varphieq} the fixed-charge on-shell  two-derivative action \eqref{WHonshell} can be rewritten as 
\be\label{onshellsaction} 
S^{\text{\tiny WH}}_{\bf q}=2\pi \int^{\tau_\infty}_{-\tau_\infty} \d\tau \,q_i\frac{\d\varphi^i}{\d\tau}= 4\pi\langle{\bf q},{\bm\varphi}_{\text{\tiny  IDB}}\rangle\,,
\ee
where ${\bm\varphi}_{\text{\tiny  IDB}}\equiv\frac12\left({\bm\varphi}(\tau_\infty)-{\bm\varphi}(-\tau_\infty)\right)$. Equation \eqref{onshellsaction} has general validity \cite{Arkani-Hamed:2007cpn}, and we have already encountered it in \eqref{WHq1}. In the case  \eqref{varphi}, one gets 
\be\label{genIDBvarphi} 
{\bm\varphi}_{\text{\tiny  IDB}}={\bm s}_*\tan\left(\frac\pi2 \sqrt{\frac{3}{k}}\right)
\ee
and \eqref{onshellsaction} becomes
\be\label{OShomS} 
S^{\text{\tiny WH}}_{\bf q}= 4\pi\langle{\bf q},{\bf s}_*\rangle \tan\left(\frac\pi2 \sqrt{\frac{3}{k}}\right)=4\pi\langle{\bf q},{\bm s}_\infty\rangle\sin \left(\frac\pi2 \sqrt{\frac{3}{k}}\right)\,.
\ee
By combining \eqref{ellsrel}, ${\bm\ell}_*=\tilde\ell_* {\bf q}$ and \eqref{ellsnorms}, \eqref{OShomS} indeed coincides with \eqref{WHonshell4}, and suggests that it is actually valid for more general $k$. 

Notice that, so far, we have neglected the contribution of the Gauss--Bonnet term \eqref{GRtheta}. Since $\gamma$ now depends on the saxions as in \eqref{gammaGB}, this term can no longer be regarded as a constant topological contribution. In the next section, we will proceed by neglecting it, or equivalently by assuming $\tilde\sfK_i=0$. Its role will then be reconsidered in Section \ref{sec:GBimpl}.  

%%%%%%%%%%%%%%%%%%%%%%%%%%%%%%%%%%%%%%%%%%%%%%%%%%%%%%%%%%%%%%%

\subsection{EFT instanton corrections}
\label{sec:EFTinstcorr}
 
As for the purely axionic wormholes discussed in Section \ref{sec:axioWH}, the divergence of the imaginary wormhole contribution may be interpreted \cite{Maldacena:2026jqd} as a coarse-grained manifestation of the underlying fundamental instanton corrections. 

In a supersymmetric context, BPS instantons are expected to dominate, or at least play an important role, among the possible non-perturbative contributions. These generate terms proportional to \eqref{caloqop} or their complex conjugate, with ${\bf q}\in \calc_{\rm I}$, where $\calc_{\rm I}$ denotes the set of BPS instanton charges defined in \eqref{CIcone}. The complex conjugate contribution can be interpreted as generated by the anti-BPS instanton of charge $-{\bf q}$. Writing \eqref{caloqop} in the form \eqref{instcorr}, the corresponding actions are given by 
\be\label{BPSinstaction} 
S^{\text{\tiny BPS}}_{\bf q}=\overline S^{\text{\tiny BPS}}_{-{\bf q}}=2\pi\langle  {\bf q},{\bm s}\rangle\ .
\ee
So, the action of BPS instantons is completely fixed by supersymmetry. Note that, as in the case of BPS strings in \eqref{stringWGC}, using \eqref{ellsrel}, \eqref{ellsnorms} and the standard Cauchy–Schwarz inequality, it is easy to see that the actions \eqref{BPSinstaction} satisfy the WGC-like upper bound
\be\label{instBPSb}
S^{\text{\tiny BPS}}_{\bf q}=\overline S^{\text{\tiny BPS}}_{-{\bf q}}\leq 2\pi\sqrt{\frac{k}2}\,\|{\bf q}\|\,,
\ee
as already observed in \cite{Etheredge:2026rio}. However, as we will see, differently from the non-supersymmetric axiverse models considered in Section \ref{sec:axioWH}, this does not necessarily imply that these instantons  can contribute to the realization of the IDB.

In order to address this question, we evaluate  \eqref{instcorr} along the wormhole -- see \eqref{axionprof}. Up to the overall constant phases $e^{\pm2\pi\ii\langle {\bf q},{\bm a}_0\rangle}$, from \eqref{BPSinstaction} we get 
\be\label{BPSinstcontr} 
e^{-S^{\text{\tiny BPS}}_{\bf q}(\tau)+2\pi\langle {\bf q},{\bm \varphi(\tau)} \rangle}=e^{-2\pi \langle{\bf q},{\bf u}_-(\tau)\rangle}\quad,\quad e^{-\overline S^{\text{\tiny BPS}}_{-{\bf q}}(\tau)-2\pi\langle {\bf q},{\bm \varphi(\tau)} \rangle}=e^{-2\pi \langle{\bf q},{\bf u}_+(\tau)\rangle}\,.
\ee
Since $\langle{\bf q},{\bm s}_*\rangle >0 $ for ${\bf s}_*\in\Delta$ -- see \eqref{saxcone} -- 
from \eqref{upmflow} we immediately conclude  that, moving away from the throat along the radial direction, either $\langle{\bf q},{\bf u}_+(\tau)\rangle$ or  $\langle{\bf q},{\bf u}_-(\tau)\rangle$ gradually decrease, and eventually vanish if the trajectory can reach the critical points $\tau=\pm\tau_{\rm cr}$.  Since $\tau_{\rm cr}\leq \tau_\infty$ only if $k\leq 3$, we conclude that, at least for $k\leq 3$, BPS fundamental instantons provide the natural candidates to realize the IDB. So, as a minimal requirement, there should exist a non-empty set  of charges $\cali\subset \calc_{\rm I}$ populated by  BPS instantons.

However, one can again invoke a stronger interpretation of the diverging imaginary wormhole  contribution, as providing  coarse-grained information on the instanton corrections  that actually appear. Indeed, the fixed-charged wormholes considered above carry only the restricted set of axion charges \eqref{CWH}, which is contained but does not necessarily coincide with the set of BPS instanton charges $\calc_{\rm  I}$:\footnote{This can be understood by recalling that the set $\calc_{\rm S}^{\text{\tiny EFT}}$ of  EFT string charges generate the closure of the saxionic cone $\Delta$. Since the corresponding string tension $\calt_{\bf e}=M^2_{\Pl}\langle{{\bm \ell},\bf e}\rangle$ -- see \eqref{stringten}  -- must be positive for any ${\bm\ell}\in\calp$, and hence also for ${\bf q}\in \calc_{\text{\tiny WH}}$.}
\be 
\calc_{\text{\tiny WH}}\subset \calc_{\rm I}\,.
\ee
The coarse-grained interpretation of these wormholes, which appear in the spectral decomposition of corresponding imaginary wormholes, therefore suggests that the full set of charges in $\calc_{\text{\tiny WH}}$, or at least an infinite subset thereof, should correspond to BPS instanton contributions realizing the IDB. Since $\calp$ is open, it is natural to extend this charge set to
\be\label{EFTinst} 
\calc^{\text{\tiny EFT}}_{\rm I}\equiv\overline\calp\cap V^*_{\mathbb{Z}}\,,
\ee
or to an infinite subset thereof,  spreading in all radial directions. 

The set \eqref{EFTinst} corresponds to the {\em EFT instantons} introduced in \cite{Lanza:2021udy} and further studied in \cite{Martucci:2024trp}. At the macroscopic level, these instantons admit a controlled backreaction preserving two supercharges. Furthermore, in analogy with what  discussed for EFT strings in Section \ref{sec:windingbreak}, it is straightforward to see that the Cauchy–Schwarz inequality underlying the WGC bound \eqref{instBPSb}, and hence the bound itself, can be saturated precisely, and only, by EFT instantons. At the microscopic level, the investigation of several string theory models indicates that EFT instantons typically correspond to “movable” Euclidean branes, which can explore the entire compactification space. For instance, in the large volume F-theory compactifications briefly introduces in \ref{sec:susyexam}, $\mathcal{C}^{\text{\tiny EFT}}_{\rm I}$ corresponds to Euclidean D3-branes   wrapping movable divisors \cite{kawamata1988crepant}; see \cite{Martucci:2024trp} for more examples.

In models with $k>3$ the above conclusions do not hold. Indeed, for any BPS-instanton charge,  either EFT or not, the maximal value of \eqref{BPSinstcontr} along the imaginary wormhole profile remains smaller than
\be 
e^{-2\pi c\langle {\bf q},{\bm s}_*\rangle }\,,
\ee
for some finite constant $c$, e.g.\ $c\simeq 9.47$ for $k=4$. Since $\langle {\bf q},{\bm s}_*\rangle>0$, for any fixed ${\bm s}_*$ only a finite number of charges ${\bf q}\in\mathcal{C}_{\rm I}$ can give unsuppressed contributions. However, even in a model where \eqref{tildePhom} is homogeneous of degree $k>3$, it may still be possible to consistently focus on a saxionic subsector with effective homogeneity $k'<k$ and $k'\leq 3$, for instance when $\tilde{\mathcal{P}}({\bm\ell})$ factorizes; see Section \ref{sec:susyexam} for further examples. Running the imaginary wormhole argument in each of these subsectors, one would conclude that there should still exist an infinite number of corresponding EFT instanton contributions. These BPS instantons could be sufficient to realize a milder form of the IDB derived from imaginary wormholes associated with $k>3$ sectors.

As a simple example, consider a model with only two dual saxions $\ell_1$ and $\ell_2$, a dual saxionic cone $\mathcal{P}={\ell_{1,2}>0}$, and kinetic potential $\mathcal{F}({\bm\ell})=\ell_1 \ell_2^3$, which has $k=4$. Recalling \eqref{CWH}, the corresponding homogeneous wormholes carry strictly positive charges $q_1,q_2\geq 1$, and the associated IDB cannot be realized by BPS instantons. On the other hand, one can consistently focus on the two one-dimensional saxionic subsectors described separately by $\ell_1$ or $\ell_2$, keeping the other saxion fixed. These subsectors have $k'=1$ and $k'=3$, respectively, and therefore allow the realization of the IDB by EFT instantons. In the original theory, these carry charges $(q_1,0)$ and $(0,q_2)$, respectively, with $q_{1,2}\geq 1$. 

These considerations lead to the following BPS tower and BPS sublattice versions of the axion WGC -- see also \cite{Martucci:2024trp}. In the BPS tower version, for any saxionic sector with homogeneity $k\leq 3$ and any ${\bf q}\in \calc^{\text{\tiny EFT}}_{\rm I}$, there exists an integer $p\geq 1$ such that an EFT instanton of charge $p\,{\bf q}$ exists. In the BPS sublattice version, for any saxionic sector with homogeneity $k\leq 3$ there exists a fixed integer $p\geq 1$ such that an EFT instanton exists for every charge of the form $p\,{\bf q}$, with ${\bf q}\in \calc^{\text{\tiny EFT}}_{\rm I}$.

On the other hand, in its strongest form, the imaginary wormhole argument should also apply to the saxionic sectors with $k>3$, implying the existence of infinite towers of instantons whose charges are still contained in $\calc^{\text{\tiny EFT}}_{\rm I}$. These instantons would necessarily be {\em non}-BPS and, moreover,  have actions smaller than those of the corresponding BPS instantons: $S^{\text{\tiny non-BPS}}_{{\bf q}}\leq S^{\text{\tiny BPS}}_{{\bf q}}$.  
More precisely, the combination $S^{\text{\tiny non-BPS}}_{{\bf q}}({\bm s}(\tau))$ should become smaller than   $2\pi|\langle {\bf q},{\bm\varphi}(\tau)\rangle |$  for sufficiently large $\tau\leq \tau_\infty$. So, imposing this condition for $\tau=\pm\tau_\infty$ and using \eqref{onshellsaction}, we get 
\be\label{nonBPSWGC} 
S^{\text{\tiny non-BPS}}_{{\bf q}}({\bm s}_\infty)\leq  2\pi\langle {\bf q},{\bm\varphi}_{\text{\tiny IDB}}\rangle = \frac12 S^{\text{\tiny WH}}_{{\bf q}}  \,.
\ee
Notice that, as in the general analysis of \cite{Arkani-Hamed:2007cpn},  for $k>3$  we have  $S^{\text{\tiny WH}}_{{\bf q}}< 2S^{\text{\tiny BPS}}_{{\bf q}}({\bm s}_\infty)$   -- see  also  \cite{Bergshoeff:2004fq,Bergshoeff:2004pg}, and  \cite{Martucci:2024trp} for a  discussion of this inverse BPS bound in the present context -- 
 which makes the inequality  $S^{\text{\tiny non-BPS}}_{{\bf q}}<S^{\text{\tiny BPS}}_{{\bf q}}$ sharper.
It would be extremely interesting to test this bound in concrete string theory models -- see, for instance, \cite{Demirtas:2019lfi,Long:2021lon} for examples of microscopic realizations of non-BPS branes.

The bound \eqref{nonBPSWGC} takes the same form as \eqref{AWGC0}, but we cannot set   $\frac12 S^{\text{\tiny WH}}_{{\bf q}}=2\pi D _{\text{\tiny IDB}}\|{\bf q}\|_\infty$ as in \eqref{AWGC1}.
Rather, by a direct computation using \eqref{Lindell} and \eqref{stau},  we get
\be\label{SDqrel} 
\frac12 S^{\text{\tiny WH}}_{{\bf q}}=4\sqrt{\frac{k}{3}}\,\sin\left(\frac\pi2\sqrt{\frac{3}{k}}\right) D _{\text{\tiny IDB}}\|{\bf q}\|_\infty<2\pi D _{\text{\tiny IDB}}\|{\bf q}\|_\infty\,.
\ee
Since ${\bm s}_\infty$ can take any value within the saxionic cone, it follows that $S^{\text{\tiny non-BPS}}_{{\bf q}}$ satisfies the axion WGC conjecture with the coefficient fixed as in \eqref{AWGC1}.
 If  $k\leq 3$, as discussed above, the relevant instantons are the EFT ones, and \eqref{instBPSb} implies that they  satisfy \eqref{AWGC1} too. This further supports the possibility  that, in any quantum gravity model,   \eqref{AWGC1} provides a universal upper bound, valid for all the relevant instantons implied by the IDB.\footnote{Interestingly, as observed in \cite{Etheredge:2026rio},  \eqref{instBPSb} implies that the same conclusion applies also to any (anti-)BPS instanton  if $k\leq 7$. This  is precisely the upper bound on the homogeneity degree satisfied by all known string theory constructions. }

Of course, it remains to understand which types of corrections are actually generated. The non-BPS instantons contributing to the realization of the IDB in  $k>3$ saxionic sectors certainly produce D-term corrections to the EFT and can, for instance, modify the K\"ahler potential. By contrast, the EFT instantons associated with wormholes of  $k\leq 3$ saxionic sectors  should instead contribute to F-terms, either through the generation of a superpotential,  or through higher F-term corrections of the type discussed in \cite{Beasley:2005iu}.

Assuming, as in Section \ref{sec:susymodels}, that $k$ is an integer reduces the possibilities with $k\leq 3$ to the three cases $k=1,2,3$, with $k=3$ representing an isolated case. Indeed, the wormholes corresponding to $k=3$ and $k=1,2$ are qualitatively rather different, and one may therefore expect them to give rise to different types of EFT instanton corrections. In particular, as we will see in the next subsection, in the special case $k=3$ additional information can be extracted directly from the wormholes themselves.

%%%%%%%%%%%%%%%%%%%%%%%%%%%%%%%%%%%%%%%%%%%%%%%%%%%%%%%%%%%%%%%%%%%%

\subsection{Fermionic zero modes, D-terms and superpotentials}
\label{sec:EFTsup}

\begin{figure}[ht]
\centering
\begin{subfigure}{0.48\textwidth}
\centering
\includegraphics[width=\textwidth]{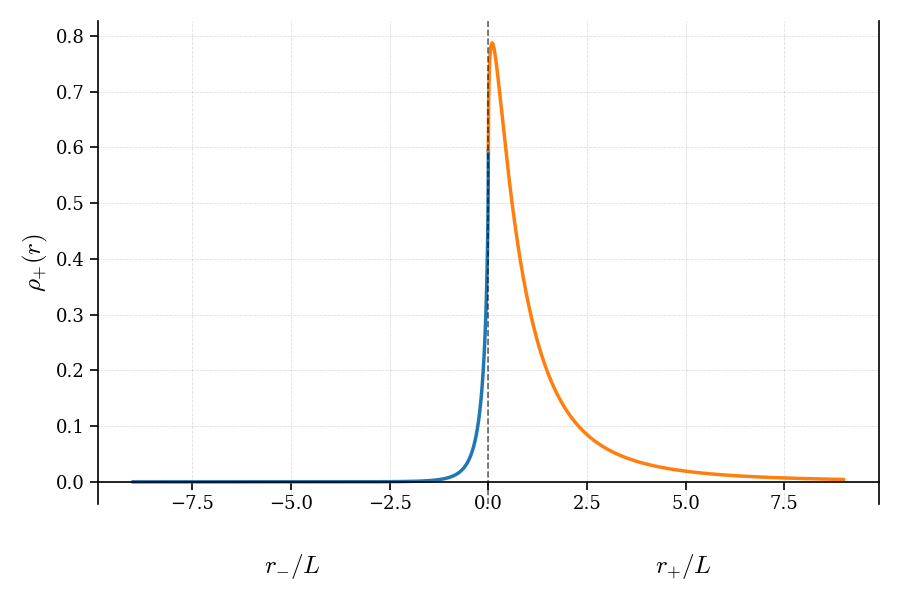}
\caption{$\rho_+$}
\end{subfigure}
\hfill
\begin{subfigure}{0.48\textwidth}
\centering
\includegraphics[width=\textwidth]{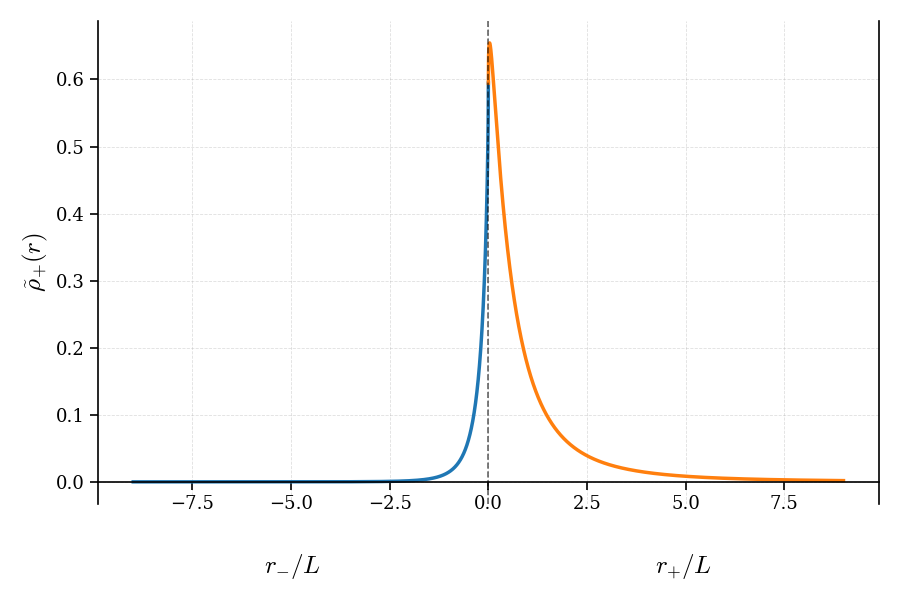}
\caption{$\tilde\rho_+$}
\end{subfigure}

\caption{
Profiles of $\rho_+$ and $\tilde\rho_+$ for $k=5$. In each plot, the left branch corresponds to the lower half-wormhole  and the
right branch to the higher half-wormhole.
}
\label{fig:rhoprofk4}
\end{figure}

All the wormholes discussed so far completely break supersymmetry.  Consider first the cases $k\geq 4$, in which supersymmetry is completely broken all over the wormhole solution.
As discussed in \cite{Martucci:2024trp},  there are eight fermionic zero modes, parametrized by eight fermionic  variables $\theta^\alpha_{\pm}$ and $\overline\theta^{\dot\alpha}_{\pm}$. Focusing on the chiral fermionic partners $\chi^i_\alpha,\bar\chi^i_{\dot\alpha}$  of the chiral fields $t^i$, on the homogeneous wormhole configurations the eight fermionic zero modes have profiles
\be\label{chizerom} 
\chi^{i}_{\pm\alpha}(\tau)=\frac1{L} s^i_*\,\rho_{\pm}(\tau)\delta_{\alpha\dot\beta}\,\bar\theta_{\pm}^{\dot\beta}\quad,\quad \overline\chi^{i\dot\alpha}_{\pm}(\tau)=\frac1{L}\, s^i_*\,\tilde\rho_{\pm}(\tau)\delta^{\dot\alpha\beta}\theta_{\pm\beta}\,.
\ee We have organized them into two separate groups of four independent zero modes, mostly localized on the upper/lower half-wormhole and labeled by $+/-$, respectively.  Each zero mode is equivalent to a global supersymmetry transformation in one asymptotic region and vanishes in the opposite asymptotic region. In order to write the more explicit form of these zero modes, it is convenient to introduce the new global radial coordinate 
\be\label{ytaur} 
y\equiv \frac{\pi}{2} \frac{\tau}{\tau_\infty}=\pm\arccos\left(\frac{L^2}{r^2_\pm}\right)\in \left(-\frac\pi2,\frac\pi2\right)\,,
\ee
in which the metric \eqref{WH} takes the form 
\be\label{ymetric} 
\d s^2=L^2\left(\frac{\d y^2}{4\cos^3 y}+\frac{\d\Omega^2}{\cos y}\right)\,.
\ee
In \eqref{ytaur} we have also indicated the relation with the radial coordinates $r_\pm\in[L,\infty)$ on the two branches $y\in[0,\frac\pi2)$ and $y\in(-\frac\pi2,0]$, respectively. The zero mode profiles have been derived in \cite{Martucci:2024trp}. We can fix their overall normalization so that 
\be
\begin{aligned}
 \rho_+(y)  & = \lq \tan \lt \frac{\pi}{4} + \sqrt{\frac{3}{k}} \, \frac{y}{2} \rt \rq^{\frac{k}{4} -1} \frac{ \left[ \cos y \, \cos \left( \frac{y}{2}-\frac{\pi}{4} \right) \right]^{3/2} }{ \cos \left( \sqrt{\frac{3}{k}} \, y\right) } \, , \\
 \tilde\rho_+(y)& = \lq \tan \lt \frac{\pi}{4} + \sqrt{\frac{3}{k}} \, \frac{y}{2} \rt \rq^{1- \frac{k}{4}} \frac{ \left[ \cos y \, \cos \left( \frac{y}{2}-\frac{\pi}{4} \right) \right]^{3/2} }{ \cos \left( \sqrt{\frac{3}{k}} \, y\right) } \\
\end{aligned}
\ee
while $\rho_-(y)=\tilde\rho_+(-y)$ and $\tilde\rho_-(y)=\rho_+(-y)$. See Figure \ref{fig:rhoprofk4} for a plot of these profiles, as functions of $r_\pm/L$ for $k=5$, which gives  a qualitative description of $\rho_+,\tilde\rho_+$ also for other values $k\geq 4$. It is clear that the  profiles of $\rho_+,\tilde\rho_+$ are very similar, and for $k=4$ they  actually coincide. 

The fermionic variables $\theta^\alpha_\pm$ and $\bar\theta^{\dot\alpha}_\pm$, together with the bosonic zero modes $x^\mu_\pm\in \mathbb{R}^4_\pm$, contribute to the gravitational path integral measure by a factor \be\label{longdist}
\int_{\mathbb{R}^4_+}\d^4 x_+\, \d^2\theta_+\d^2\bar\theta_+
\int_{\mathbb{R}^4_-}\d^4 x_-\, \d^2\theta_-\d^2\bar\theta_- \, .
\ee
As in \cite{Witten:2026twr}, from the long-distance point of view the wormhole contribution is well approximated by the insertion of bilocal terms, including contributions with opposite zero-form charges. The nature of some of these terms can be understood by inserting two $\chi$'s and two $\overline\chi$'s on each side of the wormhole. Using the relation \eqref{ytaur} between $y$ and $r_+$, one sees that the corresponding zero-mode profiles behave as fermionic propagators for $r_+/L\gg 1$. Hence, in the bilocal effective theory, the result is well approximated by the presence of an effective quartic fermionic coupling on each side of the wormhole. These couplings may  be identified with a coarse-grained manifestation of a correction  to the D-term contribution
\be\label{4ferm}
-\frac14M^2_\Pl\int \left(\calr_{i\bar\imath j\bar\jmath}
-\frac12 g_{i\bar\imath}g_{j\bar\jmath}\right)
\chi^i\chi^j \overline\chi^{\bar\imath}\overline\chi^{\bar\jmath}*1
\ee
appearing in the  $\caln=1$ supergravity action (see e.g.\ \cite{Wess:1992cp}), where $g_{i\bar\imath}$ and $\calr_{i\bar\imath j\bar\jmath}$ denote the metric and curvature on the moduli space, in chiral coordinates. This modification should correspond to a correction $\Delta K$ of the  K\"ahler potential of the original EFT. This conclusion is compatible with the interpretation of the wormhole as encoding coarse-grained information about the non-BPS instantons discussed above. It also provides a possible explanation of the inverse BPS bound
$S^{\text{\tiny WH}}_{{\bf q}}<2S^{\text{\tiny BPS}}_{{\bf q}}({\bm s}_\infty)$, 
which is necessary in order for the EFT instantons not to contribute to the realization of the IDB. Indeed, otherwise they would generate F-terms that would not be compatible with the coarse-grained information encoded by the $k\geq 4$ wormholes.

Consider now the case $k=3$, focusing for the time being on a homogeneous wormhole of fixed charge ${\bf q}$. A subtlety distinguishing this case from the others is that the on-shell action $S^{\text{\tiny WH}}_{\bf q}$ diverges, since ${\bm\ell}_\infty=0$, or equivalently $\langle{\bf q},{\bm s}_\infty\rangle=\infty$, see \eqref{OShomS}. This would make the usual exponential factor $\exp(-S^{\text{\tiny WH}}_{\bf q})$ vanish. However, because of \eqref{onshellsaction}, this divergence is precisely canceled by an equally divergent term coming from the imaginary axion shift $\Delta{\bm a}_\infty\rightarrow \Delta{\bm a}_\infty-2{\bm\varphi}_{\text{\tiny IDB}}$ associated with the Dirichlet boundary conditions of the imaginary wormhole. Even though the combination
$S^{\text{\tiny WH}}_{\bf q}-4\pi\langle{\bf q},{\bm\varphi}_{\text{\tiny IDB}}\rangle$ 
remains finite (and actually vanishes), it is convenient to regularize these divergences in order to distinguish more cleanly the kind of coarse-grained information encoded in these wormholes. As a simple possibility, one can slightly modify $P({\bm s})$ in \eqref{homP}, and hence $\tilde P({\bm\ell})$ in \eqref{tildePhom}, so that
\be\label{kreg}
k=3(1+\varepsilon)\,,\qquad 0<\varepsilon\ll 1\,,
\ee
as in \cite{Park:1990ep,Martucci:2024trp}. The precise way in which this deformation is implemented will be irrelevant.

Consider first \eqref{Lindell}. Expanding it to leading order in $\varepsilon$, and using \eqref{coordchange} and \eqref{tauinf}, one obtains
\be\label{appell}
\begin{aligned}
{\bm\ell}(r_\pm)
&\simeq {\bm\ell}_*\left[
\frac{L^2}{r^2_\pm}
+\frac\varepsilon2\arccos\left(\frac{L^2}{r^2_\pm}\right)
\sqrt{1-\frac{L^4}{r^4_\pm}}
\right]
\simeq \frac{{\bf q}}{2\pi M^2_{\Pl}r^2_\pm}+{\bm\ell}_\infty\,,
\end{aligned}
\ee
with
\be
{\bm\ell}_\infty=\frac{\varepsilon {\bf q}}{8M^2_{\Pl}L^2}\,.
\ee
In the second step of \eqref{appell}, we have used \eqref{Ltau} and the fact that the first term dominates for $r^2_\pm\lesssim L^2/\varepsilon$, while the second dominates for $r^2_\pm\gtrsim L^2/\varepsilon$. The profiles on the right-hand side of \eqref{appell} well approximate  the extremal BPS profile \cite{Lanza:2021udy} generated by the insertion of an EFT instanton of charge ${\bf q}$ on $\mathbb{R}^4_+$ and an anti-EFT instanton of charge $-{\bf q}$ on $\mathbb{R}^4_-$. Hence, on each side of the wormhole, the deviation from the extremal BPS case is  localized around the wormhole throat, and for $r_\pm$ mildly larger than $L$ the wormhole preserves one-half of the bulk supersymmetry. This conclusion is further supported by the observation that the on-shell action \eqref{OShomS} is well approximated by twice the EFT instanton action:
\be
S^{\text{\tiny WH}}_{\bf q}
\simeq 4\pi\langle{\bf q},{\bm s}_\infty\rangle\left[1+\calo(\varepsilon^2)\right]
=
\left[
S^{\text{\tiny BPS}}_{\bf q}({\bm s}_\infty)
+\overline S^{\text{\tiny BPS}}_{-\bf q}({\bm s}_\infty)
\right]\left[1+\calo(\varepsilon^2)\right]\,.
\ee

\begin{figure}[ht]
\centering
\begin{subfigure}{0.48\textwidth}
\centering
\includegraphics[width=\textwidth]{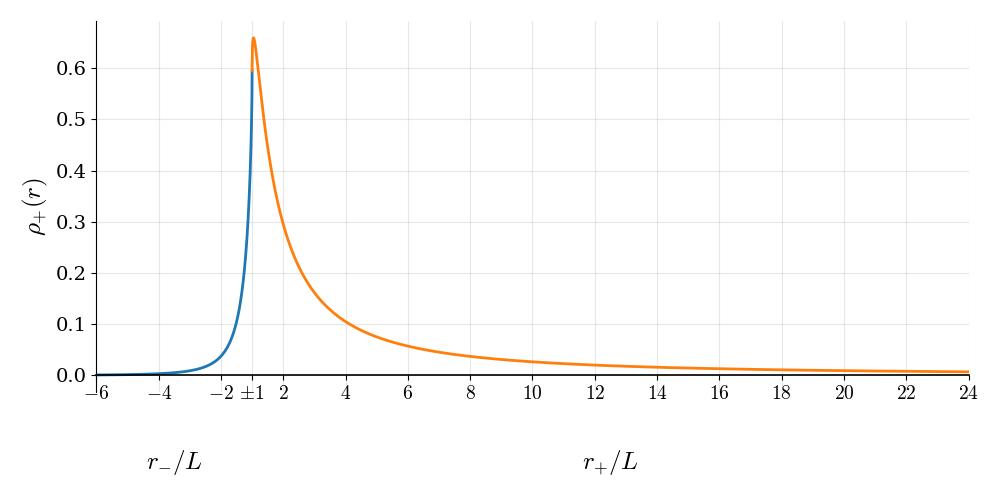}
\caption{$\rho_+$}
\end{subfigure}
\hfill
\begin{subfigure}{0.48\textwidth}
\centering
\includegraphics[width=\textwidth]{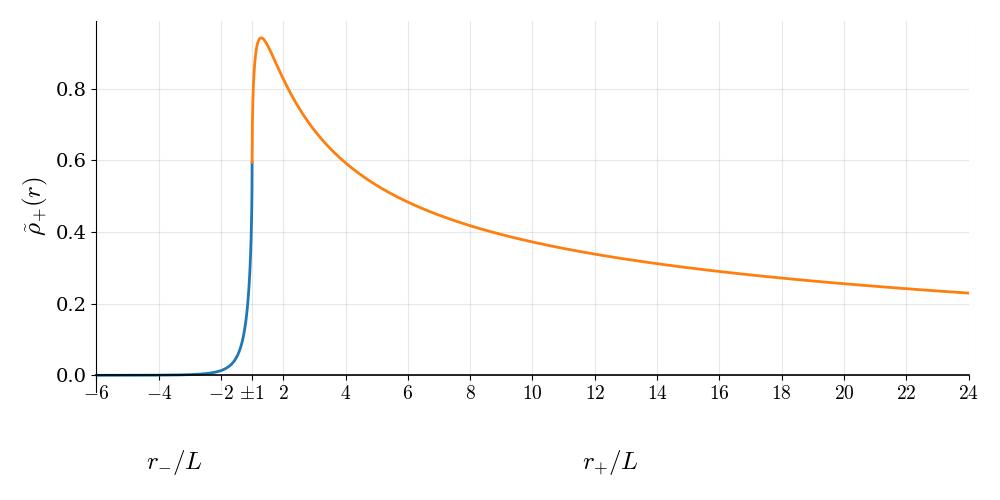}
\caption{$\tilde\rho_+$}
\end{subfigure}

\caption{ 
Profiles of $\rho_+$ and $\tilde\rho_+$ for $k=3(1+\varepsilon)$, with $\varepsilon=10^{-4}$. In each plot, the left branch corresponds to the lower half-wormhole  and the
right branch to the higher half-wormhole.
}
\label{fig:rhoprofk3}
\end{figure}

These remarks are compatible with the observation that, in the limit \eqref{kreg}, the goldstino profiles $\tilde\rho_{+}$ and $\rho_{-}$ spread over the upper/lower sides of the wormhole, while the profiles $\rho_{+}$ and $\tilde\rho_{-}$ are more localized near the throat; see Figure \ref{fig:rhoprofk3} for the plot of $\rho_{+}$ and $\tilde\rho_{+}$ for  $\varepsilon=10^{-4}$. The corresponding profiles $\tilde\rho_{-}$ and $\rho_{-}$ are obtained by reflection about the vertical axis. 

The observations above imply that $\tilde{\rho}_{+}$ and $\rho_{-}$ (that is,
$\bar{\chi}_{+}^{i\dot{\alpha}}$ and $\chi_{-}^{i\alpha}$) characterize the leading  long-distance wormhole effects, and it is  conceivable that we can  focus on these modes in order to determine the most relevant low-energy implications of the IDB. 
In \eqref{longdist} we will  therefore focus on the fermionic integration $\int \d^2\theta_+\,\d^2\bar\theta_-$.\footnote{A systematic study of the effects of the throat-localized
modes $\tilde{\rho}_{-},\rho_{+}$, and of  the corresponding  integration  $\int \d^2\bar{\theta}_+\d^2\theta_{-}$,  requires a more careful treatment of the
asymptotic degeneration of these wormhole solutions and is left for future work.} As in standard supersymmetric instanton calculations, see e.g.\ \cite{Affleck:1983mk}, its long-distance effect can be understood by inserting asymptotic local operators
$\overline\chi^{\bar\imath}_{\dot\alpha}(z_+)\overline\chi^{\bar\jmath}_{\dot\beta}(z'_+)$ and
$\chi^{i}_{\alpha}(z_-)\chi^{j}_{\beta}(z'_-)$. Adapting the calculation in \cite{Martucci:2024trp}, we conclude that, at long wavelengths, the dominant contribution is captured by the effective bilocal F-terms
\be\label{BPSbilocal} 
\int_{\mathbb{R}^4_+}\d^4 x_+\d^2\theta_+ \int_{\mathbb{R}^4_-} \d^4 x_-\d^2\bar\theta_- \sum_{{\bf q}\in\calc_{\text{\tiny WH}}}
c_{\bf q }\, e^{2\pi\ii\langle {\bf q},{\bm T}_+\rangle}e^{-2\pi\ii\langle {\bf q},\overline{\bm T}_-\rangle}\ +\ \text{c.c.}\,,
\ee
where $T^i=t^i+\chi^i\theta+\ldots $ are the chiral superfield extensions of the complex scalars $t^i$. 
In \eqref{BPSbilocal}  we have already summed over all the possible wormhole charges ${\bf q}\in\calc_{\text{\tiny WH}}$, and the CPT conjugate ones $-{\bf q}\in\calc_{\text{\tiny WH}}$, which give the c.c.\ contribution. Furthermore, in \eqref{BPSbilocal}  we have identified $t^i_+=t^i_\infty=a^i_\infty+\ii s^i_\infty$ and $\overline t^i_-=\overline t^i_{-\infty}=a^i_{-\infty}-\ii s^i_\infty$.  For instance, \eqref{BPSbilocal} generates a bilocal effective Yukawa term of the form
\be 
\Big(\sum_{{\bf q}\in \calc_{\text{\tiny WH}}}c_{\bf q}\, q_iq_jq_kq_l\, e^{2\pi\ii\langle {\bf q},{\bm t}_+\rangle  }e^{-2\pi\ii\langle {\bf q},\overline{\bm t}_-\rangle  }\Big)\left(\chi^i_+\chi^j_+\right)(x_+) \left(\overline\chi^k_-\overline\chi^l_-\right)(x_-)\,.
\ee
For real axions, the exponentially suppressed factors  guarantee the convergence of the sum. On the other hand, we are interested in the contribution of the imaginary wormhole, which has imaginary Dirichlet boundary conditions. This can be obtained by making  the imaginary axion shift   $\Delta{\bm a}_\infty\rightarrow \Delta{\bm a}_\infty-2{\bm\varphi}_{\text{IDB}}$, with ${\bm\varphi}_{\text{IDB}}$ as in \eqref{genIDBvarphi}, after which we can remove the $\varepsilon$-regularization and  the sum in \eqref{BPSbilocal} reduces to a sum of $\calo(1)$ terms,
\be 
\sum_{{\bf q}\in \calc_{\text{\tiny WH}}}c_{\bf q}\, q_iq_jq_kq_l\, e^{2\pi\ii\langle {\bf q},\Delta{\bm a}_\infty\rangle} \; ,
\ee
which badly diverges. This is just another manifestation of the divergence observed directly from the vanishing on-shell action of the  imaginary wormhole. 

From \eqref{BPSbilocal}, it is clear that the long-distance  wormhole contribution appears as an infinite sum of bilocal superpotential terms. In combination with the observations of Section \ref{sec:EFTinstcorr}, if we interpret these bilocal terms as coarse-grained information about the UV physics, this suggests that the effective field theory on each side of the wormhole should be modified by an infinite sum of superpotential terms generated by EFT instantons. If correct, in concrete string theory models this picture should admit a microscopic counterpart in terms of movable brane configurations. This possibility was already suggested in \cite{Martucci:2024trp}, and here we see how the IDB provides a natural justification thereof. As already noted in \cite{Martucci:2024trp}, such a mechanism would also give a concrete realization of the Supersymmetric Genericity Conjecture proposed in \cite{Palti:2020qlc}. In Section \ref{sec:EFTinstcorr}, we found that, also for $k=1,2$, EFT instantons are expected to provide the key UV ingredient realizing the IDC. In these cases, however, one cannot repeat the arguments just presented for $k=3$, because the corresponding wormholes degenerate and require an extension of the moduli space. It is therefore difficult to draw definite conclusions about the corresponding potential F-term contributions. In the next subsection, we will encounter another subtlety associated with the cases $k=1,2$.

\subsection{Implications of the Gauss--Bonnet term}
\label{sec:GBimpl}

In Section \ref{sec:susyIDB}-\ref{sec:EFTsup} we have neglected the contribution of the Gauss--Bonnet term \eqref{GRtheta}, where $\gamma$ has the form \eqref{gammaGB}. We also recall that, according to the bound \eqref{GBbound}, in the presence of gauge fields one {\em cannot} generically set $\tilde\sfK_i=0$. Rather, one typically has
\be
\tilde\sfK_i s^i>0\,;
\ee
see \cite{Martucci:2022krl} for a large class of string theory examples. One may therefore wonder whether the Gauss--Bonnet term can affect our conclusions. Notice, by contrast, that the Pontryagin term \eqref{aRR} does not contribute, because of the SO(4) symmetry of the wormhole solution.

An important point in the approach of \cite{DiUbaldo:2026rly,Maldacena:2026jqd} is that, in the argument leading to the IDB, wormholes with small charge ${\bf q}$ and small radius $L$ are irrelevant. This is a significant advantage over the approach followed in much of the literature on the phenomenological implications of wormholes; see, for instance, \cite{Hebecker:2018ofv} for a review. The latter typically requires sufficiently small wormholes in order to obtain a sizeable factor $\exp(-S^{\text{\tiny WH}}_{\bf q})$ and hence a not too suppressed physical effect. By contrast, in the above arguments the imaginary wormholes connecting two asymptotically flat spaces are exploited as `detectors' of an inconsistency of original EFT, and the divergence that leads to the corresponding IDB comes from the large-charge/large-$L$ sector. One may therefore expect higher-derivative corrections to be irrelevant.

We expect this conclusion to certainly hold for the wormholes corresponding to saxionic sectors with homogeneity $k\geq 4$, since the saxion vector \eqref{stau} remains finite along their entire trajectory. On the other hand, for $k\leq 3$, the trajectory \eqref{stau} diverges at the critical points $\tau=\pm\tau_{\rm cr}$, with $\tau_{\rm cr}\leq\tau_\infty$ -- see \eqref{taucr}. In these cases, the harmlessness of the Gauss--Bonnet term is therefore less obvious.

One can estimate the impact of the Euclidean counterpart of \eqref{GRtheta}, namely
\be\label{GRthetaE}
S_{\text{\tiny GB}}\equiv -\int\gamma\,E_{\text{\tiny GB}}\,*1\,,
\ee
by simply evaluating it on the homogeneous wormholes, which solve the two-derivative equations of motion. Recalling \eqref{gammaGB} and \eqref{stau}, a direct calculation gives
\be\label{onshellGB2}
S_{\text{\tiny GB}}|_{\text{\tiny WH}}
=\frac\pi{12}\langle\tilde\sfK,{\bm s}_*\rangle
\int_{0}^{1}\frac{\xi^3\d\xi}{\cos\left(\sqrt{\frac3{k}}\arccos\xi\right)\sqrt{1-\xi^2}}\,.
\ee
It is now clear that the cases $k\geq 3$ and $k=1,2$ have quite different Gauss--Bonnet contributions. Suppose first that $k\geq 3$. The integrand in \eqref{onshellGB2} is positive definite and, furthermore,
\be
\cos\left(\sqrt{\frac3{k}}\arccos\xi\right) \geq \xi\,.
\ee
Hence,
\be\label{onshellGB3}
S_{\text{\tiny GB}}|_{\rm WH}
\leq \frac{\pi}4\langle\tilde\sfK,{\bm s}_*\rangle\int_{0}^{1}\frac{\xi^2\d\xi}{\sqrt{1-\xi^2}}
=\frac{\pi^2}{16}\langle\tilde\sfK,{\bm s}_*\rangle\,,
\ee
with equality for $k=3$. Therefore, for $k\geq 3$, the Gauss-Bonnet term gives only a finite additional suppression that does not affect the conclusions of the previous subsections.

On the other hand, if $k<3$, the integrand diverges at the point $\xi_{\rm cr}$ corresponding  to $\tau_{\rm cr}$. We can split the integral \eqref{onshellGB2} into two contributions, over the intervals $(0,\xi_{\rm cr})$ and $(\xi_{\rm cr},1)$. For $k=1,2$, these two contributions diverge separately. This clearly signals that the contribution of the Gauss--Bonnet term cannot be ignored a priori and should be reconsidered.  Interestingly, however, the contribution from $(0,\xi_{\rm cr})$ gives a positive divergence, while the contribution from $(\xi_{\rm cr},1)$ gives a negative one. One may hence sum them as in a principal-value regularization, getting a finite result. It would be interesting to understand whether this may have any physical interpretation. In fact, for $k=1,2$, the two-derivative on-shell action \eqref{WHonshell3} has an even stronger divergence. However, as already observed, the formula \eqref{WHonshell4} may still be meaningful for $k=1,2$. We hope to return to these interesting questions in the future.

\section{Conclusions}
\label{sec:conclusions}

In this paper we have studied various aspects of the generalized symmetry structure, and of the associated symmetry breaking mechanisms, of four-dimensional axiverse effective field theories, including their extensions to broad classes of $\caln=1$ models.

The first part of the paper, building on previous works, most notably \cite{Choi:2022jqy,Cordova:2022ieu,Choi:2022fgx,Yokokura:2022alv,Apruzzi:2025byj}, provides a unifying framework to identify and study the invertible and non-invertible symmetries of axiverse models including an arbitrary number of periodic axions coupled to an arbitrary number of abelian gauge sectors, with arbitrary kinetic terms and axion-gauge couplings. We also included Gauss-Bonnet and Pontryagin curvature-squared terms, and commented on the effect of non-abelian gauge and other sectors.

As in \cite{Choi:2022fgx}, the non-invertible nature of the electric one-form symmetries and axion shift zero-form symmetries, and their interplay with the invertible magnetic one-form symmetries and winding two-form symmetries, implies hierarchies between the corresponding symmetry-breaking scales. While these hierarchies -- see \eqref{EEbound} and \eqref{Eabound} -- are inevitably more intricate than their simple counterpart in the axion-Maxwell model considered in \cite{Choi:2022fgx}, their physical interpretation in terms of anomalous sectors supported by axion strings and monopoles remains equally clear.

Among the possible symmetry-breaking mechanisms, we focused in particular on the breaking of the non-invertible axion shift symmetries generated by fundamental instantons. The Imaginary Distance Bound recently proposed in \cite{DiUbaldo:2026rly,Maldacena:2026jqd} plays a key role in our discussion, providing a purely bottom-up motivation for the existence of fundamental instantons. Focusing on our $\caln=1$ supersymmetric axiverse models, we obtain our strongest results. Namely,  the existence of a specific class of imaginary wormholes in saxionic sectors with homogeneity degree $k=3$ points to a non-vanishing superpotential generated by an infinite tower of EFT instantons, a possibility already suggested in \cite{Martucci:2024trp}. For saxionic sectors with $k\geq 4$, the same argument predicts D-term corrections generated by  towers of non-BPS instantons. The possible existence of such non-BPS instantons is, however, somewhat puzzling, since their action should be strictly smaller than that of BPS instantons carrying the same charge. The cases $k=1,2$ also suggest the existence of corresponding infinite towers of  EFT instantons generating F-term corrections, but in these cases the saxion profiles along the wormholes pass through infinite field-distance points, which makes the interpretation of these saddles more delicate. Moreover, we argue that precisely for $k=1,2$, the Gauss--Bonnet terms, which are generically expected to be non-vanishing \cite{Martucci:2022krl}, cannot a priori be regarded as small perturbations, and their effect on the wormhole solutions should therefore be properly taken into account.

Of course, many open questions remain. In particular,  while we have established the general framework for investigating non-invertible symmetries in axiverse models, we have discussed only a limited number of possible applications. For instance, by explicitly reintroducing non-abelian sectors, it would be interesting to investigate the potential implications of our results for the strong CP problem in the presence of an axiverse sector. Furthermore, several QFT aspects, such as fusion rules and the categorical structure of the symmetry defects, together with their possible implications, remain to be developed. Finally, the proposal of \cite{DiUbaldo:2026rly,Maldacena:2026jqd} revives the potential role of wormholes as a purely bottom-up window into phenomenologically relevant quantum gravity effects, and the categorical structure of invertible and non-invertible symmetries may act as an organizing principle for  decoding and investigating these effects.

\section*{Acknowledgments}

\noindent We thank Fabio Apruzzi, Giorgio Leone, Alberto Lerda, Salvo Mancani, Fernando Marchesano, Miguel Montero, Georges Obied, Eran Palti, Marco Peloso, Gary Shiu, Thomas Van Riet, Luca Vecchi, Timo Weigand, and Max Wiesner for useful discussions and correspondence. LM also thanks Fabio Apruzzi for collaboration on related projects. This work was supported in part by the Italian MUR Departments of Excellence grant
2023-2027 “Quantum Frontiers” and by the MUR-PRIN contract 2022YZ5BA2 - “Effective Quantum Gravity”. The work of DL is partially supported by the University of Padua under the 2023 STARS Grants@Unipd programme (GENSYMSTR – Generalized Symmetries from Strings and Branes). 

%%%%%%%%%%%%%%%%%%%%%%%%%%%%%%%%%%%%%%%%%%%%%%%%%%%%%%%%%%%%%%%%%%%%%%%%%%%%%%%%%%%%%%%%%%%

\appendix

\section{Integral coprime matrix factorization}  
\label{app:coprime fact}

In this appendix, we collect some definitions and properties regarding the factorization of rational matrices in terms of integral matrices. More details can be found in \cite{Vidyasagar2011}, which works with matrices taking values in a general principal ideal domain and in the corresponding field of fractions, while here we focus on $\bbZ$ and $\bbQ$, respectively. See also Appendix C of \cite{Kaidi:2021gbs} for another discussion of coprime matrices.

Any $n\times m$ rational matrix $\sfQ \in {\rm Mat}(n,m,\mathbb{Q})$ can be factorized in terms of pairs $(\sfP,\sfN)$ and $(\tilde{\sfP},\tilde{\sfN})$ of integral matrices, with invertible $\sfN$ and $\tilde{\sfN}$, as follows:
\begin{equation}
    \sfQ = \sfP  \sfN^{-1} = \tilde{\sfN}^{-1} \tilde{\sfP} \, .
\end{equation}
Here $\sfP$ and $\tilde{\sfP}$ are $n \times m$ matrices, while $\sfN$ and $\tilde{\sfN}$ are invertible $m \times m$ and $n \times n$ matrices, respectively.

The factorization defined by $(\sfP,\sfN)$ is called {\em right-coprime} if there exist two integral matrices $\sfX \in {\rm Mat}(m,n,\mathbb{Z})$ and $\sfY \in {\rm Mat}(m,\mathbb{Z})$ such that
\begin{equation} \label{eqn: right-cop cond}
    \sfX\sfP -\sfY\sfN = \mathds{1}_m\,,
\end{equation}
while the factorization defined by $(\tilde{\sfP},\tilde{\sfN})$ is called {\em left-coprime} if there exist two integral matrices $\tilde{\sfX} \in {\rm Mat}(m,n,\mathbb{Z})$ and $\tilde{\sfY} \in {\rm Mat}(n,\mathbb{Z})$ such that
\begin{equation} \label{eqn: left-cop cond}
    \tilde{\sfP} \tilde{\sfX}  - \tilde{\sfN} \tilde{\sfY}  = \mathds{1}_n\,.
\end{equation}
As discussed below, any rational matrix $\sfQ\in {\rm Mat}(n,m,\mathbb{Q})$ admits both a right-coprime and a left-coprime factorization.

One can prove that a right-coprime factorization $\sfQ = \sfP  \sfN^{-1}$ is also {\em weakly right-coprime} \cite{quadrat2003fractional}, which means that
\begin{equation} \label{eqn:weak right cop def}
     \forall{\bm \alpha} \in \bbQ^m \; \hbox{ such that } \; 
     \sfP {\bm \alpha} \in \bbZ^n \; \hbox{ and } \; 
     \sfN {\bm \alpha} \in \bbZ^m
     \; \Leftrightarrow \; 
     {\bm \alpha} \in \bbZ^m   \, .
\end{equation}
Similarly, a left-coprime factorization $\sfQ = \tilde{\sfN}^{-1} \tilde{\sfP}$ is also {\em weakly left-coprime}, namely
\begin{equation} \label{eqn:weak left cop def}
    \forall \tilde{{\bm \alpha}} \in \bbQ^n \; \hbox{ such that } \; 
    \tilde{\sfP}^{\rm t} \tilde{{\bm \alpha}} \in \bbZ^m \; \hbox{ and } \; 
    \tilde{\sfN}^{\rm t} \tilde{{\bm \alpha}} \in \bbZ^n
    \; \Leftrightarrow \; 
    \tilde{{\bm \alpha}} \in \bbZ^n \,.   
\end{equation}

Let us prove, for instance, the first proposition. The second can be proved in a completely analogous way. Consider a right-coprime factorization $\sfQ = \sfP  \sfN^{-1}$ and a vector ${\bm \alpha} \in \bbQ^m$ such that $\sfP {\bm \alpha} \in \bbZ^n$ and $\sfN {\bm \alpha} \in \bbZ^m$. The definition \eqref{eqn: right-cop cond} then implies
\begin{equation}
    {\bm \alpha} = (\sfX \sfP - \sfY \sfN ) {\bm\alpha} \in \bbZ^m\,,
\end{equation}
since $\sfX$ and $\sfY$ are integral matrices. Conversely, if ${\bm\alpha} \in \bbZ^m$, then $\sfP {\bm\alpha} \in \bbZ^n$ and $\sfN {\bm\alpha} \in \bbZ^m$, since $\sfP$ and $\sfN$ are integral matrices.

It is also possible to prove that a weakly left/right-coprime factorization is left/right-coprime. To see this, let us first discuss how weakly coprime factorizations are related to general factorizations. Suppose that $\sfQ = \sfP \sfN^{-1}$ is a weakly right-coprime factorization and take another factorization, not necessarily weakly right-coprime, $\sfQ = \sfP' \mathsf{N'}^{-1}$. Then
\begin{equation} \label{eqn: transf of fact}
    \sfP'= \sfP \sfT \, , \qquad \sfN'= \sfN \sfT \, ,
\end{equation}
where $\sfT= \sfN^{-1} \sfN'$ and $\det \sfT \neq 0$. Take now any vector ${\bm\alpha} \in \bbZ^m$. Since $\sfP'$ and $\sfN'$ are integral matrices, one has
\begin{equation}
    \sfP' {\bm\alpha} = \sfP \sfT {\bm\alpha} \in \bbZ^n\,,
    \qquad
    \sfN' {\bm\alpha} = \sfN \sfT {\bm\alpha} \in \bbZ^m\,.
\end{equation}
By \eqref{eqn:weak right cop def}, we deduce that $\sfT {\bm\alpha} \in \bbZ^m$. Since ${\bm\alpha} \in \bbZ^m$ is arbitrary, we conclude that
\begin{equation}
    \sfT \in {\rm Mat}(m,\bbZ) \, .
\end{equation}
If $\sfP'$ and $\sfN'$ are also weakly right-coprime, then the same argument applied to the inverse transformation gives $\sfT^{-1} \in {\rm Mat}(m,\bbZ)$. Hence $|\det\sfT| = 1$, that is,
\begin{equation}
    \sfQ=\sfP'\sfN'^{-1} \quad \text{weakly right-coprime} 
    \quad \Leftrightarrow \quad 
    \sfT \in {\rm GL}(m,\mathbb{Z}) \,.
\end{equation}

We can then notice the following two facts.
\begin{itemize}
    \item A unimodular transformation maps right-coprime factorizations to right-coprime factorizations. Indeed, consider a unimodular $\sfT$ in \eqref{eqn: transf of fact}. We can define $\sfX'= \sfT^{-1} \sfX$ and $\sfY'= \sfT^{-1} \sfY$. Then
    \begin{equation}
        \mathds{1}_m = \sfT^{-1} \sfT 
        =  \sfT^{-1}\sfX\sfP \sfT -\sfT^{-1}\sfY\sfN \sfT 
        = \sfX'\sfP' - \sfY' \sfN'  \, . 
    \end{equation}
    \item We have proven that a coprime factorization is also weakly coprime.
\end{itemize}

These two facts imply that any weakly right-coprime factorization is related by a unimodular transformation to any right-coprime factorization of the same rational matrix. Since right-coprime factorizations always exist, as shown in the next subsection, and unimodular transformations preserve right-coprimeness, we conclude that weak right-coprimeness is equivalent to right-coprimeness. Hence, one can use either \eqref{eqn: right-cop cond} or \eqref{eqn:weak right cop def} to characterize a right-coprime factorization. The same argument applies to left-coprime factorizations. In this sense, we can say that a left/right-coprime factorization is the ``minimal'' possible factorization.

\subsection{Construction of coprime factorizations}
\label{app:algor}

In the following, we provide a method for constructing left/right-coprime factorizations of a rational rectangular matrix $\sfQ \in {\rm Mat}(n,m,\bbQ)$. This also provides a constructive proof of the existence of such factorizations.

The procedure is as follows:
\begin{enumerate}
    \item Find the smallest $b\in \bbZ_{>0}$ such that $\mathsf{F} = b \sfQ \in {\rm Mat}(n,m,\bbZ)$.

    \item Find the Smith normal form of $\mathsf{F}$:
    \begin{equation}
        \mathsf{U}\mathsf{F}\mathsf{V} = \hat{\mathsf{F}} \, , 
        \qquad
        \mathsf{U} \in {\rm GL}(n, \bbZ) \, , 
        \qquad
        \mathsf{V} \in {\rm GL}(m, \bbZ) \, .
    \end{equation}
    If $\sfQ$, and hence $\mathsf{F}$, has rank $l$, then $\hat{\mathsf{F}} \in {\rm Mat}(n,m,\bbZ)$ has the form
    \begin{equation} \label{eqn: diag for rect matrix}
        \hat{\mathsf{F}} =
        \begin{bmatrix}
            \hat{a}_1 & 0 & \dots & 0 & 0 & \dots & 0 \\
            0 & \hat{a}_2 & \dots & 0 & 0 & \dots & 0 \\
            \vdots & \vdots & \vdots & \vdots & \vdots & \vdots & \vdots \\
            0 & 0 & \dots & \hat{a}_l  & 0 & \dots & 0 \\
            0 & 0 & \dots & 0 & 0 & \dots & 0 \\
            \vdots & \vdots & \vdots & \vdots & \vdots & \vdots & \vdots
        \end{bmatrix} ,
    \end{equation}
    where the $\hat{a}_i$ are the invariant factors of $\hat{\mathsf{F}}$; see Appendix B.2 of \cite{Vidyasagar2011} for more details.

    \item Obtain the Smith-McMillan form of $\sfQ$:
    \begin{equation}
        \mathsf{U}\sfQ\mathsf{V} = \hat{\mathsf{Q}}\,.
    \end{equation}
    The matrix $\hat{\mathsf{Q}}$ has a form similar to \eqref{eqn: diag for rect matrix}, but now the non-zero entries are
    \begin{equation}
        q_i \equiv \frac{a_i}{b_i} = \frac{\hat{a}_i}{b}\,,
        \qquad
        \gcd(a_i,b_i)=1\,.
    \end{equation}
    In this way, we can write
    \begin{equation}
        \hat{\mathsf{Q}} = \mathsf{A}\mathsf{B}^{-1} 
        = \tilde{\mathsf{B}}^{-1}\mathsf{A} \, ,
    \end{equation}
    where $\mathsf{A}$ is the $n\times m$ matrix of the form \eqref{eqn: diag for rect matrix} with entries $a_i$, while $\mathsf{B}$ and $\tilde{\mathsf{B}}$ are diagonal matrices of dimensions $m$ and $n$, respectively. Their first $l$ diagonal entries are $b_i$, while all remaining diagonal entries are equal to $1$.

    \item We can then obtain coprime factorizations by choosing
    \begin{equation}
        \begin{split}
            \sfP = \mathsf{U}^{-1} \mathsf{A} \, , 
            \qquad 
            \sfN = \mathsf{V}\mathsf{B}
            \quad \Rightarrow \quad & 
            \sfQ = \sfP  \sfN^{-1} \, ;
            \\
            \tilde{\sfP} = \mathsf{A} \mathsf{V}^{-1} \, , 
            \qquad 
            \tilde{\sfN} = \tilde{\mathsf{B}}\mathsf{U}
            \quad \Rightarrow \quad & 
            \sfQ = \tilde{\sfN}^{-1} \tilde{\sfP} \, .
        \end{split}
    \end{equation}
\end{enumerate}

To see that the factorizations obtained with this procedure are actually left/right-coprime, let us choose, for any $i=1,\ldots,l$, two integers $x_i, y_i \in \bbZ$ such that
\begin{equation}
    x_i a_i - y_i b_i =1 \, ,
\end{equation}
which always exist by B\'ezout's identity. Let us focus on the right factorization, since the left one can be treated in a similar way. We define
\begin{equation}
    \hat{\mathsf{X}} = 
    \begin{bmatrix}
        x_1 & 0 & \dots & 0 & 0 & \dots & 0 \\
        0 & x_2 & \dots & 0 & 0 & \dots & 0 \\
        \vdots & \vdots & \vdots & \vdots & \vdots & \vdots & \vdots \\
        0 & 0 & \dots & x_l  & 0 & \dots & 0 \\
        0 & 0 & \dots & 0 & 0 & \dots & 0 \\
        \vdots & \vdots & \vdots & \vdots & \vdots & \vdots & \vdots
    \end{bmatrix},
    \qquad
    \hat{\mathsf{Y}} = {\rm diag}(y_1,\ldots,y_l,-1,\ldots,-1)\,,
\end{equation}
where $\hat{\mathsf{X}}$ is an $m \times n$ matrix, while $\hat{\mathsf{Y}}$ is an $m \times m$ matrix. Then, by taking
\begin{equation}
    \sfX = \hat{\mathsf{X}}\mathsf{U}\,,
    \qquad
    \sfY = \hat{\mathsf{Y}}\mathsf{V}^{-1}\,,
\end{equation}
one immediately verifies that \eqref{eqn: right-cop cond} is satisfied. Hence the factorization is right-coprime.

Notice that the above procedure identifies a particular pair of coprime factorizations, while the most general one can be obtained by composing them with unimodular matrices, as discussed above.

\section{Electric one-form symmetry defects from half higher gauging}
\label{app:halfsg}

In this appendix, we derive the topological defect \eqref{Del} via a half higher-gauging construction, following and generalizing the procedure introduced in \cite{Choi:2022fgx}. Throughout this appendix, wedge and cup products are understood and will be left implicit. More precisely, we consider the
finite subgroups of $U(1)^{n_{\rm V}}_{\rm(m)} \times U(1)^{n_{\rm A}}_{\rm(w)}$ realized by the topological operators 
\begin{equation}
\cald^{\rm(m)}\lt \Sigma^{} \rt
    =
    \exp\lt \ii\,n_J(\sfL^{-1})^J{}_{I}
    \oint_{\Sigma} F^I \rt\,,
    \qquad
    \cald^{(\rm w)}\lt \gamma \rt
    =
    \exp\lt \ii\,\sfQ_{iJ}m^J
    \oint_{\gamma} \d a^i \rt\,.
\end{equation}
with $n_I\in \mathbb{Z}$ and $m^I\in\mathbb{Z}$. Recall that $\sfQ$ is defined in \eqref{sfQmatrix} and therefore depends on ${\bm\alpha}\in W_\mathbb{Q}$, and that  $\sfL$ and $\sfM$ are integral matrices appearing in the right-coprime  factorization \eqref{QPNaxion}. Clearly, these transformations depend only on the equivalence classes  $[{\bf n}]\in \Gamma^*_\sfL\equiv  W^*_\bbZ/(\sfL^{\rm t} W^*_\bbZ)$ and $[{\bf m}]\in \Gamma_\sfL \equiv W_\bbZ/(\sfL W_\bbZ)$. 

We now construct the associated condensation defect on a closed three-manifold
$\tilde\Sigma$. With a choice of discrete torsion, it takes the form
\begin{equation}
    C(\tilde\Sigma ) 
    =\frac{1}{|H^1|}
    \sum_{\substack{
        {\bm b} \in H^1(\tilde\Sigma,\Gamma_\sfL)\\
        {\bm\rho} \in H^2(\tilde\Sigma,\Gamma^*_\sfL)
    }}
    \exp\lq
    \ii(\sfL^{-1})^J{}_{I}
    \oint_{\tilde\Sigma}
    \lt
       2\pi\,  b_J\rho^I
        + b_J F^I
        - 2\pi\sfM_{iJ} \rho^I \d a^i
    \rt
    \rq .
\end{equation}
where $|H^1|\equiv |H^1(\tilde\Sigma,\Gamma_\sfL)|=|H^2(\tilde\Sigma,\Gamma^*_\sfL)|$. 
Equivalently, setting $\tilde b_J\equiv b_J-\sfM_{iJ}\d a^i$,  $\tilde \rho^I=\rho^I+\frac1{2\pi}F^I$,
one gets
\begin{equation}
\label{eqn: close cond defect}
    C(\tilde\Sigma)
    =  \frac{1}{|H^1|}
    \sum_{\substack{
        {\bm b} \in H^1(\tilde\Sigma,\Gamma_\sfL)\\
        {\bm\rho} \in H^2(\tilde\Sigma,\Gamma^*_\sfL)
    }}
    \exp\lq
    2\pi\ii(\sfL^{-1})^J{}_{I}
    \oint_{\tilde\Sigma}\tilde b_J \, \tilde\rho^I
    \rq
    \, 
    \exp\lq
    \ii\,\sfQ_{iI}
    \oint_{\tilde\Sigma} \d a^i \, F^I
    \rq .
\end{equation}
 By the equation of motion \eqref{elcons}, the second factor is trivial on any closed $\tilde\Sigma$. The first factor is also trivial: the cup product defines a perfect finite pairing between
$H^1(\tilde\Sigma,\Gamma_\sfL)$ and $H^2(\tilde\Sigma,\Gamma^*_\sfL)$, so the normalized sum gives one. Hence $C\lt \Sigma \rt =1$ for any closed three-manifold.

We can now perform half higher gauging by allowing $\tilde\Sigma$ to have boundary
$\Sigma\equiv \partial \tilde\Sigma$. The condensation defect becomes
\begin{equation}
\label{eqn: bound cond def}
    \begin{split}
    C( \tilde\Sigma,\Sigma)
    =
    &\exp\lt
    \ii\,\sfQ_{iI}\int_{\tilde\Sigma}\d a^i \, F^I
    \rt
    \\
    &\times
    \frac{1}{|H^1_\partial|}  \sum_{\substack{
        {\bm b} \in H^1_\partial(\tilde\Sigma,\partial \tilde\Sigma;\Gamma_\sfL)\\
        {\bm\rho} \in H^2_\partial(\tilde\Sigma,\partial \tilde\Sigma;\Gamma^*_\sfL)
    }}
    \exp\lt
    \ii(\sfL^{-1})^J{}_{I}
    \int_{\tilde\Sigma}
    \lt b_J-\sfM_{iJ}\d a^i\rt
    \,
    \lt \rho^I+F^I\rt
    \rt ,
    \end{split}
\end{equation}
where $H^1_\partial \equiv H^1(\tilde\Sigma,\partial\tilde\Sigma;\Gamma_\sfL)$ and  $H^2_\partial \equiv H^2(\tilde\Sigma,\partial\tilde\Sigma;\Gamma^*_\sfL)$. 
The first factor is no longer trivial, but reduces to the electric one-form symmetry insertion:
\begin{equation}
    \exp\lt
    \ii\,\sfQ_{iI}\int_{\tilde\Sigma}\d a^i \, F^I
    \rt
    =
    \exp\lt
    \ii\,\alpha^I\oint_{\Sigma}\calj_I^{\rm(e)}
    \rt .
\end{equation}
By following the very same steps of Appendix B of \cite{Choi:2022fgx}, the second factor in \eqref{eqn: bound cond def} reduces to a two-dimensional $\Gamma_\sfL$ gauge theory living on $\Sigma$:
\begin{equation}
\label{eqn: equiv teor}
    \begin{split}
    &\frac{1}{|H^1_\partial|}\sum_{\substack{
        {\bm b} \in H^1_\partial(\tilde\Sigma,\partial \tilde\Sigma;\Gamma_\sfL)\\
        {\bm\rho} \in H^2_\partial(\tilde\Sigma,\partial \tilde\Sigma;\Gamma^*_\sfL)
    }}
    \exp\lt
    \ii(\sfL^{-1})^J{}_{I}
    \int_{\tilde\Sigma}
    \lt b_J-\sfM_{iJ}\d a^i\rt
    \,
    \lt \rho^I+F^I\rt
    \rt
    \\
    &\qquad =
    \int [D\phi\,Dc]_{\Sigma}
    \exp\lg
    \ii\oint_{\Sigma}
    \lt
        \sfL^J{}_{I}\phi_J \d c^I
        + \sfM_{iJ}a^i \d c^J
        + \phi_I F^I
    \rt
    \rg .
    \end{split}
\end{equation}
Combining these two contributions precisely reproduces the defect \eqref{Del}. This proves that \eqref{Del} is topological.

\subsection{Condensation defect}
\label{app:conddef}

Using the definition \eqref{Del}, we can compute the condensation defect
\begin{equation}\label{defus}
\begin{aligned}
&\cald^{\rm(e)}_{\bm\alpha}(\Sigma)\times 
\overline\cald^{\rm(e)}_{\bm\alpha}(\Sigma)
=
\int[\cald\phi \cald \hat{\phi} \cald c \cald \hat{c}]_\Sigma
\\
&\qquad\times
\exp \left\{
\ii \oint_\Sigma
\left[
\sfL^I{}_J \lt \phi_I\d c^J - \hat{\phi}_I \d \hat{c}^J \rt
+\sfM_{iI}a^i \lt \d c^I - \d\hat{c}^I \rt
+\lt \phi_I - \hat{\phi}_I \rt F^I
\right]
\right\} \, .
\end{aligned}
\end{equation}
The fact that this fusion does not give the identity reflects the non-invertible nature of the defect. After the change of variables
\begin{equation}
    \tilde{\phi}_I=\phi_I-\hat{\phi}_I\,,
    \qquad
     \tilde c^I=c^I-\hat c^I\,,
\end{equation}
and an integration by parts, \eqref{defus} becomes
\begin{equation} \label{eqn: elec cond}
\begin{aligned}
&\int[\cald\phi \cald \tilde c]_\Sigma
\exp\left[
\ii\oint_\Sigma
\left(
\sfL^I{}_J\phi_I \, \d \tilde c^J
-\sfM_{iI} \, \d a^i \, \tilde c^I
\right)
\right]
\\
&\qquad\times
\int[\cald \tilde{\phi} \cald \hat{c}]_\Sigma
\exp\left[
\ii\oint_\Sigma
\left(
\sfL^I{}_J \tilde{\phi}_I \, \d \hat{c}^J
+\tilde{\phi}_I F^I
\right)
\right] \, .
\end{aligned}
\end{equation}
Integrating out $\phi_I$ and $\hat c^I$ one gets the constraints $\d \tilde\phi =0 = \d \tilde c$ , $\sfL^I{}_J \tilde c^J \in 2 \pi H^1(\Sigma, \bbZ)$ and $\sfL^I{}_J \tilde \phi_I \in 2 \pi H^0(\Sigma, \bbZ)$. That is, the closed one-form $\eta^I \equiv \sfL^I{}_J \tilde c^J$ and the closed zero-form $\tilde \eta_J \equiv \sfL^I{}_J \tilde \phi_I$ can be regarded, respectively, as the integral uplift of a $\Gamma_\sfL$ and a $\Gamma_\sfL^*$ cohomology class class $[{\bm \eta}] \in H^1(\Sigma, \Gamma_\sfL)$ and $[\tilde{\bm \eta}] \in H^0(\Sigma, \Gamma^*_\sfL)$.
Thus the remaining path integral reduces to a finite sum, and the two factors are precisely the winding and magnetic symmetry generators. Ignoring constant prefactors, we obtain
\begin{equation}
\begin{aligned}
\cald^{\rm(e)}_{\bm\alpha}(\Sigma)\times
\overline\cald^{\rm(e)}_{\bm\alpha}(\Sigma)
=
\sum_{\substack{
 {\bm \eta} \in H^1(\Sigma,\Gamma_\sfL)}}
&\exp \left\{
\ii \,\sfQ_{iI}
\oint_\Sigma \eta^I\d a^i 
\right\}
\\
&\times \sum_{\substack{
\tilde{\bm \eta} \in H^0(\Sigma,\Gamma^*_\sfL)
}}
\exp \left\{
\ii (\sfL^{-1})^I{}_J \oint_\Sigma \tilde\eta_I F^J
\right\} \, .
\end{aligned}
\end{equation}

\section{Invariance under integral shift}
\label{app:shiftinv}

In this appendix, we prove that the defects \eqref{Del} and \eqref{eqn: 0form defec} are invariant under an integer shift of the parameters ${\bm \alpha}$. Let us start with the 1-form symmetry defect. Consider the integer shift $ {\bm \alpha} \rightarrow {\bm \alpha} + {\bf m}$ with ${\bf m} \in W_\bbZ$. We can observe that, given a coprime factorization \eqref{QPNaxion}, for any ${\bf m}\in W_\mathbb{Z}$ the matrix $\sfQ({\bm\alpha}+{\bf m})$ admits the right coprime factorization 
\begin{equation}
\sfQ({\bm\alpha}+{\bf m})=\left[\sfM+\sfQ({\bf m})\sfL\right]\sfL^{-1}\,.
\end{equation} 
Then we can use this factorization to define $\cald^{\rm(e)}_{{\bm\alpha}+{\bf m}}(\Sigma)$ through \eqref{Del}. The term 
\begin{equation}
\ii\oint_\Sigma\sfQ_{iJ}({\bf m})\sfL^J{}_Ia^i\d c^I\,,
\end{equation} 
which appears in the path integral, can be reabsorbed by the field redefinition  $\phi_I\rightarrow \phi_I-a^i|_\Sigma\sfQ_{iI}({\bf m})$. This in turn produces a factor $\exp\left(-\ii\oint_\Sigma a^i\sfQ_{iI}({\bf m})F^I\right)$. On the other hand, by replacing ${\bm\alpha}\rightarrow{\bm\alpha}+{\bf m}$ in the first exponential in \eqref{Del}, an additional factor is obtained $\exp\big(\ii m^I\oint_\Sigma\calj_I^{\rm(e)}\big)$. These two factors combine into the trivial operator $\exp\left(\ii m^I\oint_\Sigma G_I\right)=\mathds{1}$,  whose  triviality is  the operatorial counterpart of the quantization condition $\oint_\Sigma G_I\in 2\pi \mathbb{Z}$. This shows that, indeed, $\cald_{\bm\alpha}^{\rm(e)}$ depends only on the parameters $\alpha^I\in\mathbb{Q}$ mod 1, that is, only on  ${\bm\alpha}$ regarded as an element of $ W_{\mathbb{Q}}/W_{\mathbb{Z}}$.

Now, let us move on to the 0-form symmetry defect. Consider the shift $ {\bm \alpha} \rightarrow {\bm \alpha} + {\bf m}$ with $ {\bf m} \in V_\bbZ$. Similarly to above, given a coprime factorization \eqref{0formCfact}, the matrix $\sfC({\bm \alpha} + {\bf m})$ admits the right coprime factorization
\begin{equation}
\sfC({\bm\alpha}+{\bf m})=\left[\sfP+\sfC({\bf m})\sfN\right]\sfN^{-1}\,.
\end{equation}
In this way, two additional terms appear in the path integral, $\exp \lt \ii \, m^i \oint_\Sigma \calj^{\rm(a)}_i \rt$ and \\ $\exp \lq 1/(4 \pi)(\sfN^{\rm t} \, \sfC({\bf m})  \, \sfN)_{IJ} \int_Y B^I \wedge B^J\rq$. We can notice that the second term, due to the equation of motion of $\tilde{A}_I$, can be written as $\exp \lq 1/(4 \pi) \sfC({\bf m})_{IJ} \int_\Sigma A^I \wedge F^J\rq$, once we restrict to the boundary of $\Sigma$. Then the sum of the two terms obtained is exactly $\d \calb_{2,i}$, introduced in \eqref{BHrel}, and its periods are integer multiples of $2 \pi$. Since $\sfC({\bf m})$ has integer entries, the additional phase we obtain is trivial, as we expected. So we can conclude that $\cald_{\bm\alpha}^{\rm(a)}$ depends only on the parameters $\alpha^i\in\mathbb{Q}$ mod 1, that is, only on ${\bm\alpha}$ regarded as an element of $ V_{\mathbb{Q}}/V_{\mathbb{Z}}$.

\section{Comments on non-abelian sectors}
\label{app:nonabelian}

In this appendix we discuss how the inclusion of a non-abelian sector affect the results of Section \ref{sec:noninvsymm}. It is well known \cite{Gaiotto:2014kfa} that, in the absence of charged matter transforming in non-adjoint representations, these theories can admit $Z_{\rm e}^{(1)}$ electric and $Z_{\rm m}^{(1)}$ magnetic one-form symmetries, where $Z_{\rm e}$ and $Z_{\rm m}$ are finite groups corresponding to the center of $G$ and the Pontryagin dual of $\pi_1(G)$, respectively. These extend the set of symmetries identified in the previous subsections.

More crucial for us is the impact of the non-abelian sector on the zero-form symmetries described in Section \ref{sec:ax0form}. Let us focus on a given simple gauge group factor $G$, since the generalization to more general non-abelian gauge groups is straightforward. As in Footnote \ref{foot:nonabelian}, it will generically couple to the axions through a term of the form
\be\label{nonaFF} 
-\frac1{8\pi}\widehat\sfK_i\int a^i \tr\left(F\wedge F\right)\,,
\ee
where $\widehat\sfK_i\in\mathbb{Z}$ and the trace is normalized so that $(4\pi)^{-2}\int \tr\left(F\wedge F\right)\in\mathbb{Z}$. With this normalization, instantons do not break the axion periodicity $a^i\simeq a^{i}+1$ for generic $\widehat\sfK_i$.

Suppose first that the corresponding $Z_{\rm m}^{(1)}$ magnetic one-form symmetry is trivial. Then the term \eqref{nonaFF} generically breaks the zero-form symmetries identified by \eqref{eqn: 0form defec}. The only surviving ones are those corresponding to ${\bm\alpha}\in V_{\mathbb{Q}}/V_{\mathbb{Z}}$ such that
\be\label{noBcond} 
\langle\widehat{\bf K},{\bm\alpha}\rangle\equiv \widehat\sfK_i\alpha^i\in \mathbb{Z}\,.
\ee
Roughly speaking, this condition removes the rational axionic shift symmetries along one direction, up to a possible finite group of additional surviving symmetries. More precisely, we can introduce a sublattice $\tilde V_{\mathbb{Z}}\subset V_{\mathbb{Z}}$ annihilated by the vector $\widehat{\bf K}=\{\widehat\sfK_i\}\in V^*_\mathbb{Z}$ and generated by $n_{\rm A}-1$ vectors $\tilde{\bm v}_a$. We then have
\be 
\langle \widehat{\bf K}, \tilde{\bm v}_a \rangle=0\,.
\ee 
Thus, we can identify a ``quasi-continuous'' set of zero-form symmetries preserved by the inclusion of the non-abelian sector by taking
\be\label{restralpha} 
{\bm\alpha}=\tilde\alpha^a \tilde{\bm v}_a\quad \text{mod}\  V_\mathbb{Z}\,,
\ee 
for any choice of $\tilde\alpha^a\in\mathbb{Q}$. Of course, there may also be other choices of ${\bm\alpha}$ satisfying \eqref{noBcond}, but these give only an additional finite set of possibilities. If $G$ is not simply connected, this finite set of preserved axion shift symmetries can be further enlarged by following \cite{Cordova:2022ieu}. However, these additional zero-form symmetries cannot be regarded as quasi-continuous and therefore provide much weaker constraints on the generation of potential terms.

In supersymmetric models, the constants $\widehat\sfK_i$ identify the saxionic contribution to the kinetic terms:
\be\label{nonaFFsax} 
-\frac1{8\pi}\widehat\sfK_i\int s^i \tr\left(F\wedge * F\right)\,.
\ee
Requiring this kinetic term to have the correct sign gives the condition $\langle\widehat{\bf K},{\bm s}\rangle\equiv\widehat\sfK_i s^i>0$. Recalling \eqref{saxcone}, we conclude that $\widehat{\bf K}\in \calc_{\rm I}$. This imposes further restrictions on the possible zero-form axion shift symmetries that are not broken by the non-abelian sector. In particular, we may try to restrict to saxionic directions that do not couple to the non-abelian sector. This is possible if we can formally identify the restricted saxionic cone $\tilde\Delta$ with a codimension-one facet of $\Delta$ orthogonal to $\widehat{\bf K}$.

%%%%%%%%%%%%%%%%%%%%%%%%%%%%%%%%%%%%%%%%%%%%%%%%%%%%%%%%%%
\newpage
\bibliographystyle{jhep}
\bibliography{references}

\end{document}